\documentstyle[epsf]{l-aa.pr}

\def\ha{\relax \ifmmode {\rm H}\alpha\else H$\alpha$\fi}
\def\hi{\relax \ifmmode {\rm H\,{\sc i}}\else H\,{\sc i}\fi}
\def\hii{\relax \ifmmode {\rm H\,{\sc ii}}\else H\,{\sc ii}\fi}

\def\muo{\relax \ifmmode \mu_0\else $\mu_0$\fi}
\def\mue{\relax \ifmmode \mu_{\rm e}\else $\mu_{\rm e}$\fi}
\def\re{\relax \ifmmode r_{\rm e}\else $r_{\rm e}$\fi}
\def\magarc{mag arcsec$^{-2}$}
\def\boundboxo{\epsfbox[30 190 538 540]}

\def\abst{
 The stellar and dust content of spiral galaxies as function of radius
has been investigated using near-infrared and optical broadband surface
photometry of 86 face-on spiral galaxies.  Colors of galaxies correlate
with the azimuthally averaged local surface brightness both within and
among galaxies, with the lower surface brightness regions being bluer. 
The colors formed from different passband combinations correlate
strongly indicating that they probably arise from the same physical
process. 

A 3D radiative transfer model was developed to calculate the effect of
dust absorption and scattering on the luminosity and color profiles of
galaxies.  Stellar synthesis models were used to investigate the effects
of the star formation history and the metallicity on the broadband color
profiles.  Combining all optical and near-infrared data shows that the
color gradients in this sample of face-on galaxies are best explained by
a combined stellar age and metallicity gradient across the disk, with
the outer regions being on average younger and of lower metallicity. 
Dust reddening probably plays only a minor role, as the dust models
cannot produce reddening profiles that are compatible with the
observations. 

The observed color differences implicate substantial $M/L_\lambda$
differences, both within galaxies and among galaxies.  The variations
are such that the ``missing light'' problem derived from rotation
fitting becomes even worse. Late-type galaxies (T$\ge$6) have lower
metallicities and are often of younger average age than earlier types
and have therefore an entirely different $M/L_\lambda$ in most
passbands. The near-infrared passbands are recommended for studies where the
$M/L_\lambda$ ratios should not vary too much. 

\keywords{dust, extinction -- Galaxies: evolution --Galaxies:
fundamental parameters -- Galaxies: photometry -- Galaxies: spiral --
Galaxies: stellar content}
}

\begin{document}

 \thesaurus{03(09.04.1; 11.05.2; 11.06.2; 11.16.1; 11.19.2; 11.19.5)}
 \title{
Near-infrared and optical broadband surface photometry of 86 face-on
disk dominated galaxies. 
 \thanks{Based on observations with the Jacobus Kapteyn Telescope and the
Isaac Newton Telescope operated by the Royal Greenwich Observatory at
the Observatorio del Roque de los Muchachos of the Instituto de
Astrof\'{\i}sica de Canarias with financial support from the PPARC (UK) and NWO
(NL) and with the UK Infrared Telescope at Mauna Kea operated by the
Royal Observatory Edinburgh with financial support of the PPARC.
\newline
The luminosity/color profiles of all galaxies are available in
electronic form at the CDS via anonymous ftp 130.79.128.5 as part of Paper~I.
 }
} 
 \subtitle{IV. Using color profiles to study stellar and dust content of galaxies.}
 \author{Roelof S.~de Jong \inst{ } }
 \offprints{
R.S.~de Jong, University of Durham, Dept.~of
Physics, South Road, Durham DH1 3LE, United Kingdom, e-mail:
R.S.deJong@Durham.ac.uk} 
 \institute{
Kapteyn Astronomical Institute, P.O.Box 800, NL-9700 AV
Groningen, The Netherlands}
 \date{received, accepted}
 \maketitle
 \markboth{R.S.~de Jong}{Near-IR and
 optical broadband observations of 86 spirals. IV Color profiles}

\section{Introduction}
\label{Intro4}

For many years broadband colors have been used to obtain a basic insight
into the contents of galaxies.  Broadband photometry is relatively easy
to obtain and gives an immediate impression of the spectral energy
distribution (SED) of an object.  Broadband colors are particularly
efficient when used for statistical investigations such as this one. 
Colors have been used to estimate the stellar populations of galaxies
(e.g.\ Searle et al.~\cite{Sea73}; Tinsley~\cite{Tin80};
Frogel~\cite{Fro85}; Peletier~\cite{PelPhD}; Silva \&
Elston~\cite{SilEls94}) and it has been suggested that colors can give
information about the dust content of galaxies (Evans~\cite{Eva94};
Peletier et al.~\cite{Pel94}, \cite{Pel95}).  In this paper I use radial color
profiles to investigate the stellar and dust content of galaxies. 

The problem of determining the stellar content of galaxies from
integrated SEDs has been approached from two sides, often called the
empirical and the evolutionary approach (for a review, see
O'Connell~\cite{OCon87}).  In the first method, stellar SEDs are fitted
to the observed galaxy SEDs (Pickles~\cite{Pic85};
Peletier~\cite{PelPhD}).  This method works only if one has spectral
(line) information.  Generally, the broadband colors of a galaxy can be
explained by a combination of the SEDs of two or three types of stars
(Aaronson~\cite{Aar78}; Bershady~\cite{Ber93}).  In the second, more
theoretical approach, stellar SEDs are combined, using some knowledge of
initial conditions and evolutionary time scales of different stellar
populations, to produce evolutionary stellar population synthesis models
(for reviews Tinsley~\cite{Tin80}; Renzini \& Buzzoni~\cite{RenBuz86};
more recent models are e.g.\ Buzzoni~\cite{Buz89}; Bruzual \&
Charlot~\cite{BruCha96}; Worthey~\cite{Wor94}). 

The papers of Disney et al.~(\cite{DDP89}) and Valentijn~(\cite{Val90})
have renewed the debate on whether spiral galaxies are optically thick
or thin. Broadband colors of galaxies can be used to examine this
problem, because the dependence of dust extinction on wavelength causes
reddening. This can be used to measure extinction at a
certain point through the disk using a galaxy or another object behind it
(Andredakis \& van der Kruit~\cite{AndKru92}) or to measure extinction
within a galaxy, for instance across a spiral arm dust lane (Rix \&
Rieke~\cite{RixRie93}; Block et al.~\cite{Blo94}). To measure the
global dust properties of a galaxy by reddening one can use the color
profile. If one assumes that dust is more concentrated towards the
center (just like the stars), the higher extinction in the center
produces a color gradient that makes galaxies redder inwards
(Evans~\cite{Eva94}; Byun et al.~\cite{Byun94}).

Most previous studies investigating stellar population and dust
properties of galaxies used their integrated broadband colors.  The use
of surface photometry colors is less common, as it is easier to compare
integrated photometry than surface photometry for large samples of
galaxies.  Integrated photometry samples the bulk properties of
galaxies, but because the light distribution of galaxies is strongly
concentrated, one effectively measures the colors of the inner regions
of galaxies.  The half total light radius of an exponential disk is
$\sim$1.7 scalelengths, while luminosity profiles are easily traced out
to 4-6 scalelengths.  Therefore, half of the light in integrated colors
comes from an area that is less than $1/5$ of the area commonly observed
in galaxies (say within $D_{25}$). 

Our knowledge of the star formation history (SFH) and the dust content
of galaxies improves when we start looking at local colors instead of
integrated colors.  A first improvement is obtained by using the radial
color distribution (i.e.\ the color profile) of a galaxy.  This has been
common practice for elliptical galaxies (e.g.\ Peletier et
al.~\cite{Pel90a}; Goudfrooij et al.~\cite{Gou94}), but not for spiral
galaxies, because elliptical galaxies are assumed to have a simple SFH
and low dust content (but see Goudfrooij~\cite{GouPhD}) opposed to
spirals.  Color studies of spiral galaxies have been concentrated on
edge-on systems, in the hope to be better able to separate the dust and
stellar population effects (e.g.~Just et al.~\cite{Jus96}).  Even more
detailed information about galaxies can be obtained by the use of
azimuthal profiles (Schweizer~\cite{Sch76}; Wevers et al.~\cite{Wev86})
and color maps, but these techniques require high resolution, high
signal-to-noise observations and are hard to parameterise to global
scales, which means that they cannot be used in statistical studies. 

Due to the large variety of galaxies, statistical studies of galaxies
require large samples.  The introduction of CCDs into astronomy made it
possible to obtain for large samples of galaxies accurate optical
surface photometry in reasonable observing times (Kent~\cite{Kent84}). 
Very large data sets of CCD surface photometry have recently become
available (Cornell~\cite{Cor87}; Han~\cite{Han92}; Mathewson et
al.~\cite{Mat92}; Giovanelli et al.~\cite{Gio94}).  Unfortunately, most
of these samples are observed in only one or two passbands. 
Furthermore, the surface photometry is often reduced to integrated
magnitudes and isophotal diameters to study extinction effects with an
inclination test or to study the Tully-Fisher relation (Tully \& Fisher
\cite{TulFis77}, hereafter TF-relation).  Fast plate measuring machines
have also produced surface photometry of large sets of galaxies
(e.g.~Lauberts \& Valentijn~\cite{ESO-LV}, hereafter ESO-LV), but again
only in one or two passbands. 

Since near-infrared (near-IR) arrays have become available only in the
late eighties, near-IR surface photometry is available for only a few
somewhat larger samples of spiral galaxies.  Terndrup et
al.~(\cite{Ter94}) observed 43 galaxies in $J$ and $K$, which was
complemented with $r$ passband photometry of Kent~(\cite{Kent84},
\cite{Kent86}, \cite{Kent87}).  They explained the observed colors
mainly by population synthesis and invoked dust only for the reddest
galaxies.  Peletier et al.~\cite{Pel94} imaged 37 galaxies in the
$K$-passband and combined their data with the photometry of the ESO-LV
catalog. They explained their surface photometry predominantly in terms of
dust distributions and concluded that spiral galaxies are optically
thick in the center in the $B$ passband, under the assumption that there
are no population gradients across the disk. 

There are two sets of observations that allow a direct physical
interpretation of color gradients in spiral galaxies: 1) The current
star formation rate (SFR) as measured by the \ha\ flux has a larger
scalelength than the underlying older stellar population (Ryder \&
Dopita~\cite{RydDop94}). There are relatively more young stars in the
outer regions of spiral galaxies than in the central regions. This will
be reflected in broadband colors of spiral galaxies. 2) From
metallicity measurements of \hii\ regions it is known that there are
clear metallicity differences in the gas among different galaxies and
that there are metallicity gradients as function of radius within
galaxies (Villa-Costas \& Edmunds~\cite{VilEdm92}; Zaritsky et
al.~\cite{Zar94}). If the metallicity gradients in the gas are also
(partly) present in the stellar components, the effects might be
observable in the broadband colors. Stellar population synthesis
models incorporating both age and metallicity effects are needed in the
comparison with observations of radial color gradients.

Broadband photometry is often assumed to trace baryonic mass, and the
transformation from light to mass is performed by postulating a
mass-to-light ratio ($M/L_\lambda$).  Both dust extinction and
differences in stellar populations will influence $M/L_\lambda$ ratios,
most notably in the bluer optical passbands.  Color differences, among
galaxies and locally within galaxies, will translate in different
$M/L_\lambda$ values; one can expect that this will influence studies
involving rotation curve fitting and the TF-relation. 

In this paper I concentrate on the use of color profiles as a diagnostic
tool to investigate dust and stellar content of spiral galaxies.  Other
processes that may contribute to the broadband colors (e.g.\ emission
from hot dust in the $K$ passband) are ignored as they are expected to
be small in most cases.  The structure of this paper is as follows.  In
Sect.~\ref{data4} the data set is described and the color profiles of
the 86 spiral galaxies using the $B, V, R, I$ and $K$ passband data are
presented.  Section~\ref{colprofsec} describes the extinction models and
the stellar population models used in this paper and then compares these
models to the data.  In Sect.~\ref{struccol}, I investigate the relation
between the color properties of the galaxies and the structural galaxy
parameters derived in the previous papers of this series.  Implications
of the current measurements are discussed in Sect.~\ref{discus4} and the
paper is summarized in Sect.~\ref{concl4}.

\section{The data}
 \label{data4}

\begin{figure*}
\def\epbox{\epsfxsize=8.6cm\epsfbox[60 452 560 768]}
\mbox{\epbox{pcolsku508n2.tbl.ps}}
\mbox{\epbox{pcolsku628.tbl.ps}}
\mbox{\epbox{pcolsku1305.tbl.ps}}
\mbox{\epbox{pcolsku1455.tbl.ps}}
\mbox{\epbox{pcolsku1551.tbl.ps}}
\mbox{\epbox{pcolsku1559.tbl.ps}}
\mbox{\epbox{pcolsku1577.tbl.ps}}
\mbox{\epbox{pcolsku1719.tbl.ps}}
\caption[]{
 Observed color profiles for 8 out of the 86 galaxies.  The dashed lines
indicate the maximum error due wrong sky background subtraction, the
arrow indicates one disk scalelength in the $K$ passband. 
\label{colprof}
 }
\end{figure*}
%======================================================================

In order to examine the parameters describing the global structure of
spiral galaxies, 86 face-on systems were observed in the $B, V, R, I, H$
and $K$ passbands.  A full description of the observations and data
reduction can be found in de Jong \& van der Kruit (\cite{deJ1},
hereafter Paper~I) of which only the essentials are repeated here.  The
galaxies in this statistically complete sample of undisturbed spirals
were selected from the UGC (Nilson \cite{Nilson}) to have red diameters
of at least 2\arcmin\ and axis ratios larger than 0.625.  The galaxies
were imaged along the major axis with a GEC CCD on the 1m Jacobus
Kapteyn Telescope at La Palma in the $B, V, R$ and $I$ passbands and
with a near-IR array on the United Kingdom Infra-Red Telescope at Hawaii
in the $H$ and $K$ passbands.  Standard reduction techniques were used
to produce the images, which were calibrated using globular cluster
standard star fields.  The sky brightness was determined outside the
galaxy in areas free of stars and its uncertainty constitutes one of the
main sources of error in the derived parameters. 

\begin{figure*}
 \mbox{\epsfxsize=18.5cm\epsfbox[85 450 580 780]{mucolrbk.ps}}
\caption[]{
 The average $B$--$K$ colors of all galaxies at different radii as
function of the azimuthally averaged $R$ passband surface brightness
measured at these radii.  The galaxies are divided into the four
indicated RC3 morphological type index (T) bins.  The dashed lines give
a common reference in all four bins, but have no physical meaning.  The
lines have a $B$--$K$ color gradient of 1/7 mag per $R$-\magarc. 
 }
 \label{mucol}
 \end{figure*}
%======================================================================

The ellipticity and position angle (PA) of each galaxy were determined
from the $R$ passband image at an outer isophote (typically at 24
$R$-\magarc).  The radial surface brightness profiles were determined
for all passbands by calculating the average surface brightness on
elliptical annuli of increasing radius using the previously determined
ellipticity and PA.  This method ensures that the luminosity profile
reflects the average surface brightness at each radius, independent of
passband.  Free ellipse fitting at each surface brightness interval
(e.g.~Kent~\cite{Kent84}) will give a disturbed representation of the
radial luminosity distribution in face-on galaxies, due to spiral arms,
bars and bright \hii\ regions.  The surface brightness profiles were
used to calculate the integrated luminosity of the galaxies.  Internal
and external comparisons showed that the derived parameters are well
within the estimated errors. 

The decomposition of the light of the galaxies into its fundamental
components (bulge, disk and sometimes a bar) is described by de Jong
(\cite{deJ2}, hereafter Paper~II). An exponential light distribution
was assumed for both the bulge and the disk and these were fitted to
the full 2D image. An extensive error analysis of the determination of
the fundamental galaxy parameters was performed and this revealed that
the dominant source of error is the uncertainty in the sky background.
 
The color profiles of the galaxies were calculated by subtracting the
radial surface brightness profiles of the different passbands from one
another.  Some typical profiles are presented in Fig.~\ref{colprof}, the
profiles of all galaxies can be found in de Jong~(\cite{deJPhD}) and are
available in electronically readable format.  The dashed lines indicate
the maximum errors due to the uncertainty in the sky surface brightness. 
One should be cautious in interpreting the colors in the inner few
seconds of arc, because the profiles were not corrected for the
differences in seeing (Paper~I) between the different passbands. 

A quick inspection shows that almost all galaxies show color gradients.
They become bluer going radially outward, even when taking the sky
background subtraction uncertainties into account.  The color gradients
extend over several disk scalelengths. Note that bulges leave no clear
signature in the color profiles. From the color profiles alone one can
not tell which part is bulge dominated and which part is disk dominated.

The profiles in Fig.~\ref{colprof} are the observed profiles.  The
corrections needed to translate observed quantities into more physical
quantities are discussed by de Jong (\cite{deJ3}, hereafter Paper~III). 
In the remainder of this paper, only the photometric observations are
used, corrected for Galactic extinction using the precepts of Burstein
and Heiles (\cite{BurHei84}) and the extinction curve of Rieke and
Lebofsky (\cite{RieLeb85}).

\section{Color gradients}
 \label{colprofsec}

The color gradients of Fig.~\ref{colprof} have been put on a common
scale in Fig.~\ref{mucol}, where I have plotted for all galaxies the
average $B$--$K$ color at each radius as function of the azimuthally averaged
$R$ passband surface brightness at the corresponding radii.  The galaxies
are divided in four bins based on their morphological type, using the
RC3 (de Vaucouleurs et al.~\cite{rc3}) type indices T (see also Papers~I
and III).  There is a clear correlation between average surface
brightness at a radius and the average color at that radius; the lower
surface brightness regions are bluer.  This indicates the relation
between Hubble type, surface brightness and integrated color: since
late-type galaxies have on average a lower central surface brightness
(Paper~III), they are bluer.  This is not the whole story, since for
each morphological type at each surface brightness there is considerable
scatter.  Furthermore, even at the same average surface brightness,
late-type galaxies are on the average bluer than early-type galaxies. 

The two most straightforward explanations for the color gradients are 1)
radial changes in stellar populations and 2) radial variations in
reddening due to dust extinction.  For both possibilities, I investigate
a range of models to limit the acceptable parameter space.  The
extinction models have a range in relative distributions of dust and
stars. The colors of the stellar synthesis population models depend on
the star formation history (SFH) and the metallicity of the stars.

The colors and color gradients of the galaxies formed from the different
passband combinations are correlated and the models should be fitted in
a six-dimensional ``passband space''.  Predicting the right color
gradient in one combination of passbands, but a wrong one in an other
combination makes a model at best incomplete and therefore undesirable. 
Color--color plots will be used to show as much information as possible
in one plot.  The $H$ passband data are not shown, as the differences
between $H$ and $K$ predicted by the models (both the population and the
extinction models) are smaller than the measurement errors.  In the
remainder of this section I first discuss the extinction models and the
stellar population synthesis models used in this paper and then compare
the models with the data.

\subsection{Extinction models}
 \label{dusmod}

In this
section, I present my new dust models, show the predicted luminosity
profiles, color profiles and color--color diagrams, and compare the
results with existing dust models. 

\subsubsection{Modeling dust effects}

Numerous researches have investigated the effects of dust extinction on
the observed light distributions of galaxies.  The primary goal of most
of the studies is to investigate the inclination dependent effects of
the total magnitude of galaxies (e.g.\ Huizinga~\cite{HuiPhD}).  In some
studies the extinction effects on the observed (exponential) light
profile of galaxies is studied.  The most detailed are the Triplex
models by Disney, Davies \& Phillipps (\cite{DDP89}, hereafter DDP; see also
Huizinga~\cite{HuiPhD}; Evans~\cite{Eva94}). 

The effects of reddening on the observed colors and color profiles has
been examined in a number of studies.  The simplest model to predict the
reddening of a galaxy is to use directly the standard (Galactic)
extinction law, but this is of course a gross oversimplification.  DDP
have named this the Screen model, which has all dust placed between us and the
galaxy.  In reality the dust is mixed between the stars, so that on the
near side of the galaxy a considerable fraction of stars will be only
slightly obscured.  For the same amount of dust, the observed reddening
is considerably less than predicted by the Screen model, especially
since the most reddened stars are also the most obscured stars and
therefore the ones that contribute less to the overall color of the
system. 

As soon as the dust is mixed with the stars one has to take both
absorption and scattering into account.  Intuitively one expects that
for face-on galaxies at least as much light gets scattered into the line
of sight as out of it, especially since there are more photons traveling
in the plane of a galaxy which can be scattered into face-on directions
than the other way around.  As only the absorbed photons really
disappear, it is better to use relative absorption rather than relative
extinction between different passbands to estimate reddening effects in
face-on galaxies.  It is essential to incorporate both absorption and
scattering into extinction models to make accurate predictions of the
effects of dust on colors and color gradients of galaxies. 

A number of studies have investigated the effect of reddening on
integrated colors of galaxies (Bruzual et al.~\cite{Bru88}; Witt et
al.~\cite{Witt92} and references therein).  In these studies scattering
is included and stellar and dust distributions are used that allow
approximations to reduce computing time; e.g.\ Bruzual et al.\ use plane
parallel distributions and Witt et al.\ use spherically symmetric
distributions.  Color profiles produced by dust models are not often
presented.  Evans (\cite{Eva94}) investigates the effects of extinction
as function of radius in face-on galaxies for a non-scattering medium. 
Byun et al.~(\cite{Byun94}) also investigate the effects of dust on
luminosity and color profiles, using the method of Kylafis \& Bahcall
(\cite{KylBah87}).  Their method includes first order scattering and
approximates multiple scattering.  The results of Byun et
al.~\cite{Byun94} are compared with the results presented here in
Sect.~\ref{dusprofsect}. 

To estimate to what extent the color gradients can be attributed to
reddening by dust extinction, Monte Carlo simulations were made
of light rays traveling through a dusty medium. The models are
described in full detail in Appendix~\ref{apmodel}. 

The distributions of stellar light and dust in these models were
described by exponential laws in both the radial and vertical
directions.  In the radial direction these distributions were
parameterised by the scalelength of the stars ($h_{\rm s}$) and the dust
($h_{\rm d}$), and in vertical direction by the scaleheight of stars
($z_{\rm s}$) and dust ($z_{\rm d}$).  In all models I used $h_{\rm
s}/z_{\rm s}\!=\!10$ and for simplicity no bulge component was added to
the stellar light distribution. 

Since the effects of dust on the color profiles is the main interest of this
study, absolute calibration of the amount of starlight is
arbitrary.  Only the relative effect of dust from one passband to the
other is important.  The amount of dust in the models is parameterised
by {\em the} optical depth of a system, $\tau_{0,V}$, defined as the
optical thickness due to dust absorption and scattering in the $V$
passband through the disk from one pole to the other along the symmetry
axis (Eq.~(\ref{optdepth})).

\begin{figure}
 \mbox{\epsfxsize=8.6cm\boundboxo{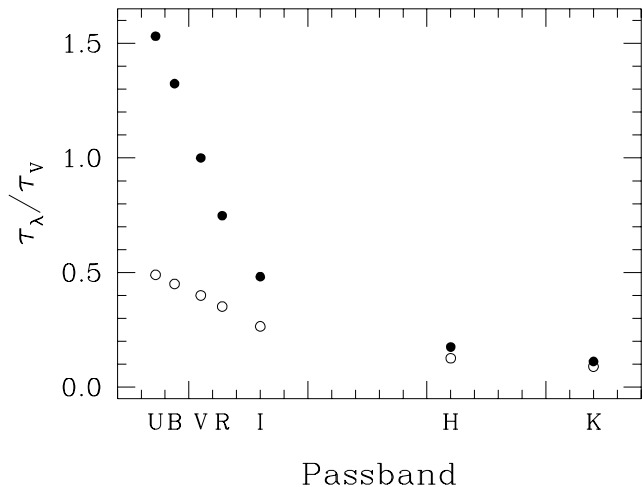}}
\caption[]{
 The relative extinction ($\tau_\lambda/\tau_V$) and the relative
extinction ($(1-a_\lambda)\tau_\lambda/\tau_V$) used in the dust models. 
 }
 \label{extabscur}
 \end{figure}
%======================================================================

Three dust properties were incorporated into the dust model to describe
the wavelength dependent effects of dust extinction: the relative
extinction ($\tau_\lambda/\tau_V$), the albedo ($a_\lambda$), and the
scattering asymmetry parameter ($g_\lambda$).  The relative extinction
was adopted from Rieke and Lebofsky (\cite{RieLeb85}); the other two
parameters were drawn from Bruzual et al.~(\cite{Bru88}).  These values
are listed in Table~\ref{dusprop} and Fig.~\ref{extabscur} shows the
relative extinction and the relative absorption
($(1-a_\lambda)\tau_\lambda/\tau_V$).  Note that for the optical
passbands the change in relative absorbtion is much smaller than the
change in relative extinction. 

{\tabcolsep=4mm
\begin{table}[t]
%\begin{center}
 \begin{tabular}{cccc}
\hline
\hline
\ \ passband & $\tau_\lambda/\tau_V$ & $a_\lambda$ & $g_\lambda$\\
\hline
$U$     & 1.531 & 0.68  & 0.67\\
$B$	& 1.324	& 0.66	& 0.59\\
$V$	& 1.000 & 0.60	& 0.50\\
$R$	& 0.748 & 0.53	& 0.40\\
$I$	& 0.482 & 0.45	& 0.29\\
$H$	& 0.175 & 0.28	& 0.04\\
$K$	& 0.112	& 0.20	& 0.00\\
\hline
\hline
\end{tabular}
 \caption[]{Values used in the scattering model for the dust properties
relative extinction ($\tau_\lambda/\tau_V$), albedo ($a_\lambda$) and
scattering asymmetry ($g_\lambda$) as function of photometric passband. }
\label{dusprop}
%\end{center}
\end{table}
}
%======================================================================

Before presenting the model results, a few words of caution are in
order.  First, extragalactic dust properties are poorly known.  There
are only a few measurements of extinction laws in extragalactic systems
(e.g.\ Knapen et al.~\cite{Kna91}; Jansen et al.~\cite{Jan94}) other
than for the Magellanic Clouds (see Mathis~\cite{Mat90} for references). 
All measurements seem to be consistent with the Galactic extinction law,
except for a few measurements in the Small Magellanic Cloud.  It is well
known that the Galactic extinction curve is not the same in all
directions, but the one adopted here is appropriate for the diffuse
interstellar medium (for discussion see Mathis~\cite{Mat90}).  The
parameters $a_\lambda$ and $g_\lambda$ have never been measured in
extragalactic systems and are poorly known even for our own Galaxy.  The
adopted values for these parameters stem, especially for the longer
wavelengths, from model calculations.  Still, no large variations are
expected in the extinction properties, unless the dust in other galaxies
is made of totally different material (see also the discussion in
Bruzual et al.~\cite{Bru88}). 

As a second word of caution, the models presented here describe only
smooth diffuse dust.  The effects of non-homogeneous dust distributions
should be considered.  A large ensemble of optically thick clouds has
only a reddening effect if the clouds have a large filling factor, but
such a configuration becomes comparable to the presented models with
high $\tau_{0,V}$.  The reddening effect of a clumpy medium will be
smaller than the effect predicted by the diffuse dust models for the
same amount of dust, but the direction of the reddening vectors will be
the same as long as the dust properties in the clouds are more or less
the same as used here.  If clouds are optically thick at all wavelengths
one has the case of gray dust and no color gradients at all.  Model
calculations using a clumpy dust medium in the absence of scattering are
presented in Huizinga (\cite{HuiPhD}, Chapter~5). 

As a final word of caution, a young population of stars probably has a
smaller scaleheight than an old population of stars.  It might be more
appropriate to use a smaller stellar scaleheight in the blue than in the
near-IR.  The relative contributions from young and old populations are
difficult to estimate however, and for simplicity one stellar
scaleheight is used for all passbands.  These models do not include the
dust shells around the extremely luminous stars in the final stages of
their life.  Even though such shells will make these stars redder, they
will not produce a radial effect (unless the shell properties depend on
galactic radius).  Effectively these shells will only make the total
underlying population redder at all radii and they are of no further
concern here. 

\subsubsection{Resulting profiles}
\label{dusprofsect}
\def\xsize{5.64cm}
\begin{figure*}
\mbox{\epsfxsize=\xsize\epsfbox[0 190 515 660]{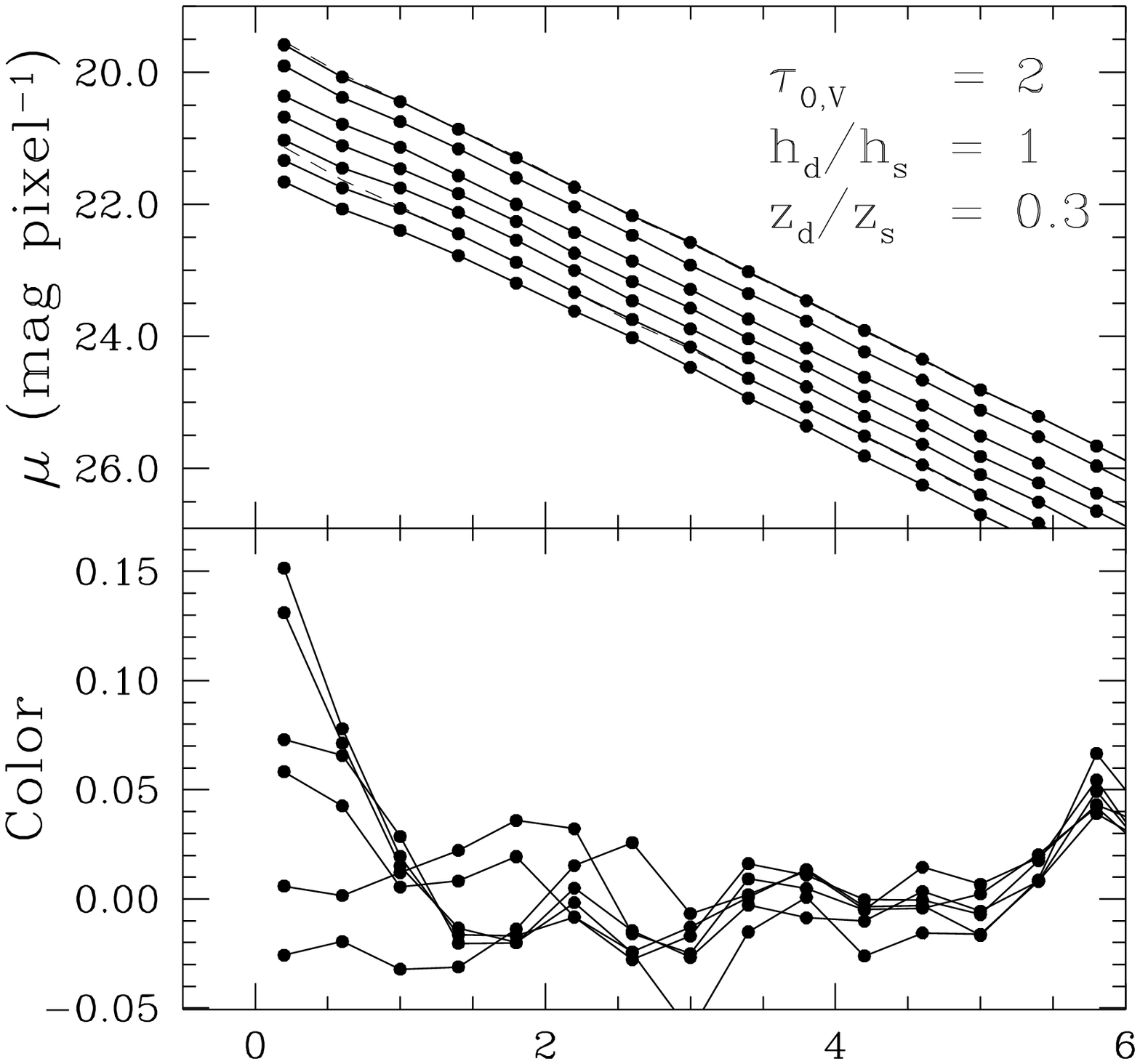}}
\mbox{\epsfxsize=\xsize\epsfbox[0 190 515 660]{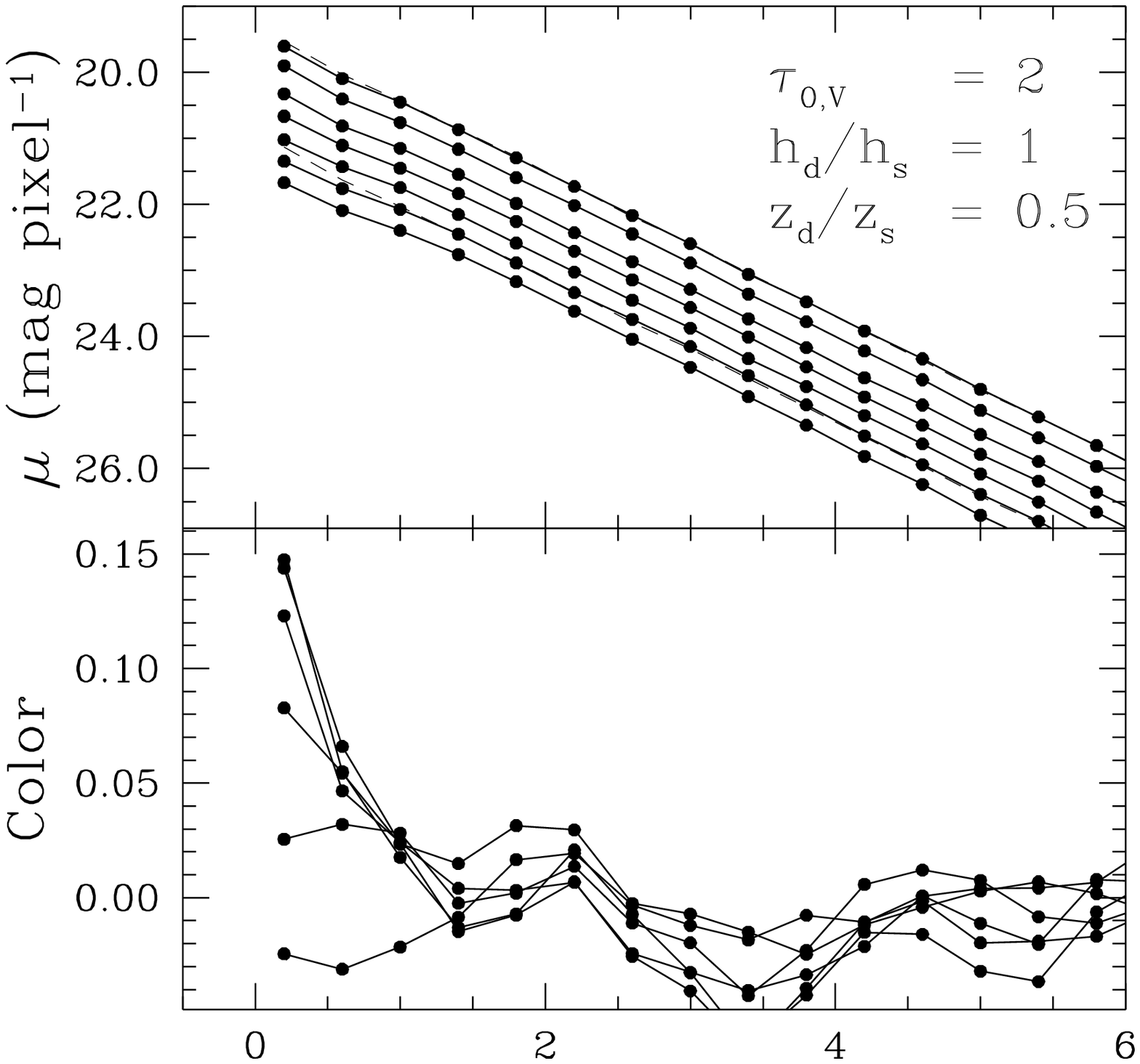}}
\mbox{\epsfxsize=\xsize\epsfbox[0 190 515 660]{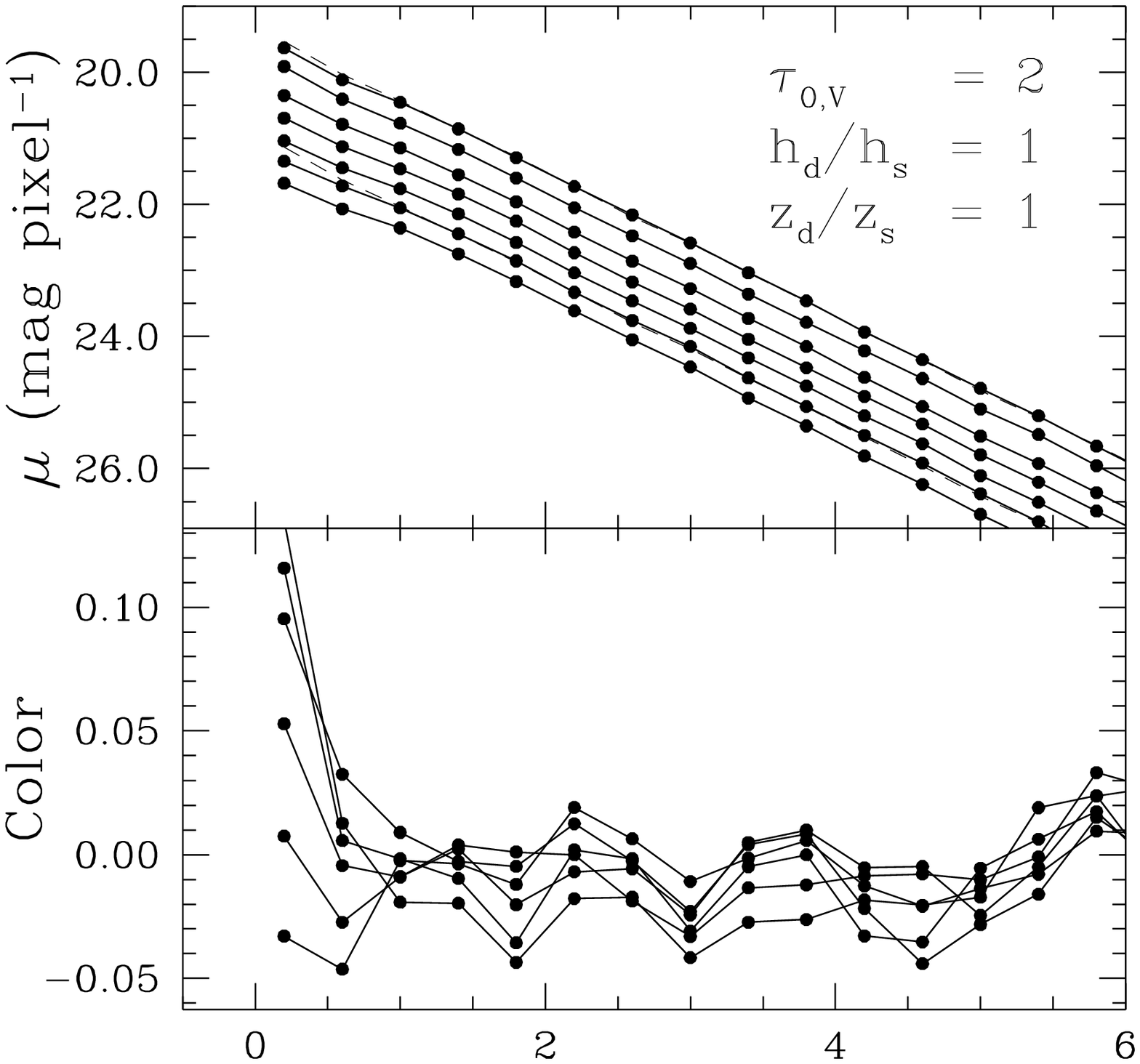}}

\mbox{\epsfxsize=\xsize\epsfbox[0 190 515 660]{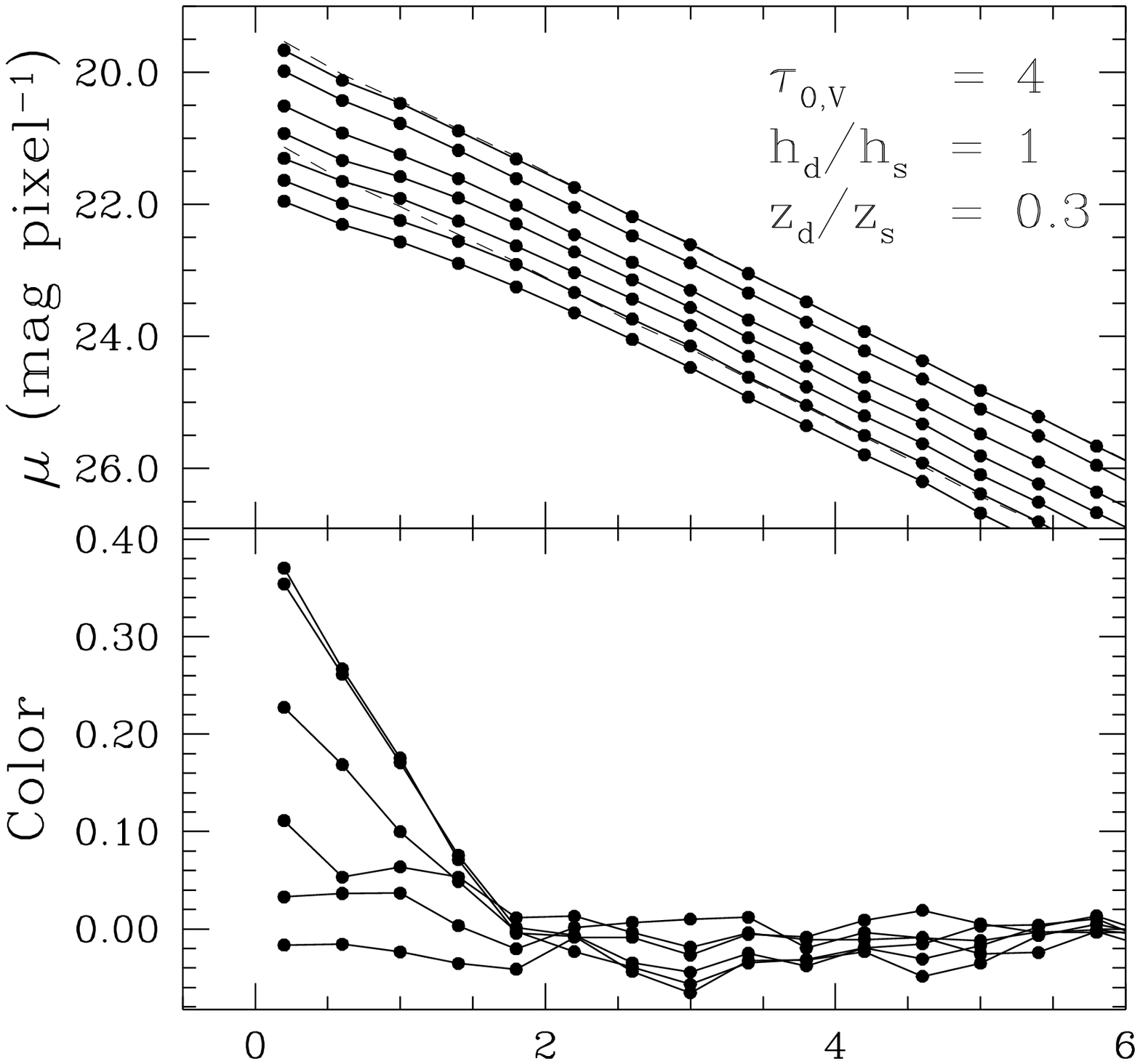}}
\mbox{\epsfxsize=\xsize\epsfbox[0 190 515 660]{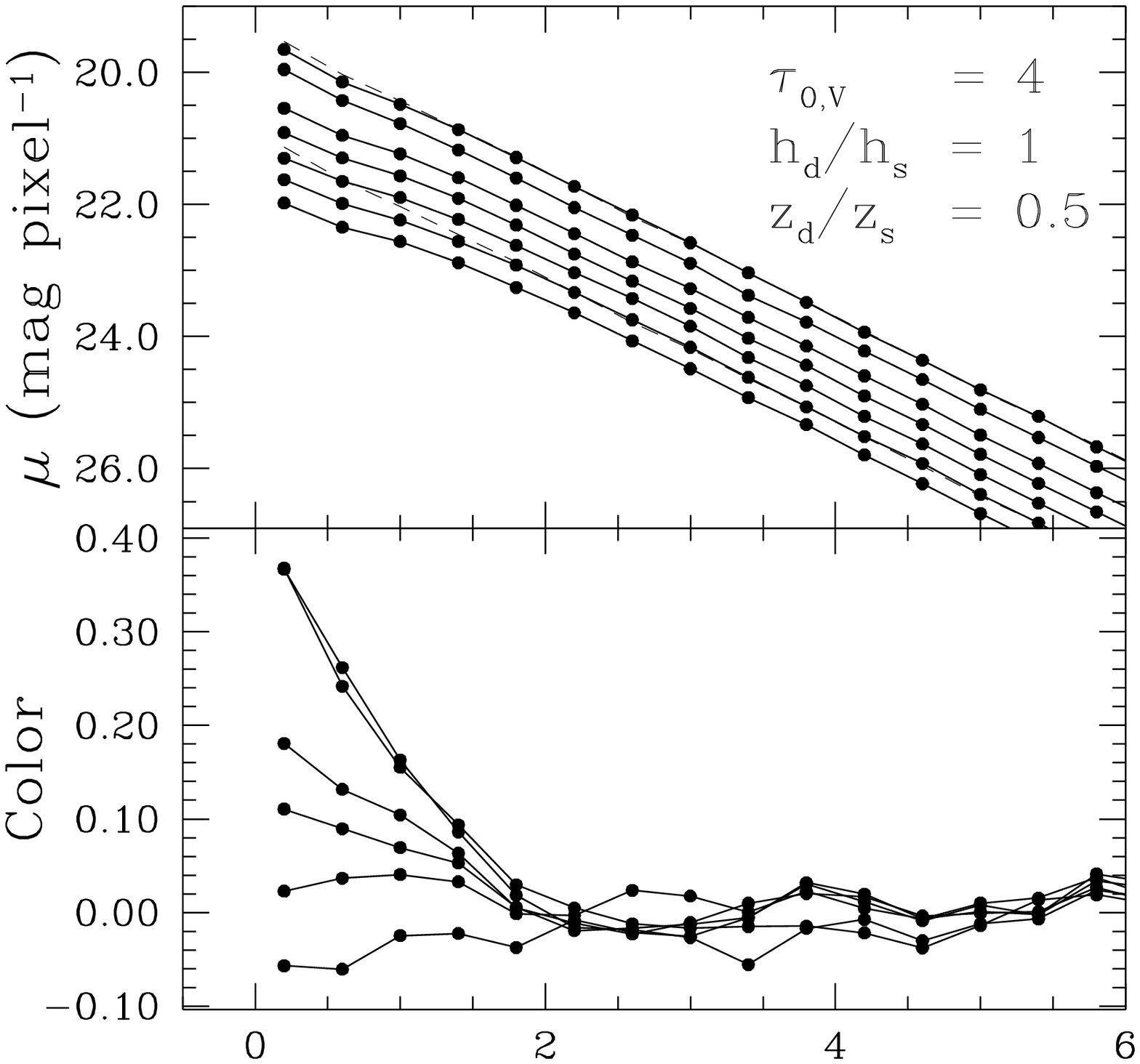}}
\mbox{\epsfxsize=\xsize\epsfbox[0 190 515 660]{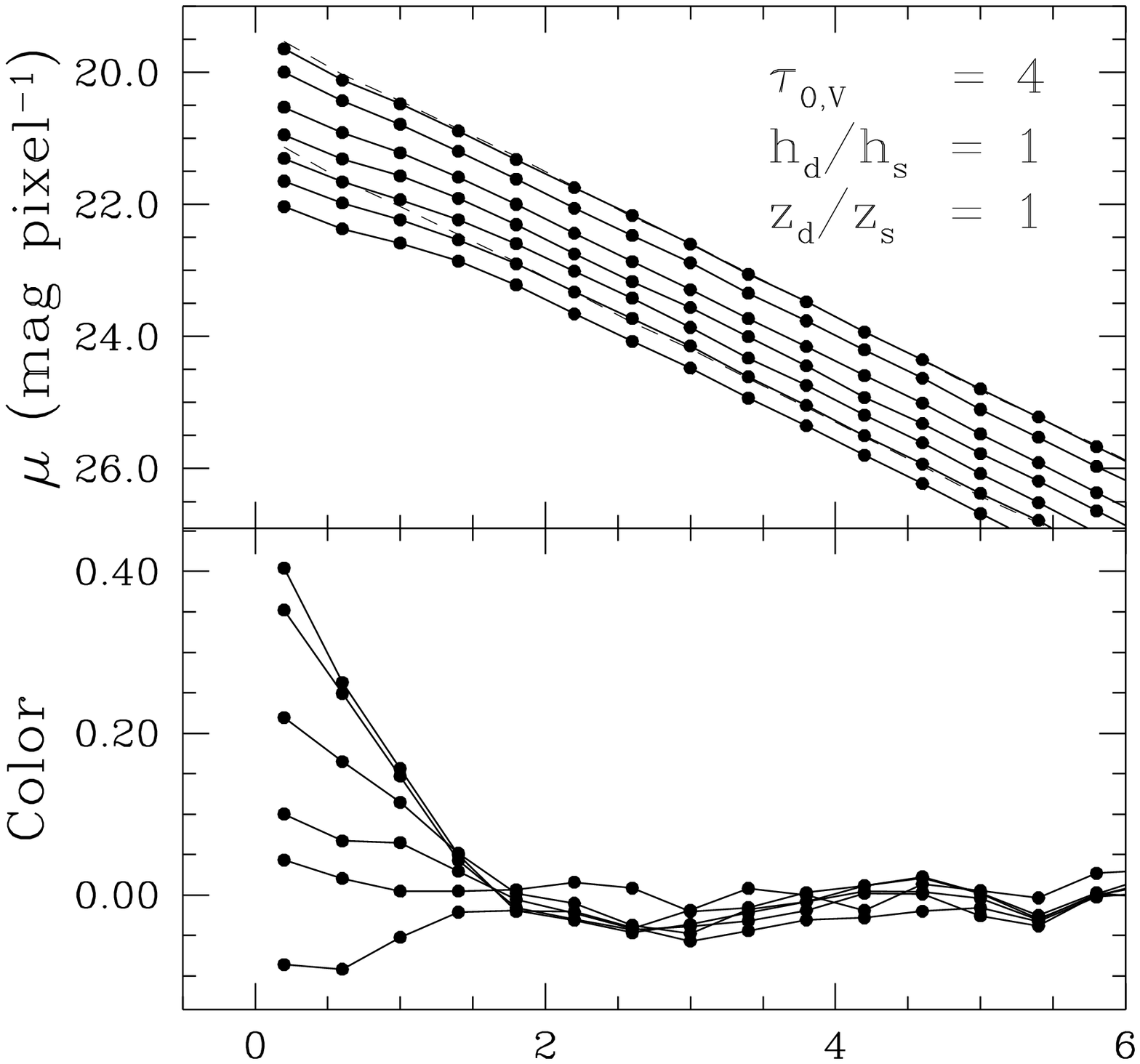}}

\mbox{\epsfxsize=\xsize\epsfbox[0 190 515 660]{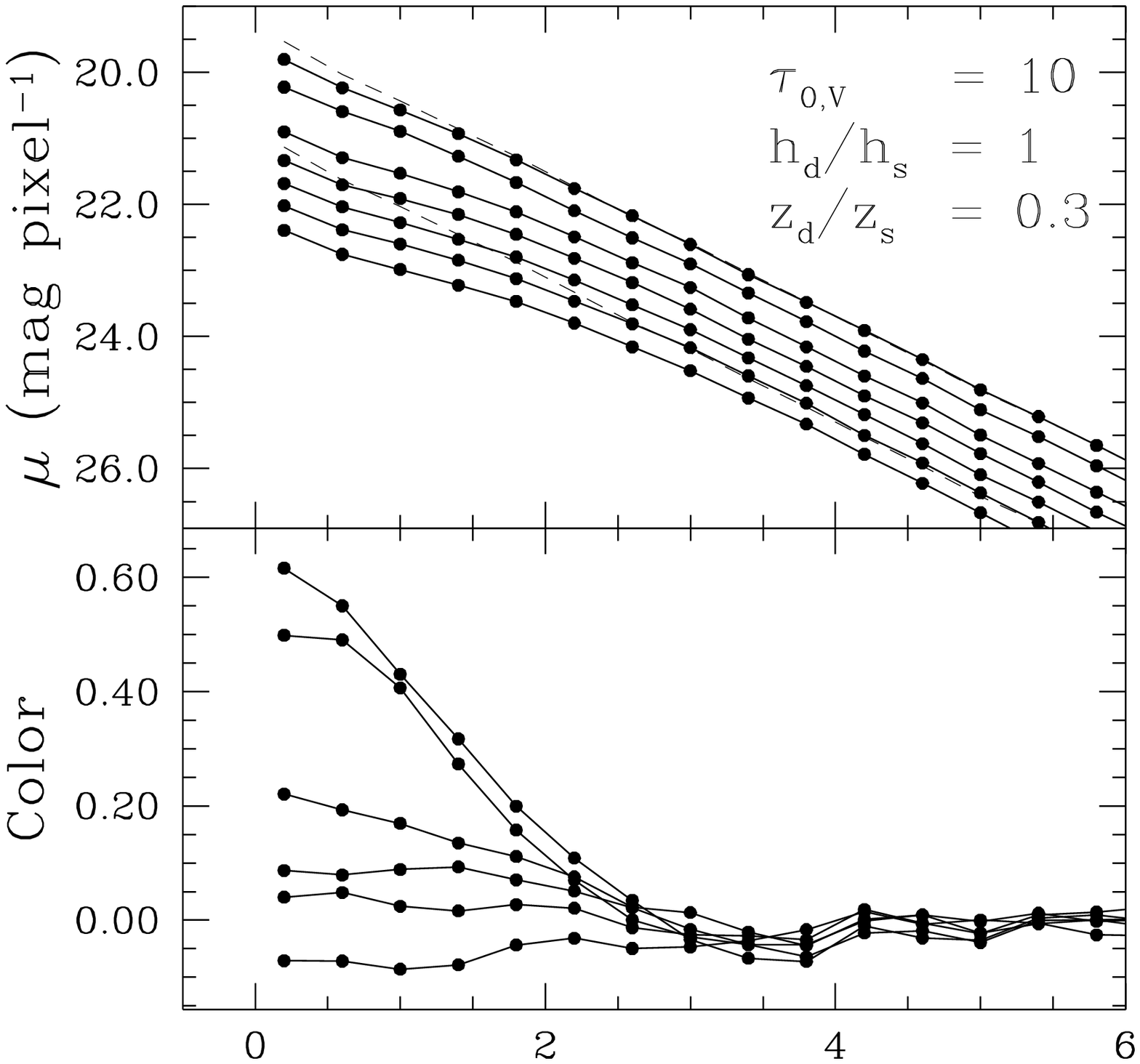}}
\mbox{\epsfxsize=\xsize\epsfbox[0 190 515 660]{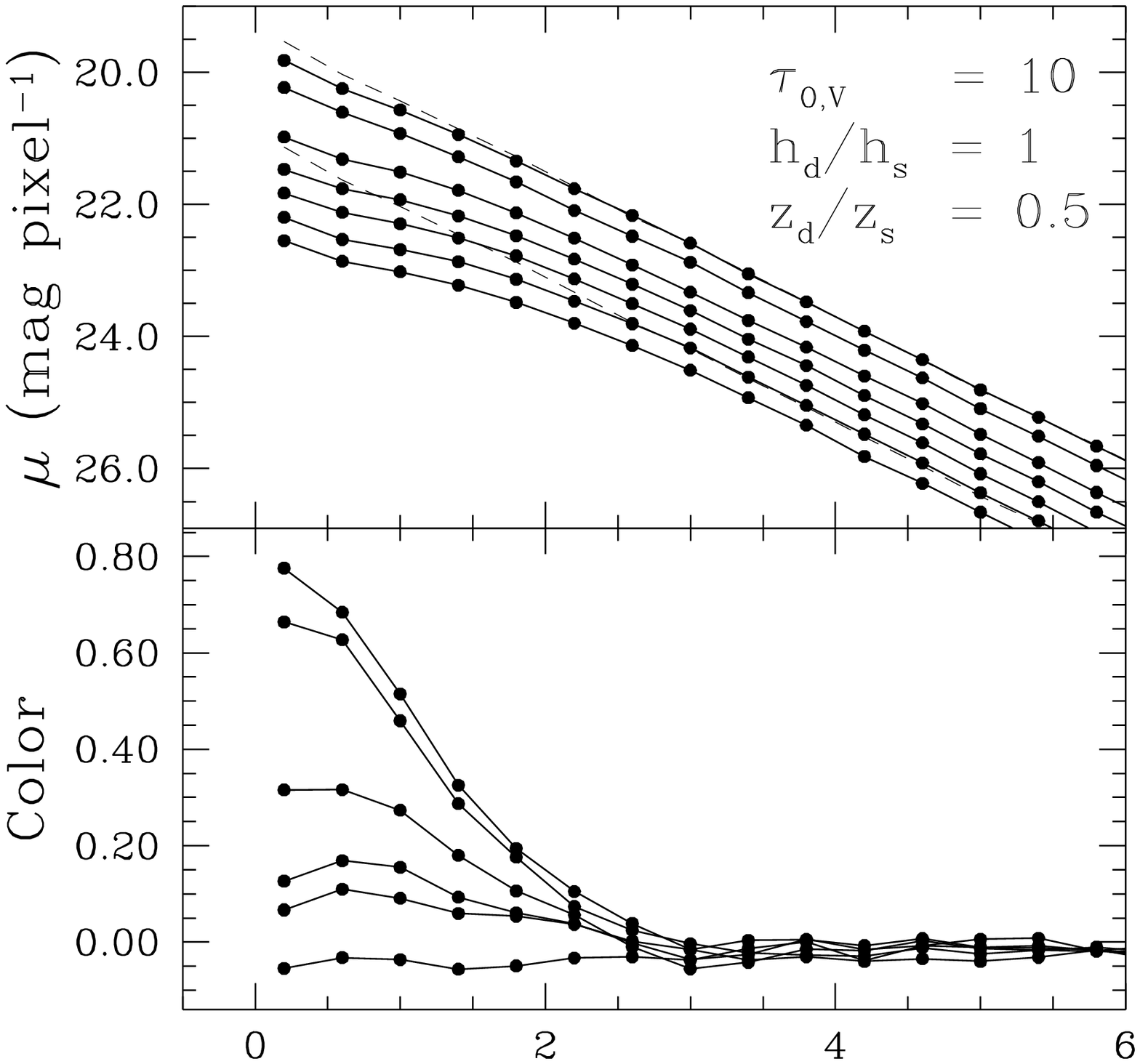}}
\mbox{\epsfxsize=\xsize\epsfbox[0 190 515 660]{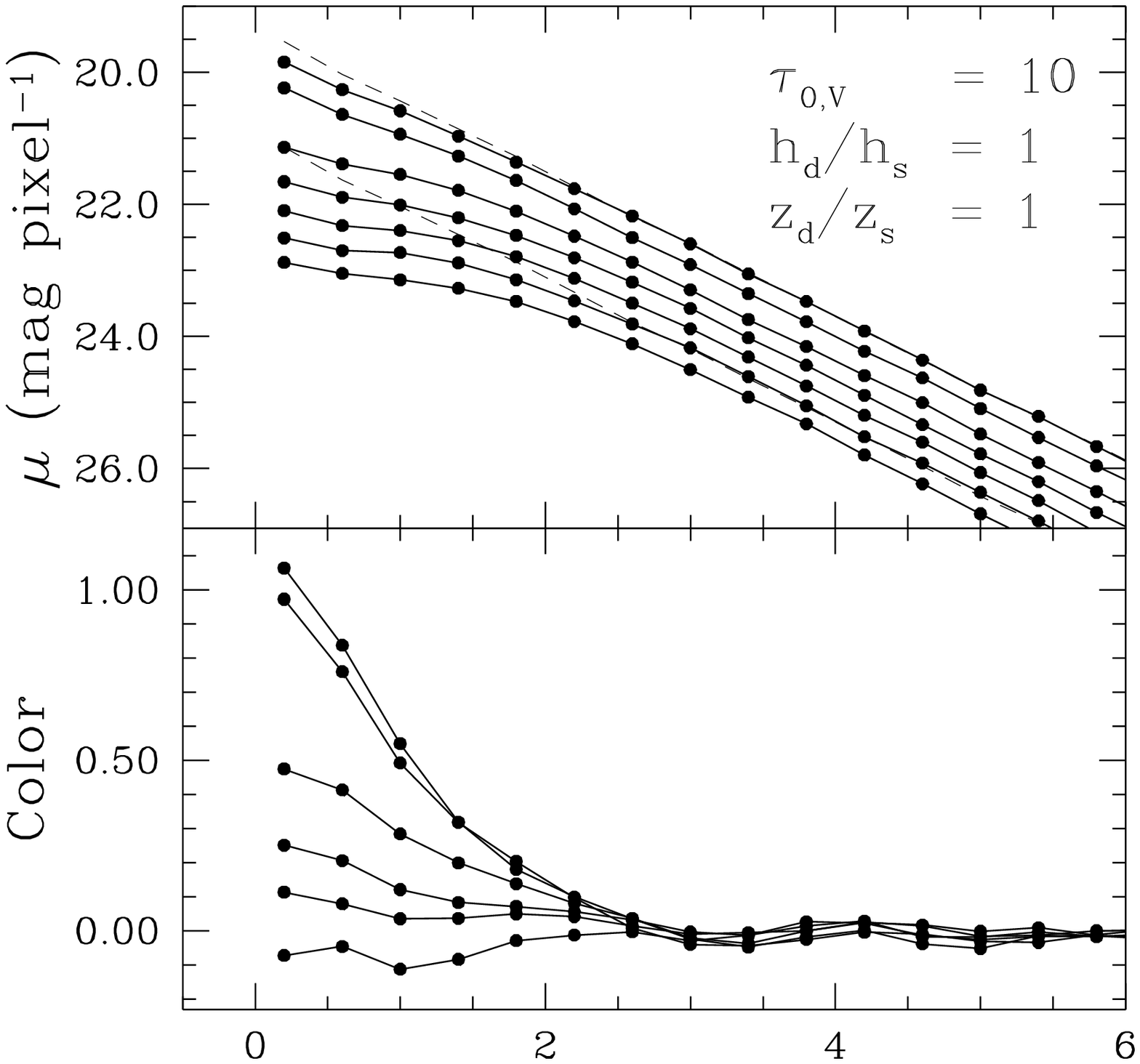}}

\mbox{\epsfxsize=\xsize\epsfbox[0 190 515 660]{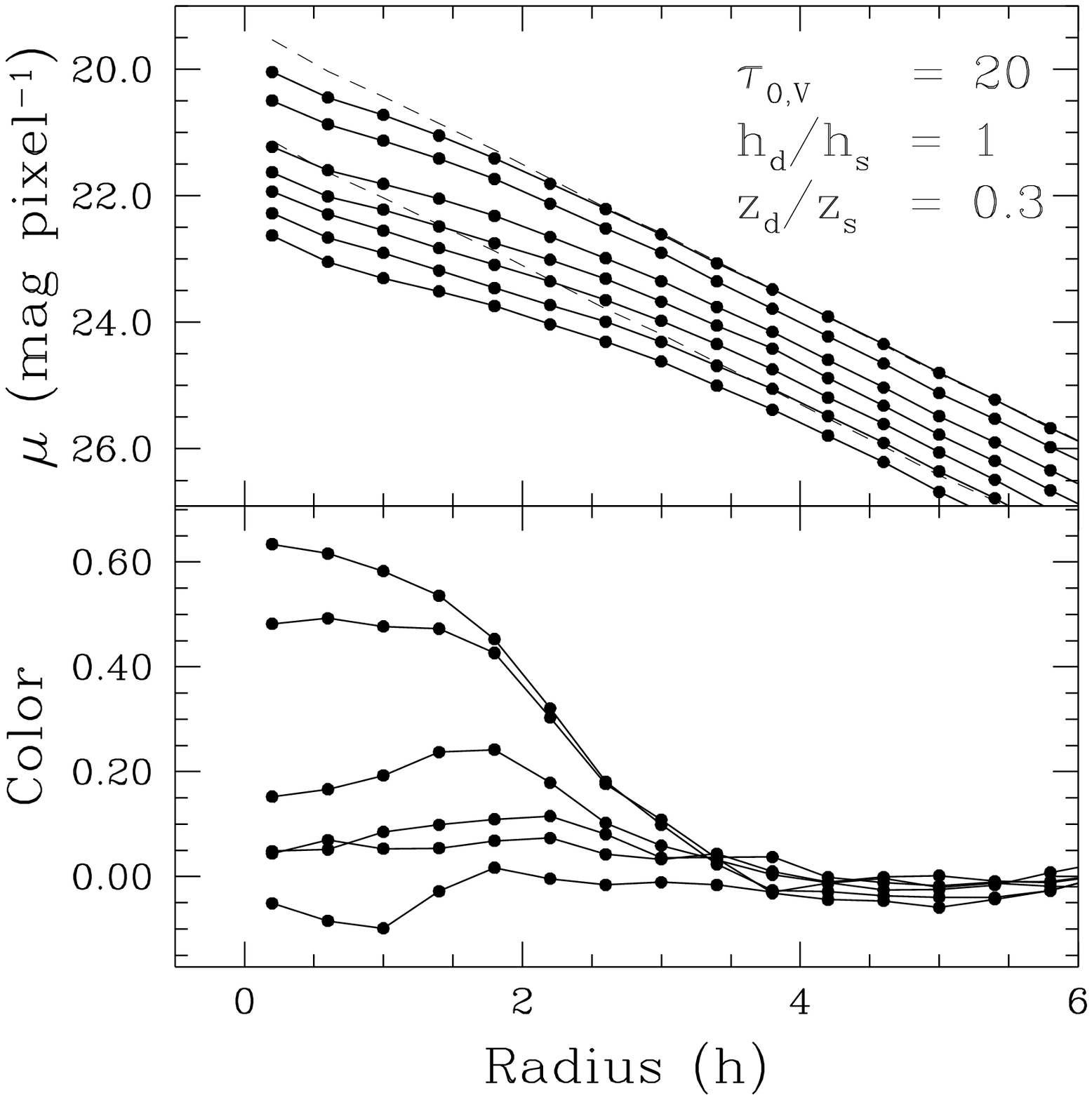}}
\mbox{\epsfxsize=\xsize\epsfbox[0 190 515 660]{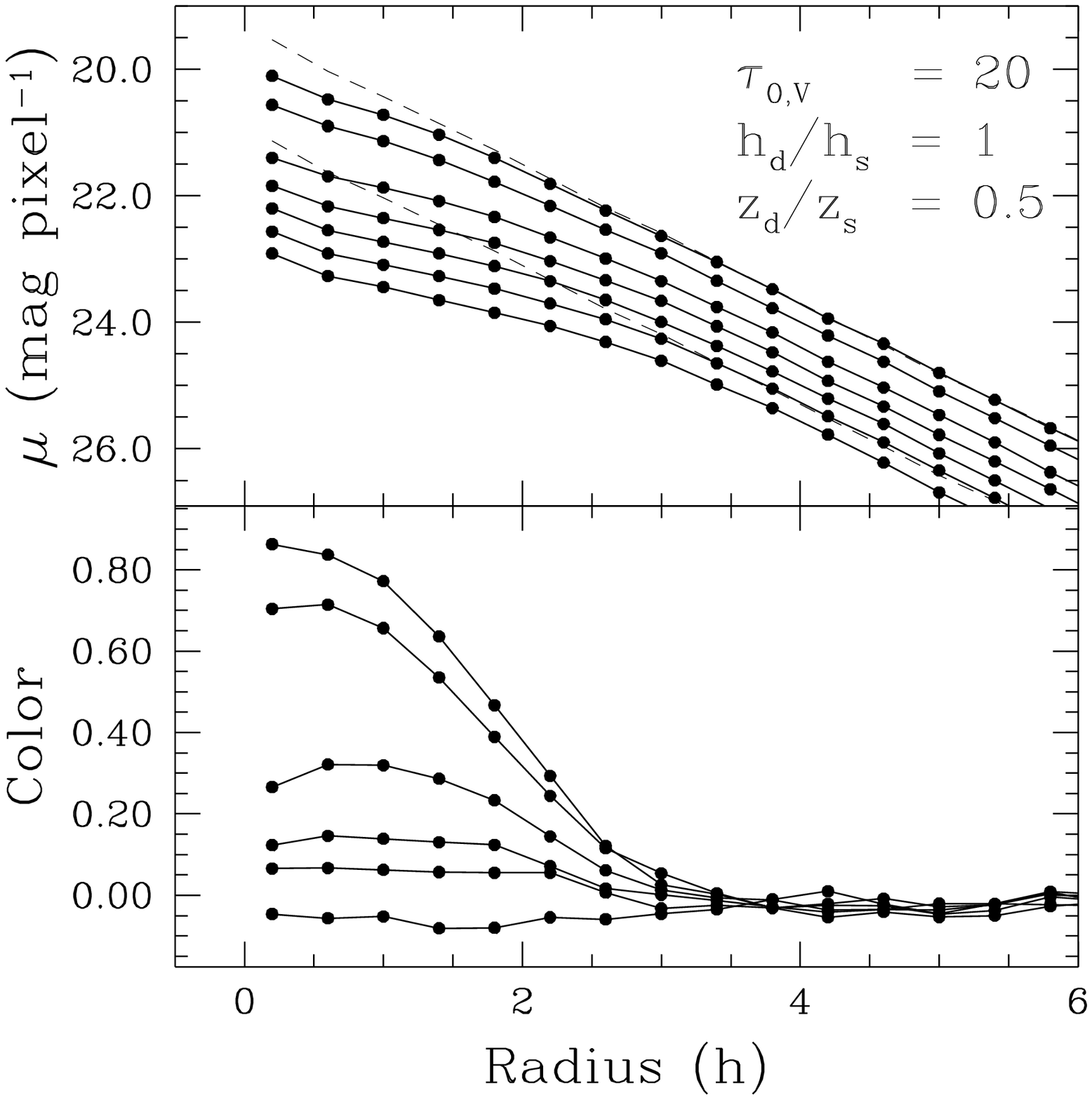}}
\mbox{\epsfxsize=\xsize\epsfbox[0 190 515 660]{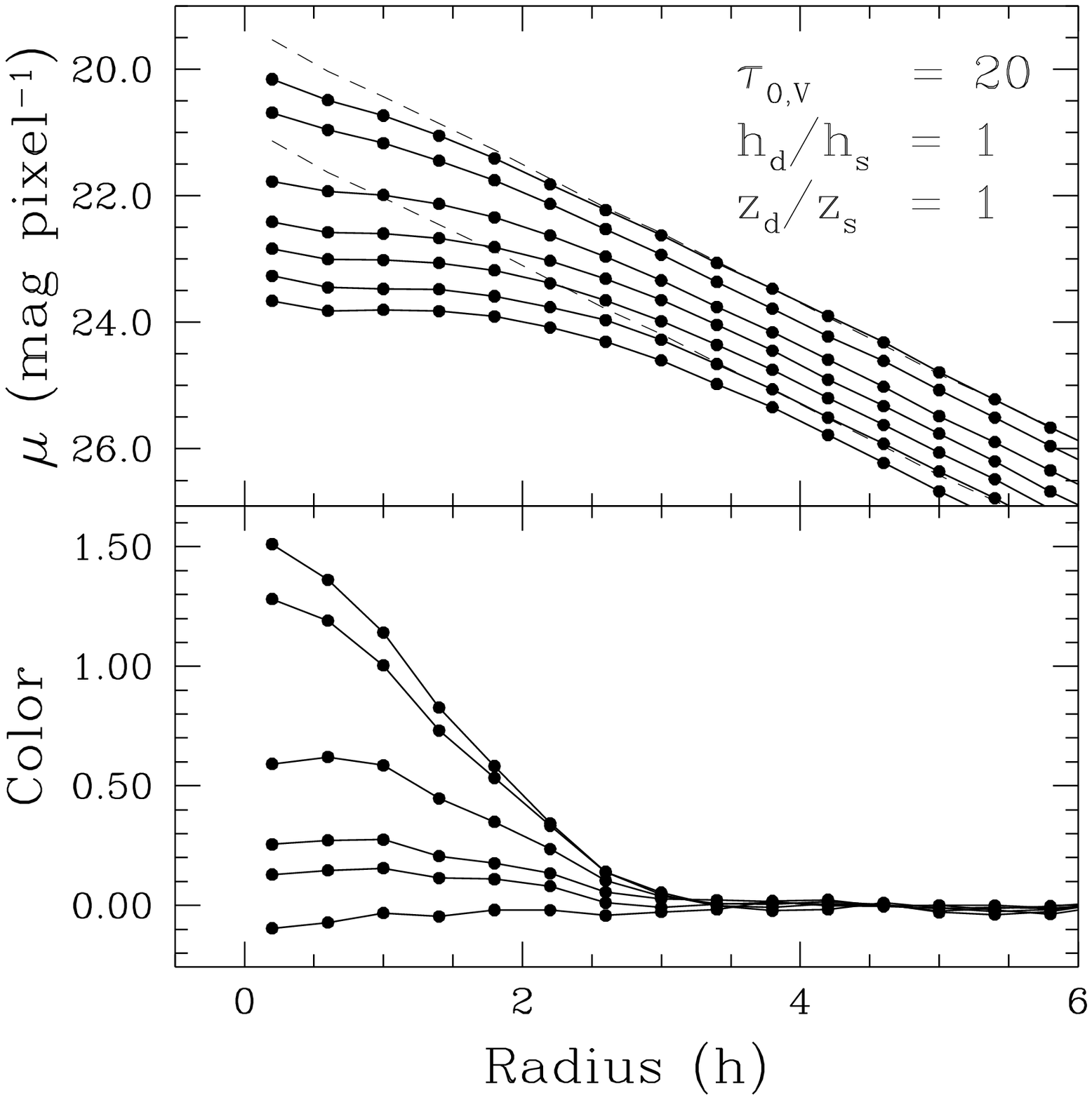}}
 \caption[]{
 The surface brightness (top) and color (bottom) as function of radius
resulting from the Monte Carlo dust simulations of face-on galaxies. 
The radius is units of disk scalelength.  The central optical depth and
dust to stellar scalelength and scaleheight ratios are indicated top
right.  The dashed lines indicate unobscured $B$ and $K$ passband
profiles.  The luminosity profiles have an arbitrary offset and are from
top to bottom the $K$, $H$, $I$, $R$, $V$, $B$ and $U$ passband
profiles.  The color profiles are plotted under the assumption that the
underlying color indices are zero, and are from top to bottom
($B$--$K$), ($B$--$H$), ($B$--$I$), ($B$--$R$), ($B$--$V$) and
($B$--$U$). 
 }
 \label{dusprof}
 \end{figure*}

\addtocounter{figure}{-1}

\begin{figure*}
\mbox{\epsfxsize=\xsize\epsfbox[0 190 515 660]{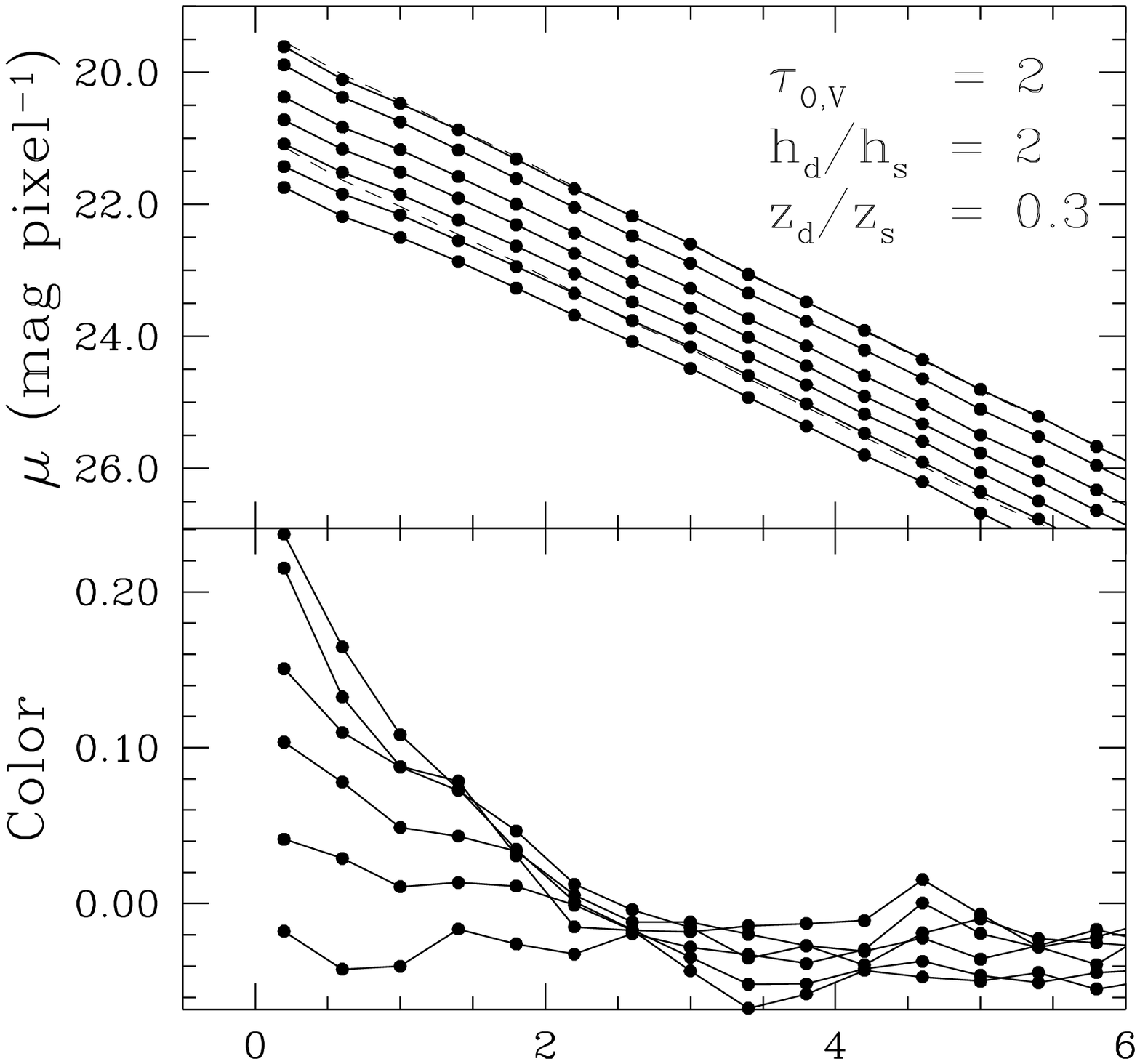}}
\mbox{\epsfxsize=\xsize\epsfbox[0 190 515 660]{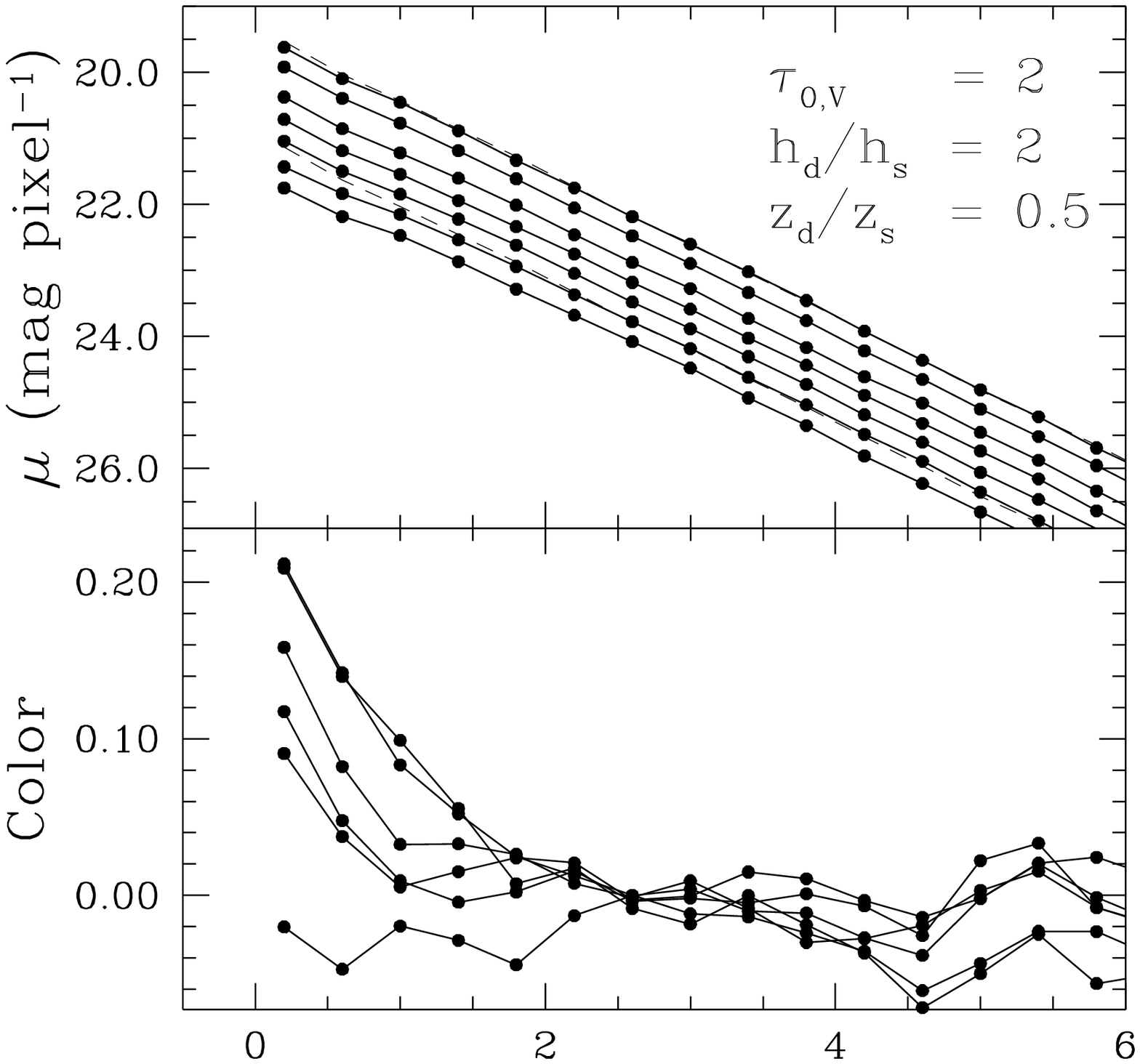}}
\mbox{\epsfxsize=\xsize\epsfbox[0 190 515 660]{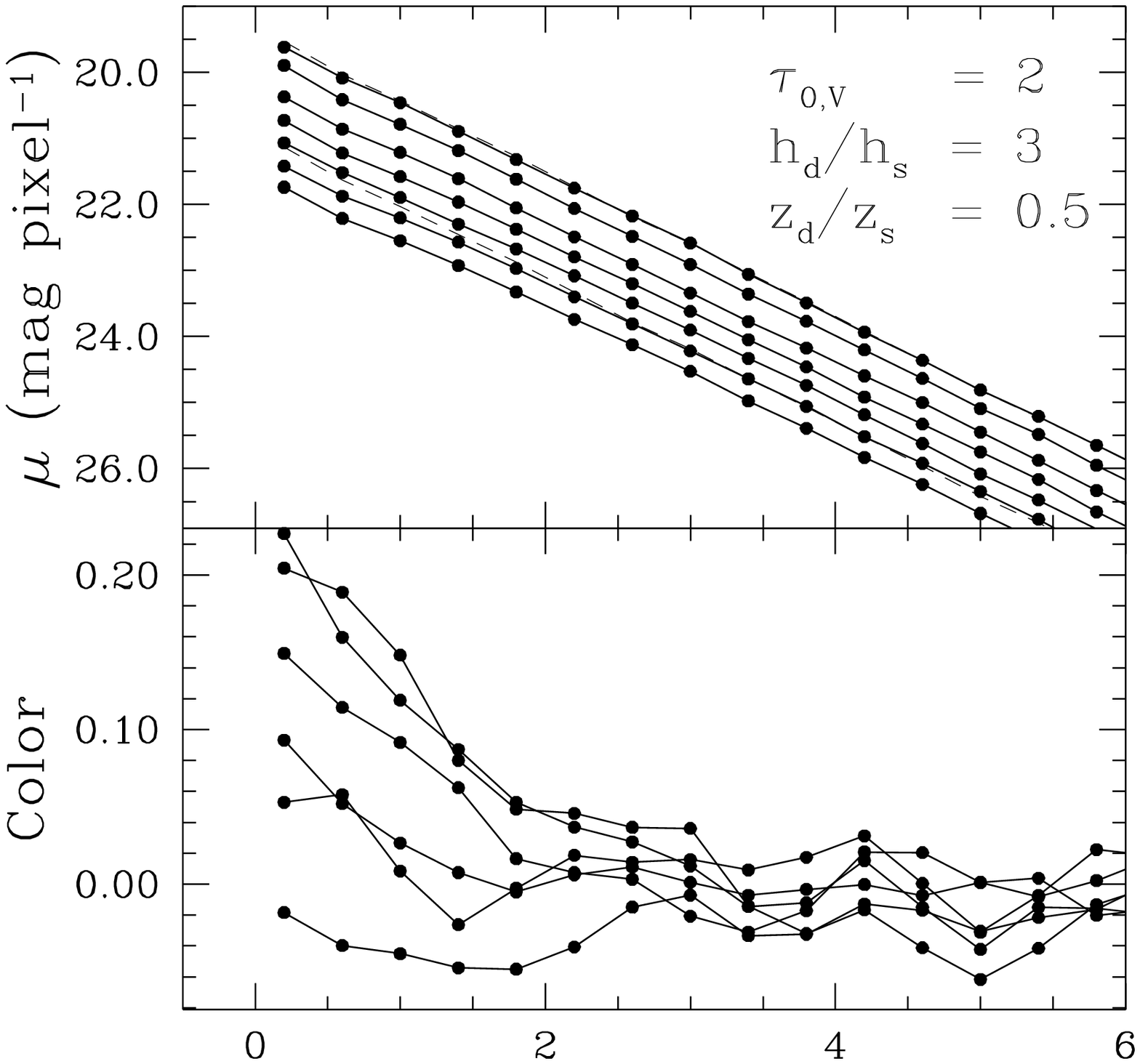}}

\mbox{\epsfxsize=\xsize\epsfbox[0 190 515 660]{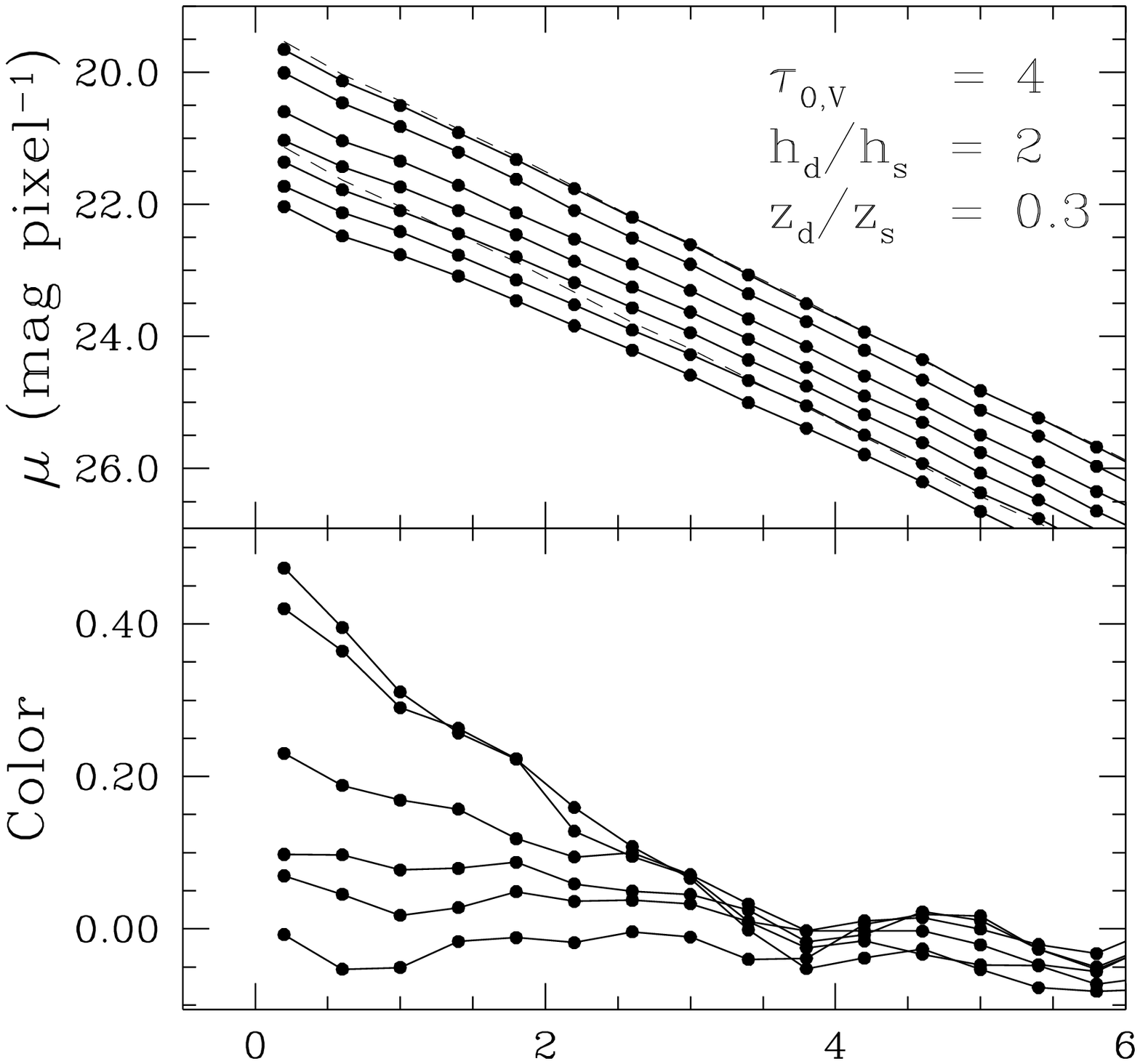}}
\mbox{\epsfxsize=\xsize\epsfbox[0 190 515 660]{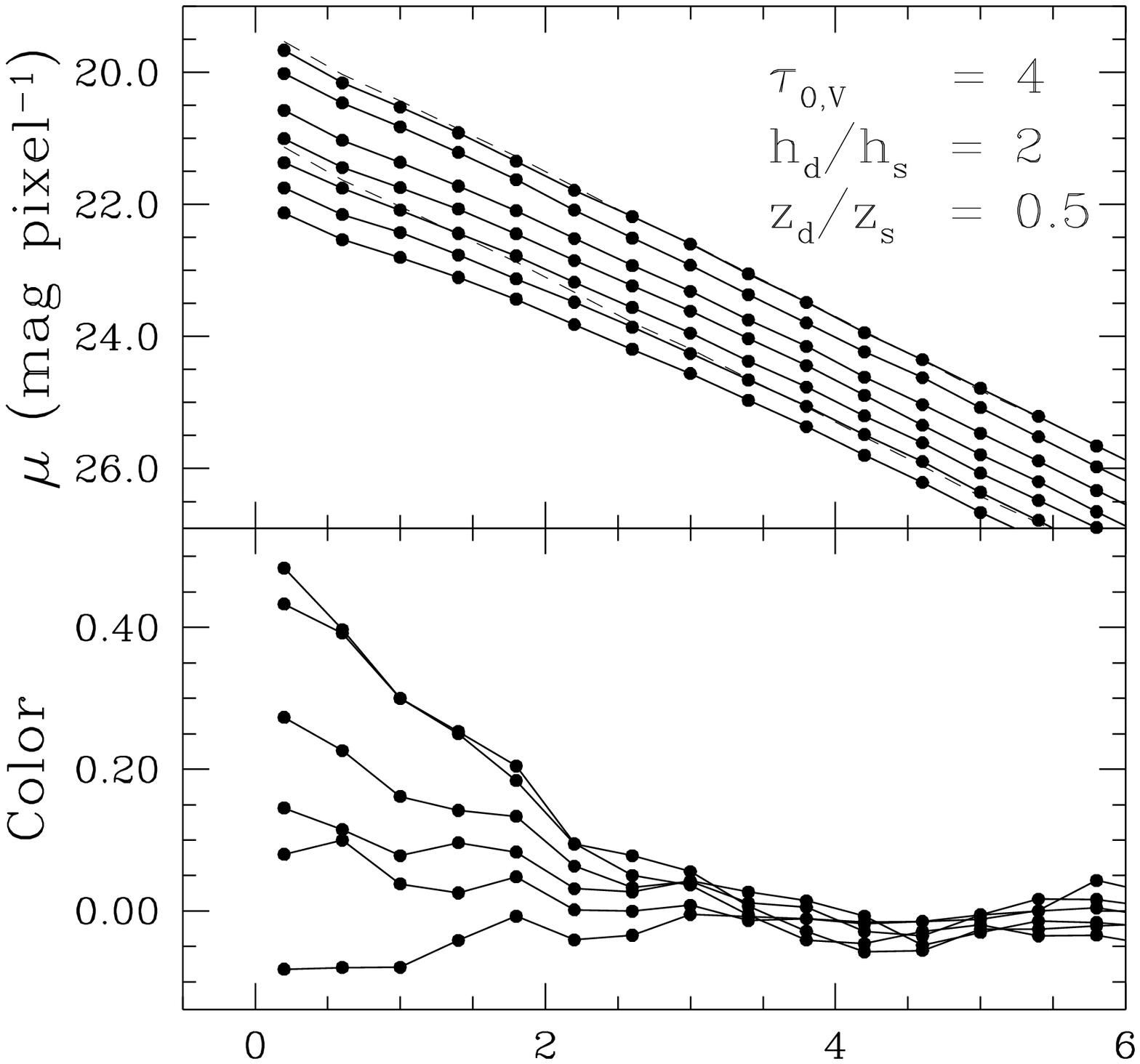}}
\mbox{\epsfxsize=\xsize\epsfbox[0 190 515 660]{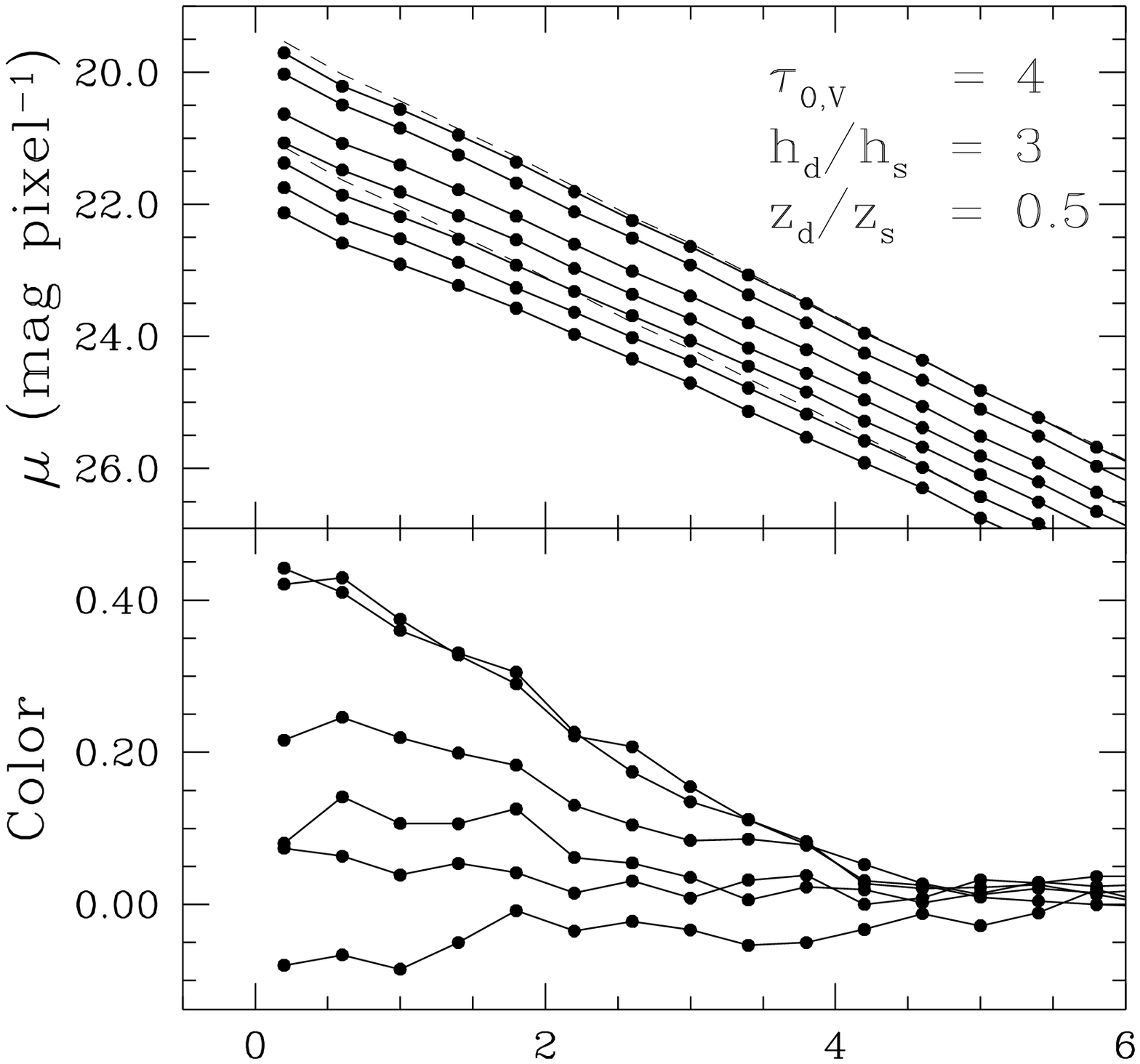}}

\mbox{\epsfxsize=\xsize\epsfbox[0 190 515 660]{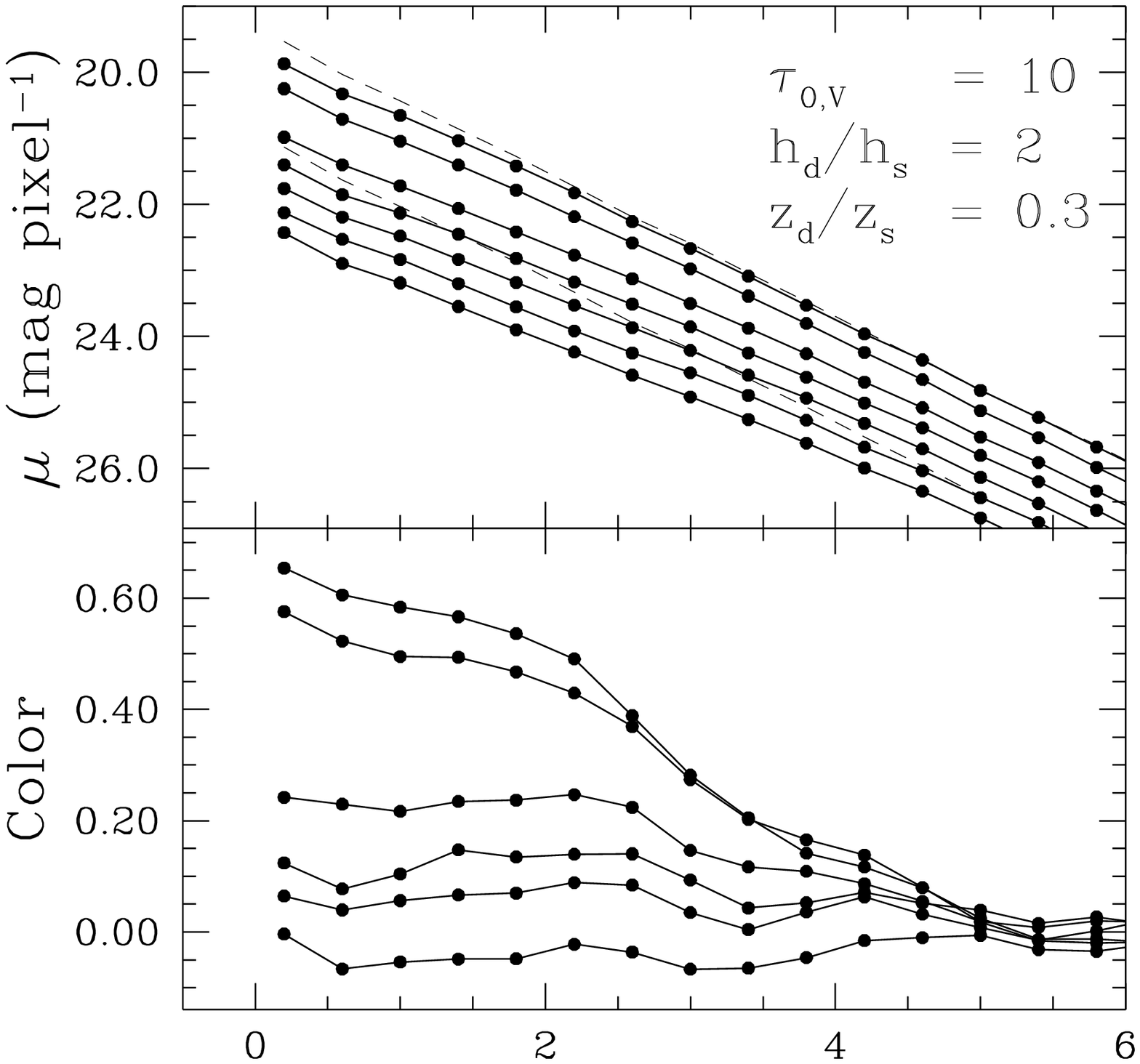}}
\mbox{\epsfxsize=\xsize\epsfbox[0 190 515 660]{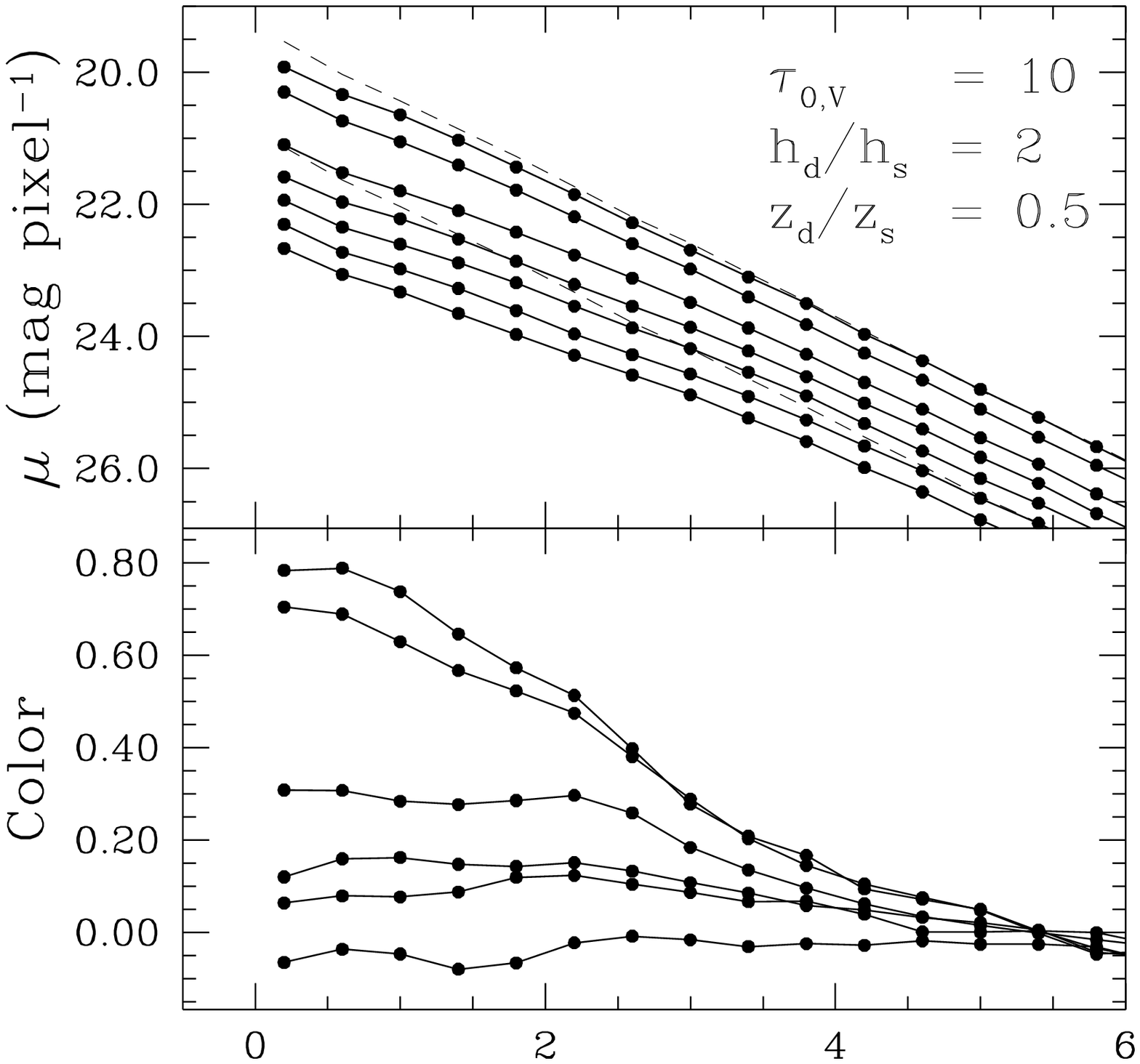}}
\mbox{\epsfxsize=\xsize\epsfbox[0 190 515 660]{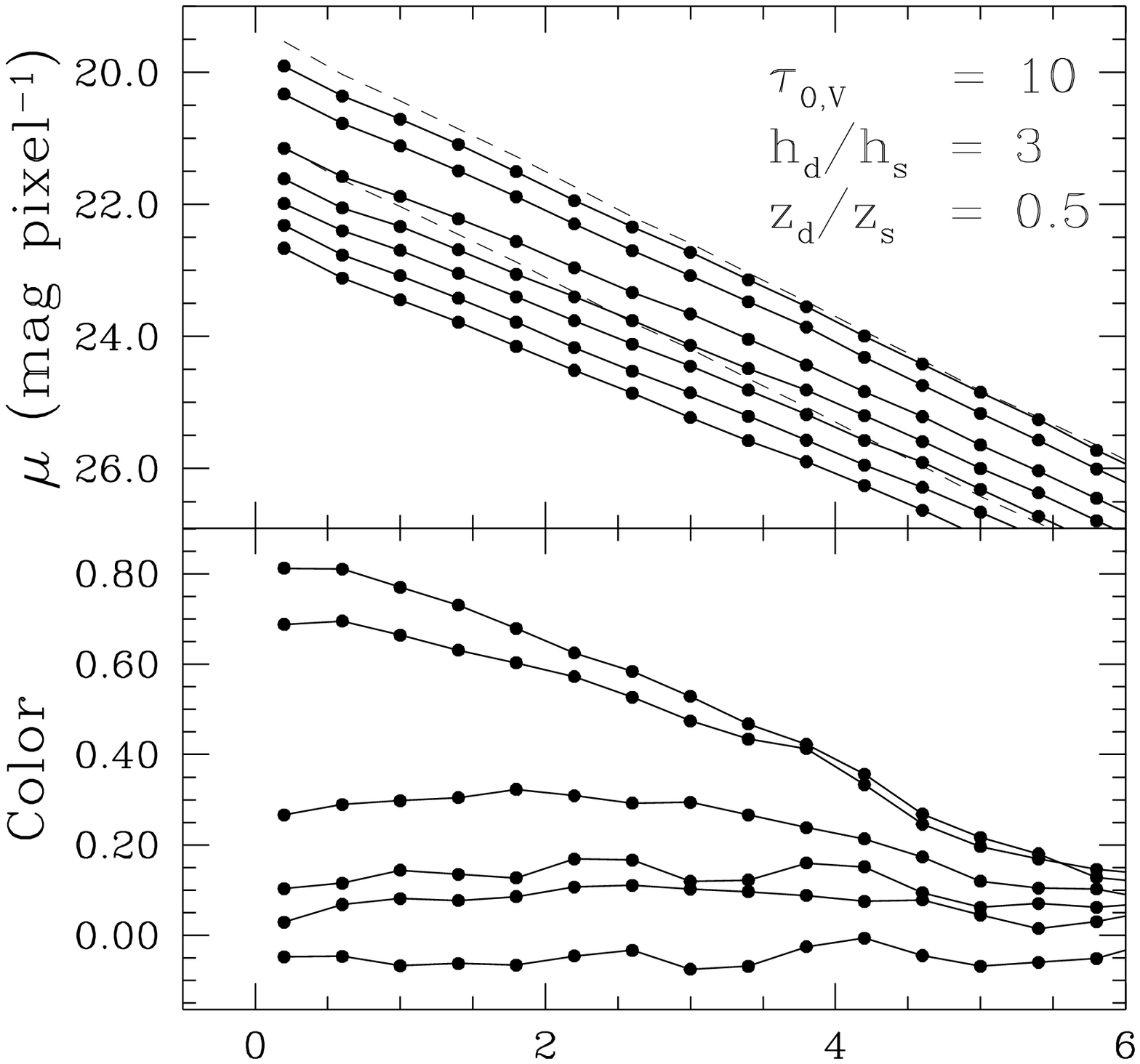}}

\mbox{\epsfxsize=\xsize\epsfbox[0 190 515 660]{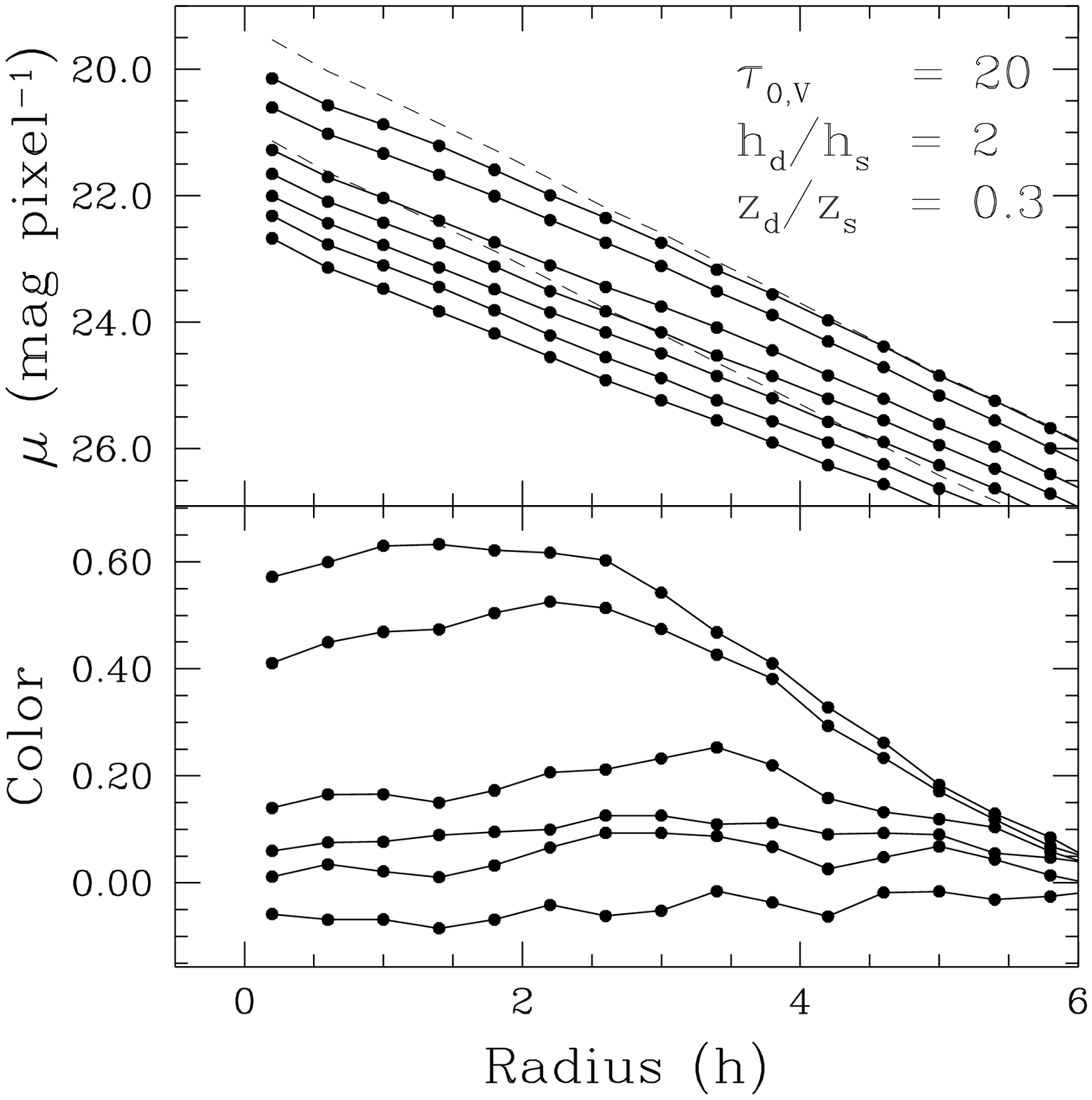}}
\mbox{\epsfxsize=\xsize\epsfbox[0 190 515 660]{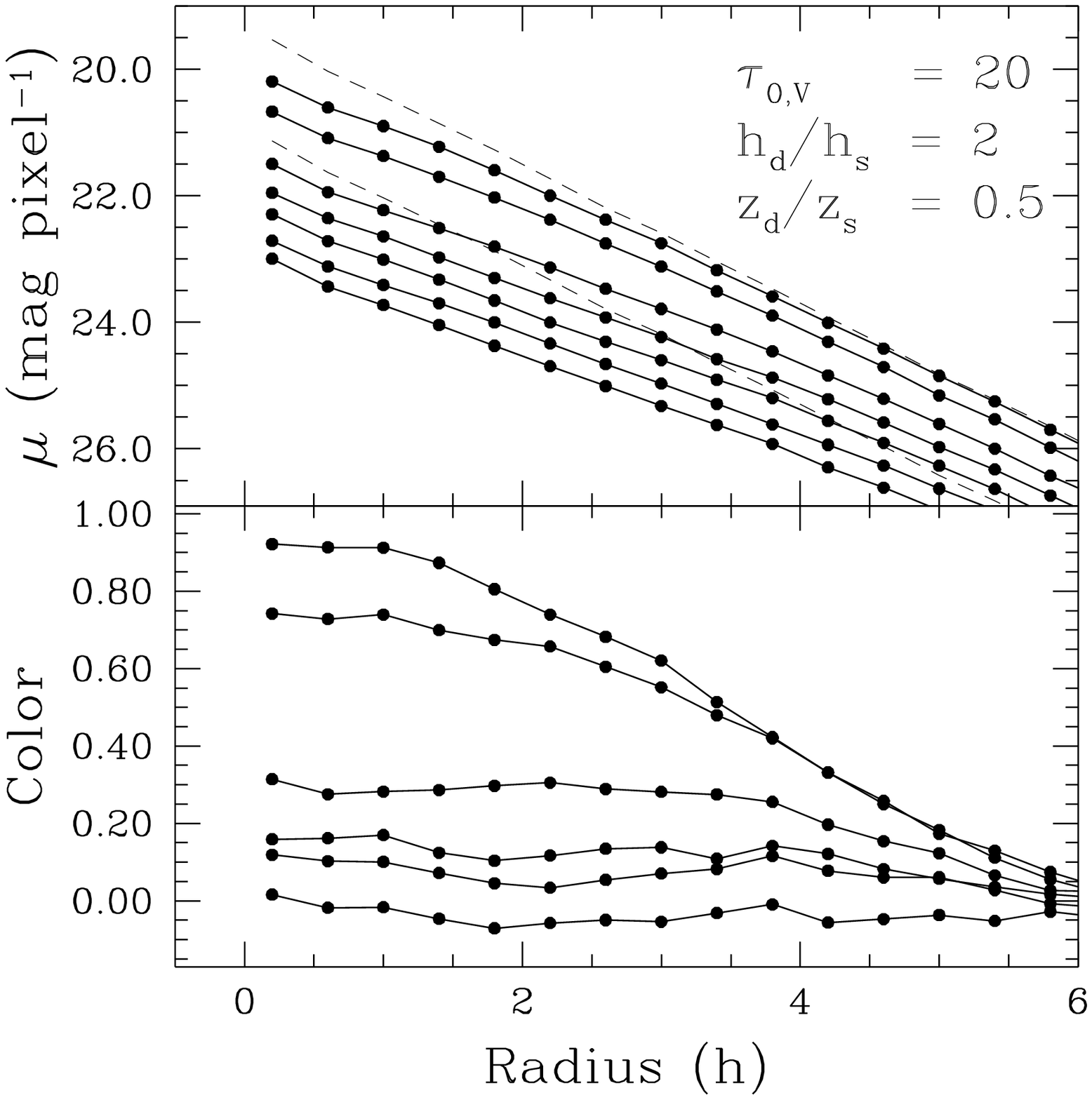}}
\mbox{\epsfxsize=\xsize\epsfbox[0 190 515 660]{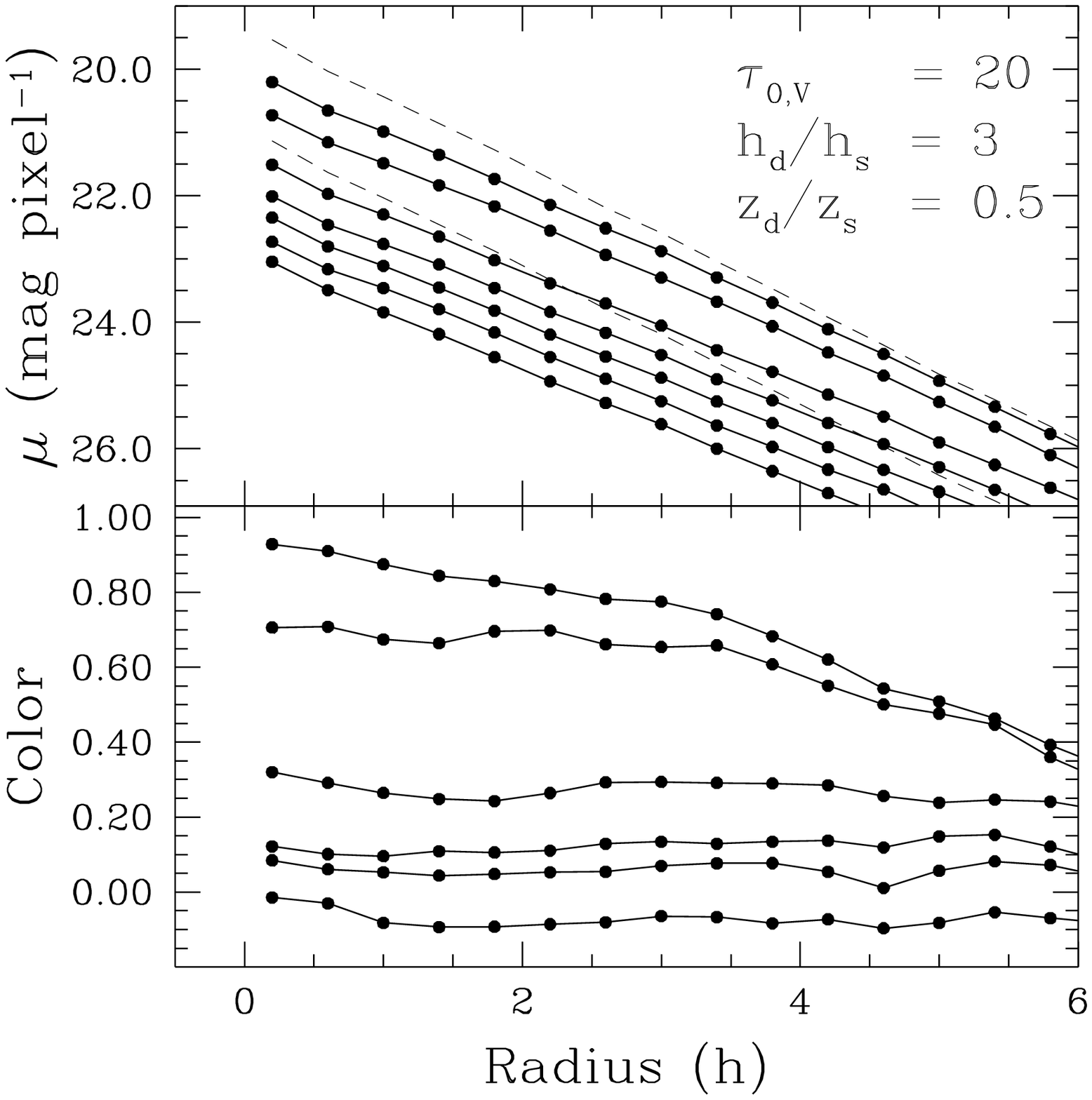}}

\caption[]{-continued.}
\label{dusprof2}
\end{figure*}
%======================================================================

Figure~\ref{dusprof} shows luminosity and color profiles resulting from
the Monte Carlo simulations. The luminosity profiles for the different
passbands have been given an arbitrary offset and the dust free cases
of the $B$ and the $K$ passbands are indicated by the dashed lines. The
color profiles have been plotted under the arbitrary assumption that
the underlying stellar populations have color indices of zero in all
passband combinations. The noise in the color profiles is due to the
statistical processes inherent to Monte Carlo simulations. 

 The luminosity profiles of the $h_{\rm d}/h_{\rm s} \!= \!1$ models
show deviations from the unobscured profiles only at the inner two
scalelengths.  These deviations are quite small except for the highest
$\tau_{0,V}$ values.  The differences between the different $z_{\rm
d}/z_{\rm s}$ models are also quite small and are only apparent for the
high $\tau_{0,V}$ values.  The gradients can be large in $B$--$H$ and
$B$--$K$, but are in general smaller than 0.3 mag in the other color
combinations.  The color gradients are small in the wavelength range
from the $U$ to the $R$ passband, because the absorption properties do
not differ very much among these passbands.  The change in scattering
properties causes the differences in extinction in this wavelength
range. 

The luminosity profiles for the $h_{\rm d}/h_{\rm s} \!= \!2$-3 models
are affected over several scalelengths by dust extinction.  In fact, the
$h_{\rm d}/h_{\rm s} \!= \!3$, $\tau_{0,V} \!= \!10$-20 models are
optically thick over almost the entire disk.  The result is that the
profiles stay exponential, but with a lower surface brightness and a
slightly different scalelength from the unobscured case.  The color
profiles show no color gradients for these models, only color offsets. 
The typical surface brightness is produced at $\tau_\lambda \!= \!1$
over the entire disk, and the color offsets reflect that different
wavelengths probe different depths into the galaxy. 

To my knowledge, the models of Byun et al.~\cite{Byun94} are the only
models in the literature that have exponential light and dust
distributions, and include scattering to calculate luminosity and color
profiles.  These models can be compared to the models presented, but
only indirect, because Byun et al.\ defined the optical depth of a
system differently.  They parameterize the optical depth of a system as
the {\em absorption} in the $V$ passband through the whole disk of a
face-on galaxy along the symmetry axis, while here the {\em extinction}
is used (Eq.(~\ref{optdepth})).  Furthermore, they use the Galactic {\em
extinction} law to translate the {\em absorption} coefficient from one
passband to another.  Using Table~\ref{dusprop} one can calculate that
their $\tau_V(0)$ models corresponds to my \mbox{$\tau_{0,V} \!=
\!\tau_V(0)/(1\!-\!a_\lambda)$} models, which is 2.9$\tau_V(0)$ for the
$B$ passband and 1.8$\tau_V(0)$ for the $I$ passband.  It is probably
most meaningful to compare the $\tau_{0,V} \!= \!20$, $z_{\rm d}/z_{\rm
s} \!= \!0.3$, $h_{\rm d}/h_{\rm s} \!= \!1$ $B$ passband profile of
Fig.~\ref{dusprof} with the bulgeless (BT0.0), face-on $\tau(0) \!=
\!5.0$ profile of their Fig.\,7.  The central extinction of slightly
more than 1\,mag and the general shape of the luminosity profile (which
is unaffected by extinction for radii larger than 2-3 scalelengths) are
comparable.  Their $B$--$I$ color profiles are physically not very
plausible, because they use the Galactic extinction curve instead of an
absorption curve to translate their absorption coefficients from one
passband to another.  A comparison of the color profiles is therefore
not meaningful.  Their luminosity profiles can be used, but note that
their $\tau_V(0)$ should be divided by (1--$a_V$) to get the extinction
suffered by a point source behind the galaxy, as done here. 

What is the range of plausible model parameter values? Quite high
$\tau_{0,V}$ and/or $h_{\rm d}/h_{\rm s}$ values are needed to explain
the observed $B$--$K$ color gradients of one magnitude over five
scalelengths (Fig.~\ref{colprof}) by dust reddening alone.  On the other
hand, models with equal scaleheight for dust and stars need an
additional dust component, because these models do not produce a clear
dust lane in edge-on galaxies.  It is also unlikely on dynamical grounds
that the dissipational dust and the dissipationless stars have the same
scaleheight.  Kylafis \& Bahcall (\cite{KylBah87}) find a dust-to-star
scaleheight ratio of 0.4 in their best fitting model of edge-on galaxy
NGC\,891 and the dominant dust component is expected to have a $z_{\rm
d}/z_{\rm s}$ ratio between 0.3 and 0.5. 

The high $h_{\rm d}/h_{\rm s}$ models are favored by Valentijn
(\cite{Val90}, \cite{Val94}), who concluded from inclination tests that
Sb-Sc galaxies have a $\tau_B \!\sim \!1$ through the disk at $D_{25}$. 
This extinction at about 3-4 stellar scalelengths translates to
$\tau_{0,V} \!\sim \!20$ models, if the dust density is distributed
exponentially with $h_{\rm d}/h_{\rm s} \!= \!1$.  The edge-on
extinction from the center out to 3-4 stellar scalelengths gives at
least $\tau_V \!\approx \!50$ in such a model, which is in conflict with
observations of edge-on galaxies.  A few edge-on galaxies have been
imaged in the near-IR indicating $A_V\!\approx\!8$-10 (Wainscoat et
al.~\cite{Wai89}; Aoki et al.~\cite{Aoki91}), and the Galactic center
can be seen in the $K$ passband (Rieke \& Lebofsky~\cite{RieLeb85},
$A_V\!\approx\!30$ and $A_K\!\approx\!3$).  Thus $\tau_{0,V}\!=\!20$
models represent extremely dusty galaxies, and certainly no galaxies
with $\tau_{0,V}$$\ge$20 are expected.  If one increases $h_{\rm
d}/h_{\rm s}$ to 3, $\tau_{0,V}$ has to be 2-4 to get $\tau_V \!= \!1$
through the disk at 3 stellar scalelengths.  This then gives an edge-on
$\tau_V \!= \!20$-40 from the center out to 3 stellar scalelengths when
using large $z_{\rm d}/z_{\rm s}$ values, which are the most favorable
for these models.  Therefore, the edge-on extinction values of the $\tau_{0,V}
\!= \!2$-4, $h_{\rm d}/h_{\rm s} \!= \!3$ models are marginally
consistent with the observations, but these models do not produce very
large color gradients. 

\subsubsection{Resulting color--color diagrams}

\begin{figure*}[tbh]
 \mbox{\epsfxsize=8.8cm\epsfbox[20 122 575 670]{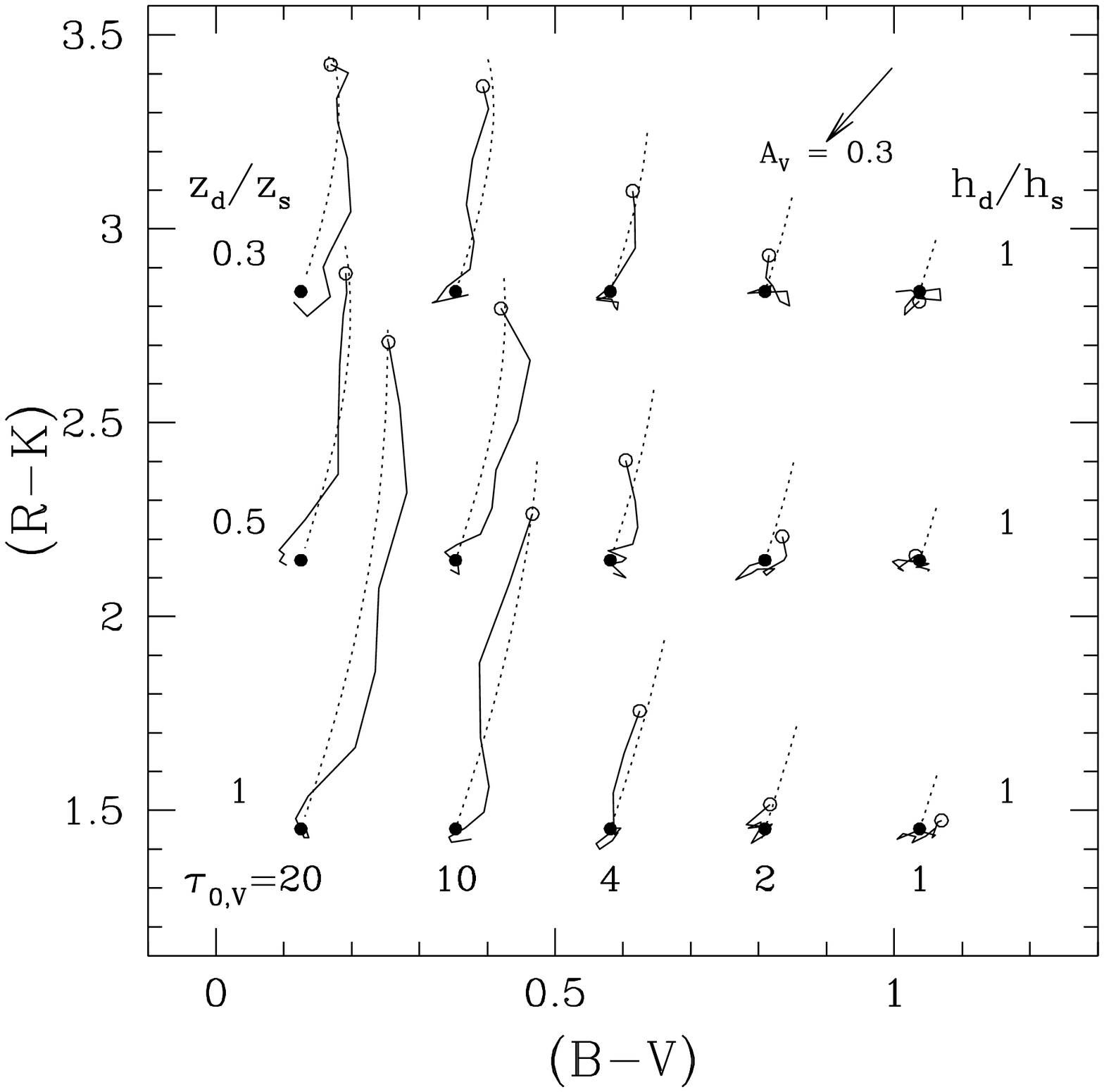}}
 \mbox{\epsfxsize=8.8cm\epsfbox[20 122 575 670]{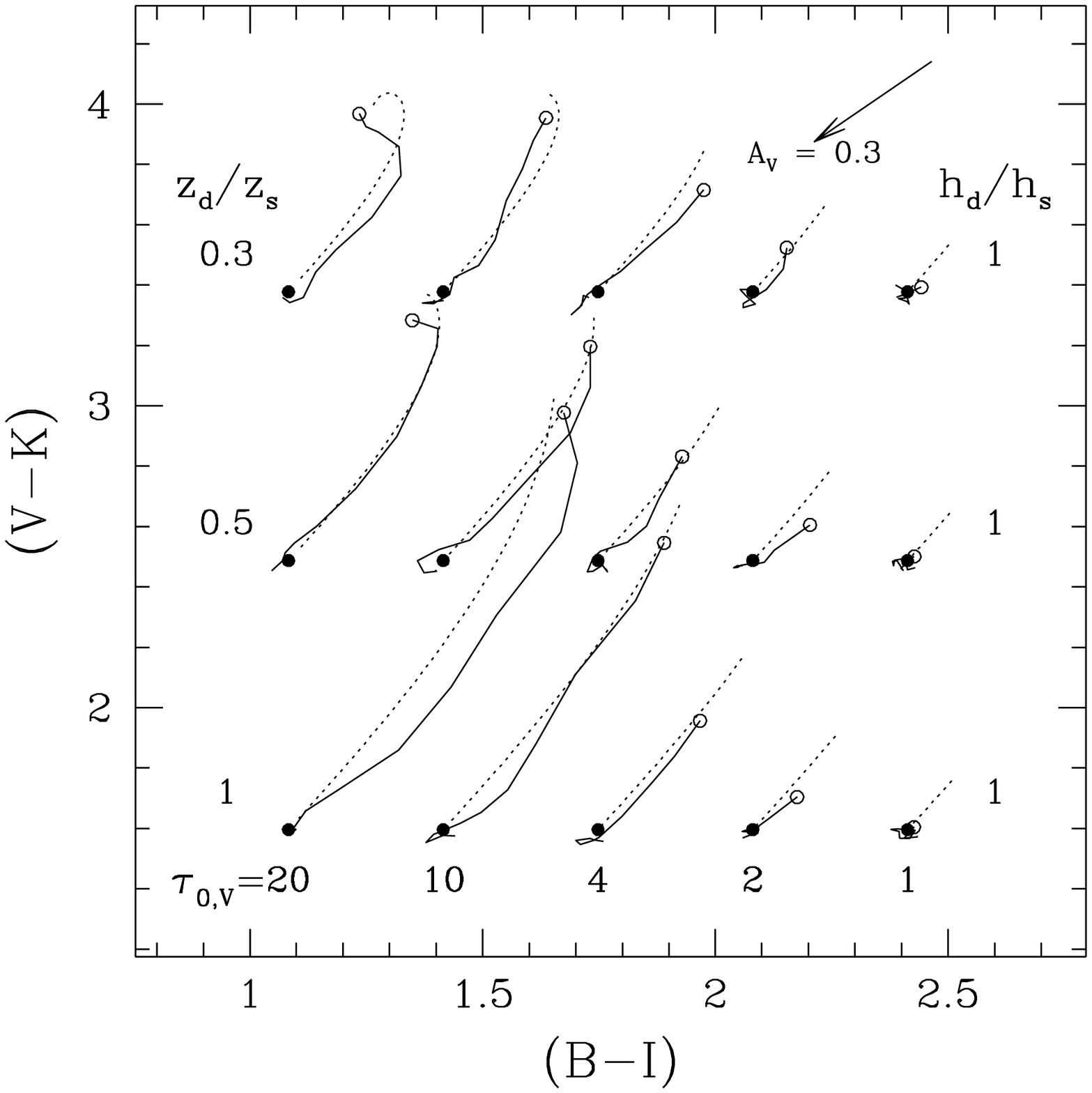}}
 \mbox{\epsfxsize=8.8cm\epsfbox[20 163 575 670]{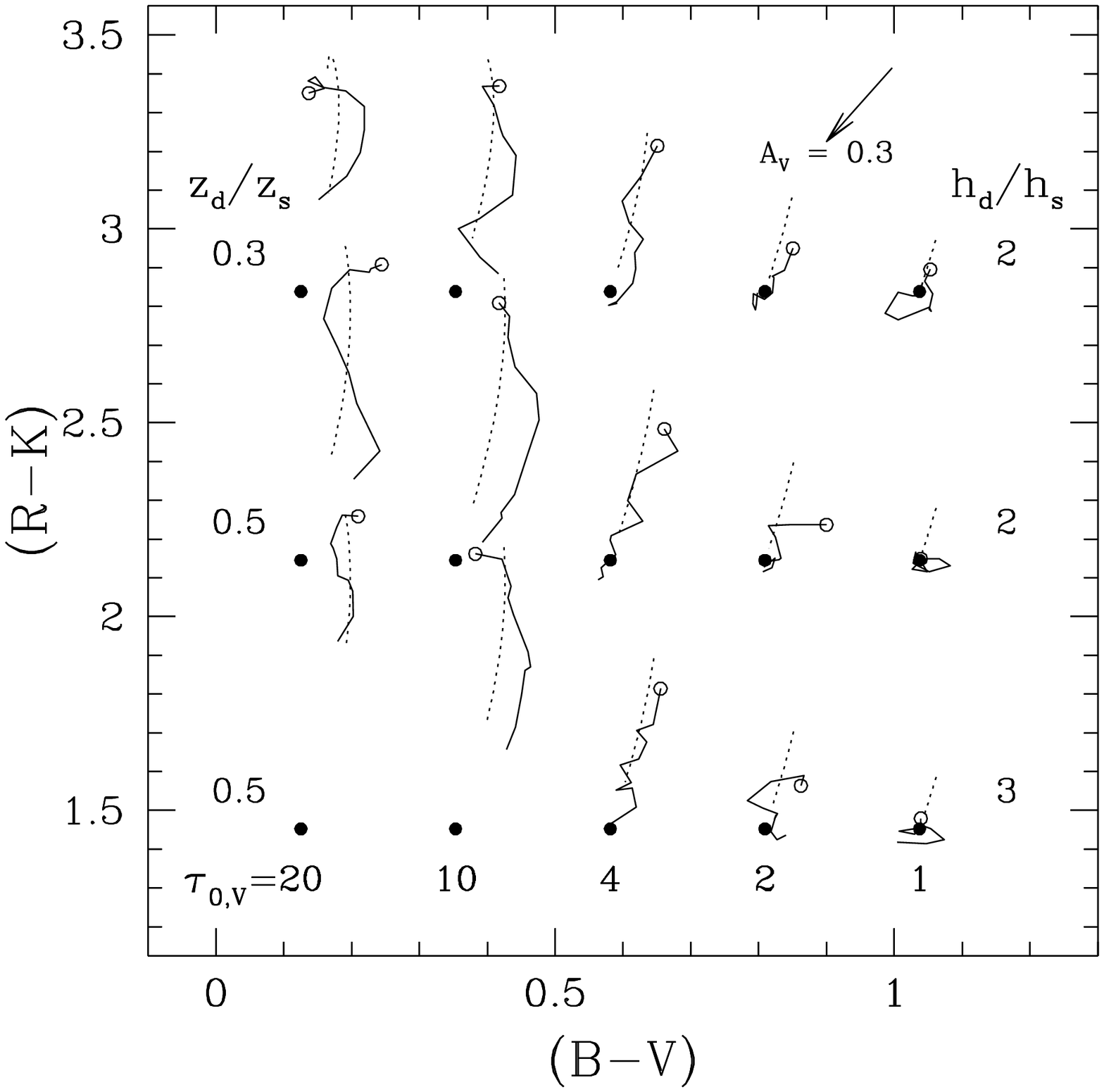}}
%\ \ \ 
 \mbox{\epsfxsize=8.8cm\epsfbox[20 163 575 670]{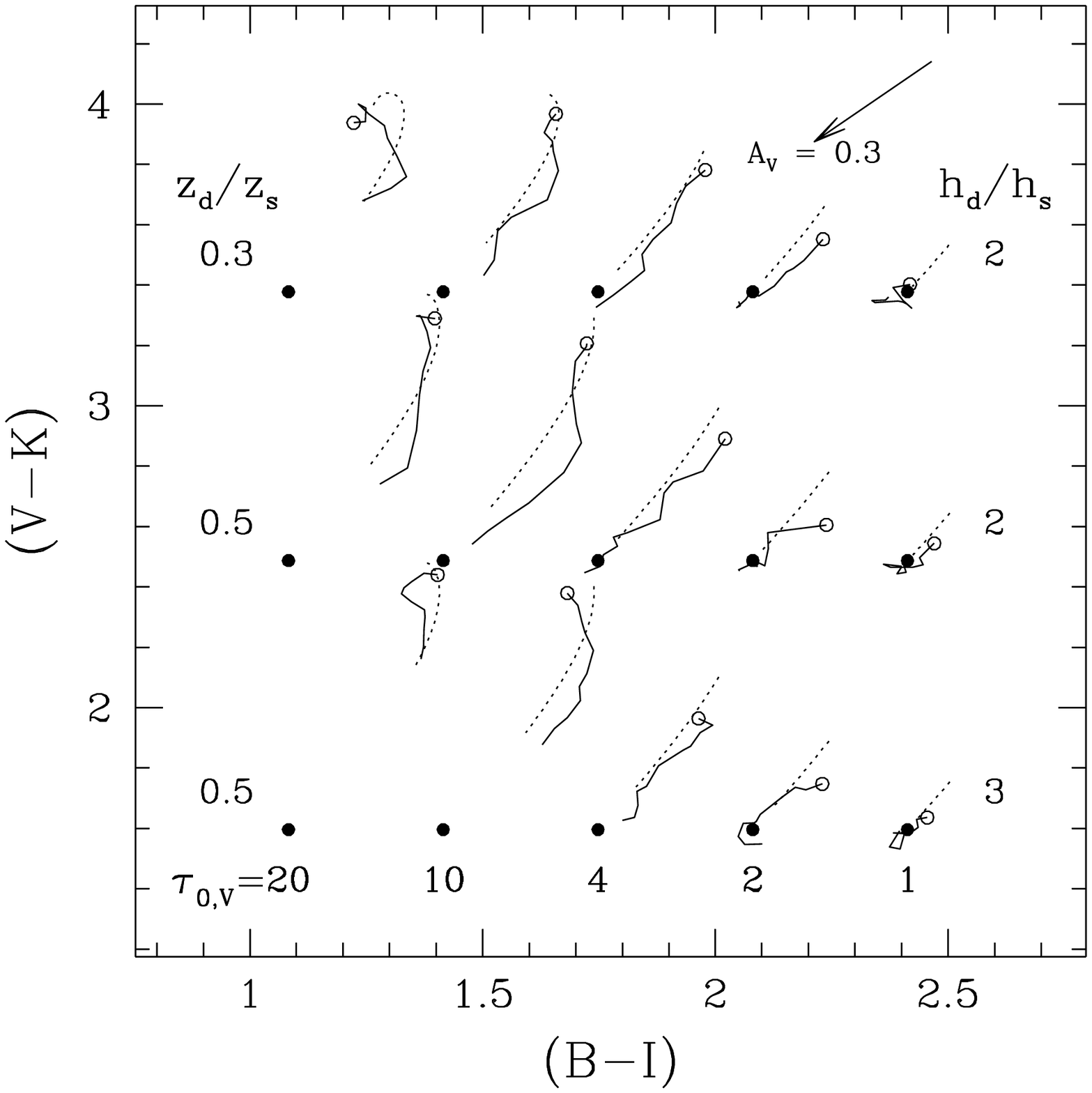}}
 \caption[]{
 Color--color plots resulting from dust models of face-on galaxies.  The
color zero-points are arbitrary (depending on the color of the
underlying stellar population) and are indicated by the filled circles. 
The centers of the model galaxies are indicated by the open circles and the
colors are followed in radial direction for four and a half scale
lengths.  The solid lines are the results from the Monte Carlo
simulations, described in Appendix~\ref{apmodel}.  The dotted lines are
the results from the DDP models, using the Galactic absorption law
rather than the extinction law between the different passbands.  The
reddening vectors in the top right corners indicate the Galactic
extinction law (i.e.\ Screen model). 
\vspace{-2ex}
 }
 \label{duscolcol}
\end{figure*}

%======================================================================

Four color--color plots of dust models are presented in
Fig.~\ref{duscolcol}.  Because extinction is a relative measurement, the
zero-point can be chosen freely in these plots; only the shapes of the
profiles are fixed.  The solid lines show the results from the Monte
Carlo simulations.  The dotted lines are the result of the Triplex
models of DDP.  To calculate the $\tau_0$ for the DDP models in the
different passbands, the indicated $\tau_{0,V}$ values were multiplied
by $(\tau_V/\tau_\lambda)(1-a_\lambda)/2$ (Table~\ref{dusprop}).  This
is equivalent to using an absorption law rather than an extinction law
between the different passbands.  The factor 2 arises because the DDP
models are characterized by the optical depth from the galaxy center to
the pole and not by the optical depth through the whole disk. 

Comparing the results from the Monte Carlo simulations with the DDP
models, one can see that the models agree remarkably well for high
optical depths.  The intuitive idea that just as many photons are
scattered out of the line of sight as are scattered in seems correct for
face-on galaxies.  Photons are only lost due to absorption.  For low
optical depths the reddening almost completely disappears.  Once a
photon gets scattered into the line of sight, the chances of it getting
absorbed or scattered again are minimal; even bluing instead of
reddening can occur.  Even though the amount of reddening is a strong
function of the dust configuration, it seems that the direction of the
reddening vector is largely dependent on the dust properties.  Obviously
the reddening produced by these models is different from the reddening
produced by a Screen model with the Galactic {\em extinction} law (also
indicated in Fig.~\ref{duscolcol}). 

In conclusion, dust can produce color gradients in face-on galaxies, but
this requires quite high central optical depths and preferably long dust
scalelengths.  The reddening vectors of realistic dust models that
include both absorption and scattering are completely different from the
often-used Screen model extinction models.

\subsection{Evolutionary stellar population synthesis models}
\label{popmod}

Ever since the invention of the concept of stellar populations in
galaxies (Baade~\cite{Baa44}), numerous models have been made to predict
the integrated light properties of such populations and thus of galaxies
as a whole.  Initially the empirical approach was often followed,
in which the different contributions of the stellar populations are added
to match the observed galaxy SED.  Later, knowledge about initial
conditions and stellar evolution were added to make evolutionary
synthesis models.  This section contains a brief description of
population synthesis models, concentrating on the 
synthesis models used here, followed by a description of the effects
that star formation history (SFH) and metallicity have on the colors
produced by synthesis models.

\subsubsection{Modeling stellar populations}
\begin{figure*}
 \mbox{\epsfxsize=8.8cm\epsfbox[20 160 575 670]{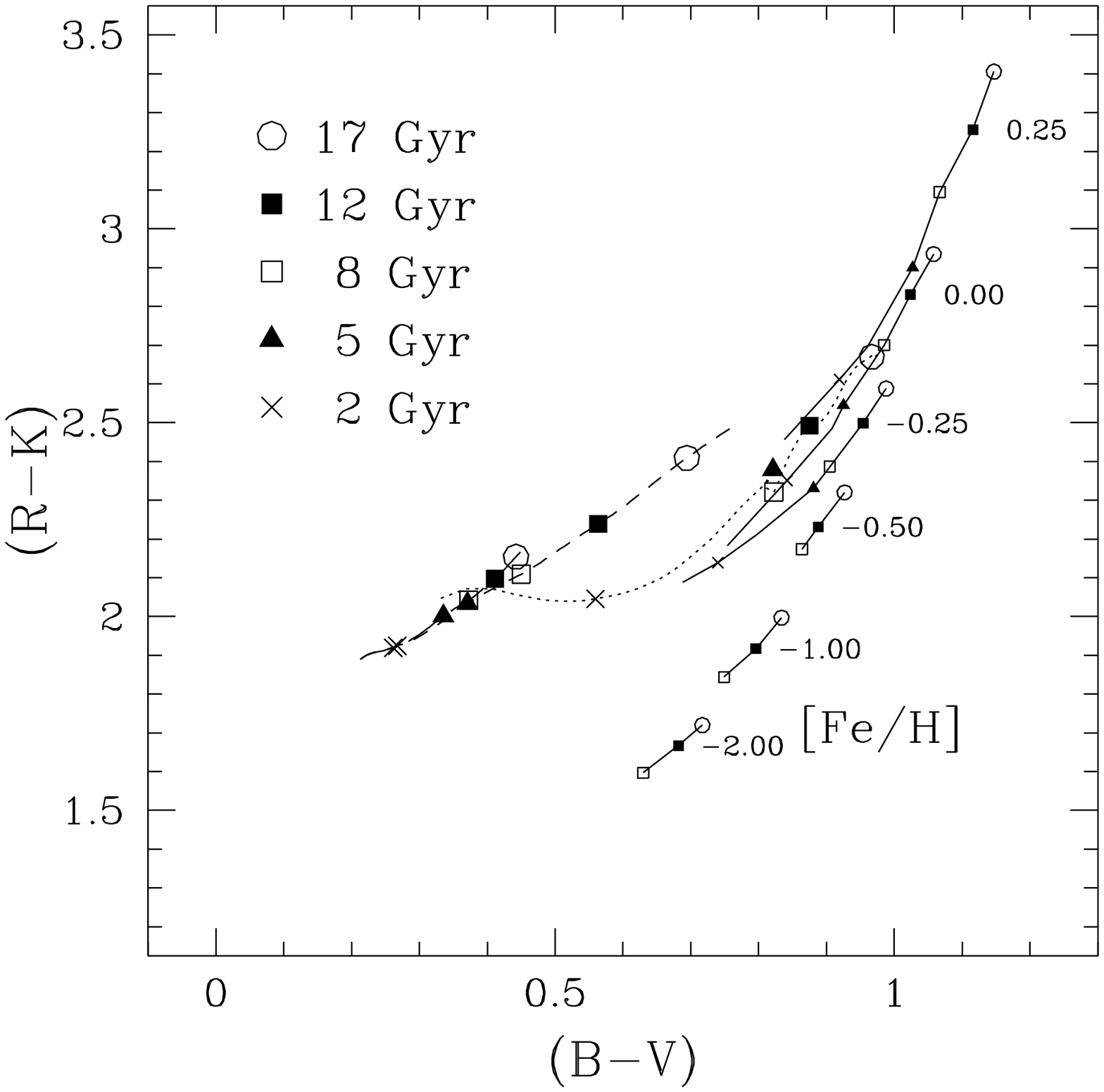}}
 \mbox{\epsfxsize=8.8cm\epsfbox[20 160 575 670]{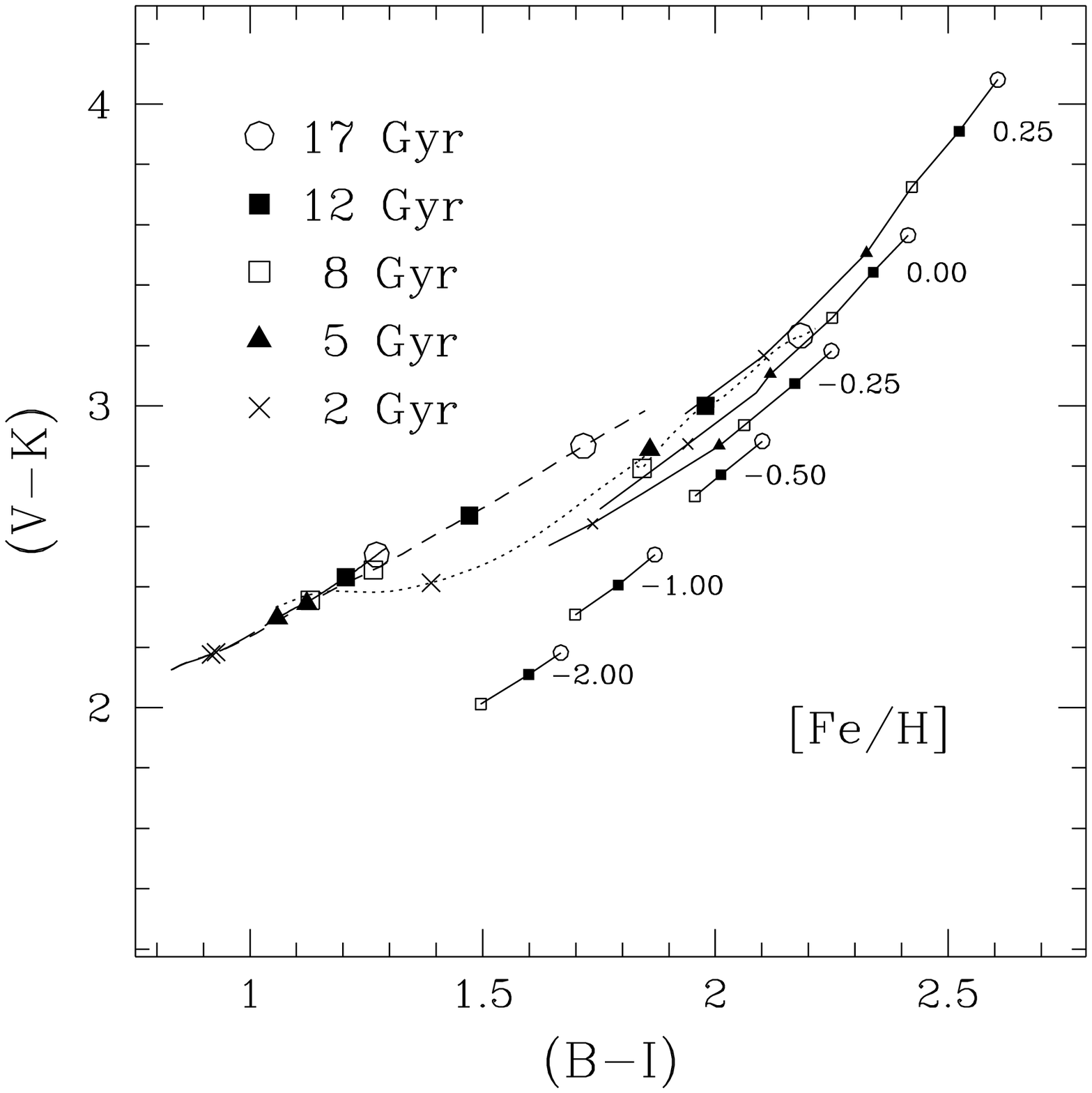}}

\caption[]{
 Evolutionary color--color plots of stellar synthesis models.  The
symbols indicate the number of years after creation of this population. 
To the right in each panel, the different ages connected by solid lines,
are the single burst models of Worthey (\cite{Wor94}) for different
metallicities.  The corresponding [Fe/H] values are indicated next to
them.  To the left in each panel are the solar metallicity models of
Bruzual \& Charlot (\cite{BruCha96}).  The dotted line indicates the
single burst evolution.  The dashed line is a model with an exponentially
declining star formation rate.  The leftmost dot-dashed line,
overlapping the blue part of the exponentially declining SFR model, indicates a
model with constant star formation.  Bruzual \& Charlot used the Johnson
$R$ and $I$ passbands which were here converted to Kron-Cousins $R$ and
$I$ passbands using the equations of Bessell~(\cite{Bes79}). 
\vspace{-0.5ex}
 }
 \label{popcolcol}
 \end{figure*}

In recent years a large number of synthesis methods have appeared in the
literature, all of which require many input parameters
(Tinsley~\cite{Tin80}; Renzini \& Buzzoni~\cite{RenBuz86};
Worthey~\cite{Wor94} and references therein).  The simplest models use a
initial mass function (IMF) to create a single burst of stars, whose
evolution in time is then followed.  Slightly more complicated models
describe the star formation rate (SFR) in time.  In the most complicated
models, the gas, stellar, and chemical evolution are linked and
described in a self-consistent way.  This last type of model has not
been used here, since to date only a small range of SFHs have been
investigated with these models.  The description of the evolution alone
does not produce an SED and therefore the models are linked to a stellar
library to calculate the evolution in time of the integrated passband
fluxes or of the integrated spectrum. 

The results from the models in the literature are not all in agreement. 
The disagreements arise mainly from differences in the treatment of the
late stages of stellar evolution.  The models do agree on the two main
parameters determining the integrated colors of a synthesized galaxy. 
First of all the colors are strongly determined by the colors of the
youngest population, thus by the SFH, and secondly the colors are
considerably affected by the metallicities of the populations.  In
Sect.~\ref{Intro4} it was noted that both these SFH and metallicity
changes have been observed on a radial scale in spiral galaxies. 
Furthermore, the radial age and metallicity gradients in our own Galaxy
are well known and have been extensively studied (Gilmore et
al.~\cite{Gil89}; Matteucci~\cite{Mat89}, \cite{Mat92} and references
therein).  Synthesis models incorporating both age and metallicity
effects are needed in the comparisons with spiral galaxy data. 

The population synthesis models of Bruzual \& Charlot (\cite{BruCha96},
BC96 models hereafter, see also Charlot \& Bruzual~\cite{ChaBru91};
Bruzual \& Charlot (\cite{BruCha93}) and of Worthey~(\cite{Wor94}, W94
models hereafter) are used in the remainder of this paper; they are
shown in the color--color diagram of Fig.~\ref{popcolcol}.  The BC96
models used here are based on the isochrone tracks of the Padova group
(Bressan et al.~\cite{Bre93}) and on an empirical stellar flux library. 
The BC96 isochrone synthesis approach makes calculation of the very
early stages of evolution of a population possible.  The W94 models are
constructed from the isochrones of VandenBerg~(\cite{Van85}) and the
Revised Yale Isochrones (Green et al.~\cite{Gre87}) and use a
theoretical stellar flux library.  A different approach was followed in
the BC96 and W94 models to calculate the integrated spectra of an
evolving stellar population, but the main difference of interest here is
the regions of age-metallicty parameter space that were investigated. 
The BC96 models were only calculated for solar metallicity, but give
colors of populations as young as 1.26$\times 10^5$\,yr.  The W94 models
span a wide range in metallicity, but the youngest population is 1.5\,Gyr. 

The models used here were calculated with the standard Salpeter IMF
(Salpeter~\cite{Sal55}), with lower mass cutoff at 0.1\,$M_\odot$ and
upper mass cutoff at 2\,$M_\odot$ (W94) or 100\,$M_\odot$ (BC96). 
The use of other solar neighborhood type IMFs or other
cutoffs has only small effects compared to the main factors determining
the colors of a synthesized population, namely age (or SFH) and
metallicity. The main result will be some shifts of the colors in
color--color space. Using IMFs deviating strongly from the solar
neighbourhood IMF will result in substantial changes, but without strong
evidence these types of IMFs are hard to justify.

A comparison study of several stellar population synthesis models has
been performed by Charlot, Worthey \& Bressan~(\cite{CWB95}, hereafter
CWB).  They find substantial discrepancies among the models, especially
in $V$--$K$, which are larger than the typical observational errors for
galaxies.  They conclude that age determinations from galaxy colors are
about 35\% uncertain at a given metallicity and 25\% uncertain in
metallicity at a given age, even for a single burst population.  The
main source of disagreement stems from differences in the stellar
evolution prescription, with a smaller component resulting from the
spectral calibrations.  For comparison, to see the effect of other
stellar evolutionary tracks, de Jong (\cite{deJPhD}, \cite{deJeso})
presents a similar color--color analysis as presented here, but using
Bruzual \& Charlot (\cite{BruCha93}) with Maeder \& Meynet
(\cite{MaeMey91}) evolutionairy tracks instead of BC96 with Padova
tracks.  

The W94 and B96 models were chosen to cover a large age-metallicity
space, but given the uncertainties in the models, any other set of
up-to-date set of models could have been chosen (e.g.~Bressan, Chiosi \&
Fagotto~\cite{Bre93}; von Alvensleben \& Gerhard~\cite{AlvGer94}).  The
particular choice of models will not change the main results of this
study. The model uncertainties will mainly result in the uncertainty of
absolute calibration of colors, but the relative color trends are
reasonably correct.  With the current state of modeling, I will only
indicate which regions in galaxies and what type of galaxies are
younger/older or have higher/lower metallicity.

\subsubsection{Star formation history in color--color diagrams}

Three evolutionary tracks of the BC96 models are shown in
Fig.~\ref{popcolcol}.  The simplest is the single burst model, in which
the color evolution of one initial starburst is followed in time.  In
the other two tracks this single burst model has been convolved in time
to yield color evolution for different SFHs.  In one model an
exponentially declining SFR with a time scale of 5\,Gyr was used, in the
other model the SFR was held constant.  These models show the importance
of the very early stages of stellar evolution.  The constant SFR model
at 17\,Gyr is as blue in $B$--$V$ and $B$--$I$ as the exponentially
declining model at 8\,Gyr and the single burst model at 1.5\,Gyr! Still,
a solar metallicity starburst cannot produce colors to the right of the
indicated single burst line in Fig.~\ref{popcolcol}, because the colors
of very young populations (age $<$1.5\,Gyr, not shown) all lie to the
left of (or in the direction of) the single burst trend.

\subsubsection{Age and metallicity in color--color diagrams}

Of the W94 models, only the single burst models of different
metallicities are shown in Fig.~\ref{popcolcol}. The synthesis method
of W94 prevents calculation of the very early evolution stages of a
stellar population. Young populations are especially hard to synthesize
for the lower metallicities, simply because there are no young, low
metallicity stars in the solar neighborhood that can be used as input
stars for the models. 

The lower metallicity populations of the W94 models are clearly bluer in
all color combinations.  Age and metallicity are not complete
degenerate.  The offset in optical--near-IR colors is slightly larger
than the offset in optical--optical colors.  A low metallicity system
can be recognized by its blue $V$--$K$ or $R$--$K$ color with respect to
its $B$--$V$ and $B$--$I$ colors.  When mixing populations of different
age and metallicity the model grid points can be more or less added as
vectors and then age and metallicity are degenerate.  An individual
galaxy can never be pinpointed to a certain age and metallicity using
broadband colors. 

\begin{figure*}
 \mbox{\epsfxsize=18.6cm\epsfbox[24 163 600 675]{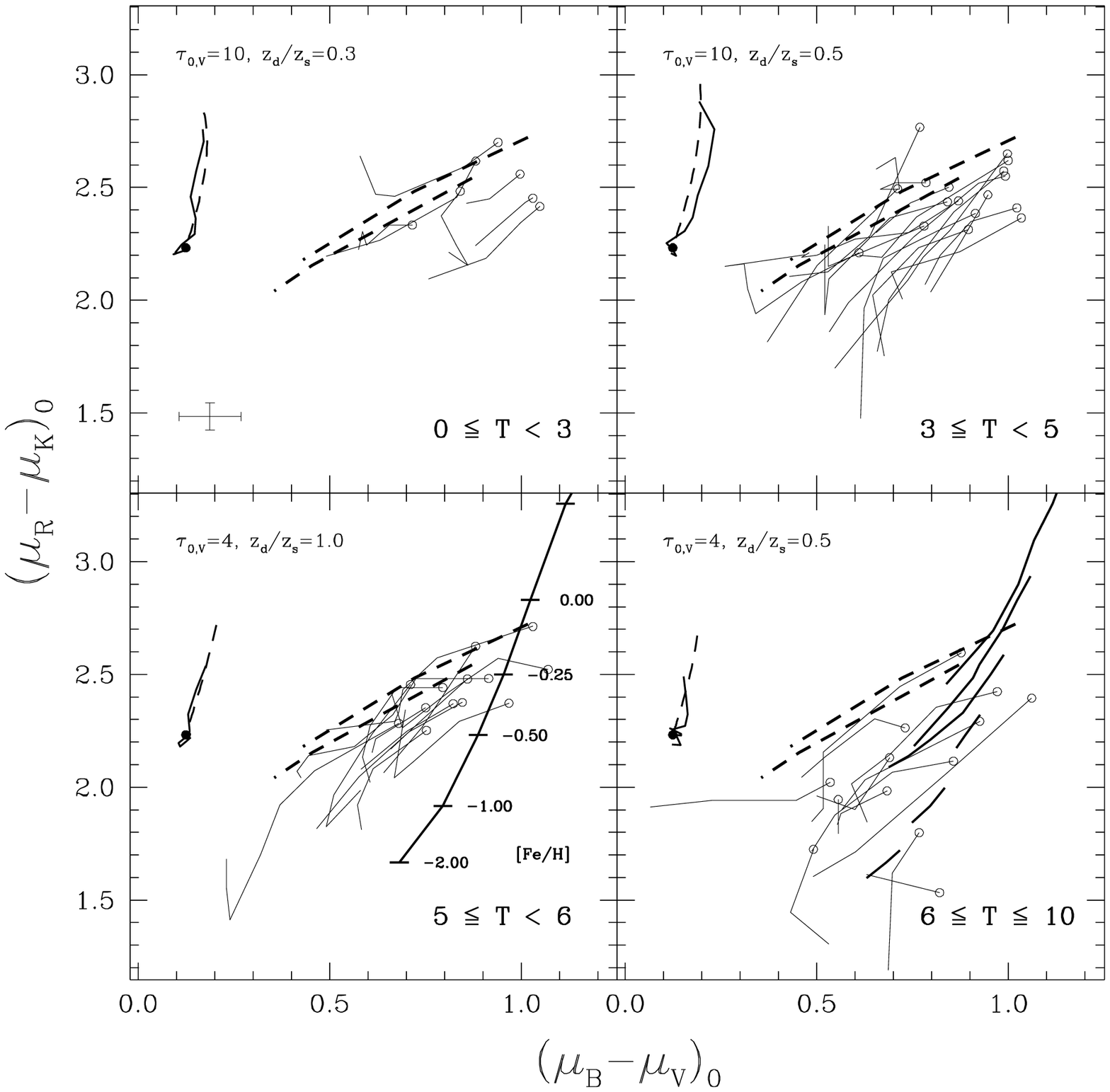}}
 \caption[]{
 The $B$--$V$ versus $R$--$K$ color--color plots of the program galaxies
divided into the four morphological type bins indicated at the bottom-right
of the panels.  The centers of the galaxies are indicated by the open
circles; the thin lines follow the color profiles in radial direction in
steps of 1\,$K$ passband scalelength.  At the bottom-left of the
0$\le$T$<$3 panel, the typical rms error in the zero-point calibration
is indicated.  Some of the dust models of Sect.~\ref{dusmod} are plotted
in the top-left of the panels.  The thick dashed lines in the center of
the panels connect the 17\,Gyr and the 8\,Gyr points of the BC96 models
presented in Sect.~\ref{popmod}.  In the 5$\le$T$<$6 panel the 12\,Gyr
single burst W94 models for different metallicities are connected,
the marks indicate the [Fe/H] values.  The thick solid lines in the
6$\le$T$\le$10 panels represent the W94 models.  Points of equal
metallicity, but different age are connected. 
\vspace{-0.5ex}
 }
 \label{datcolcol1}
\end{figure*}
\begin{figure*}
 \mbox{\epsfxsize=18.6cm\epsfbox[24 163 600 675]{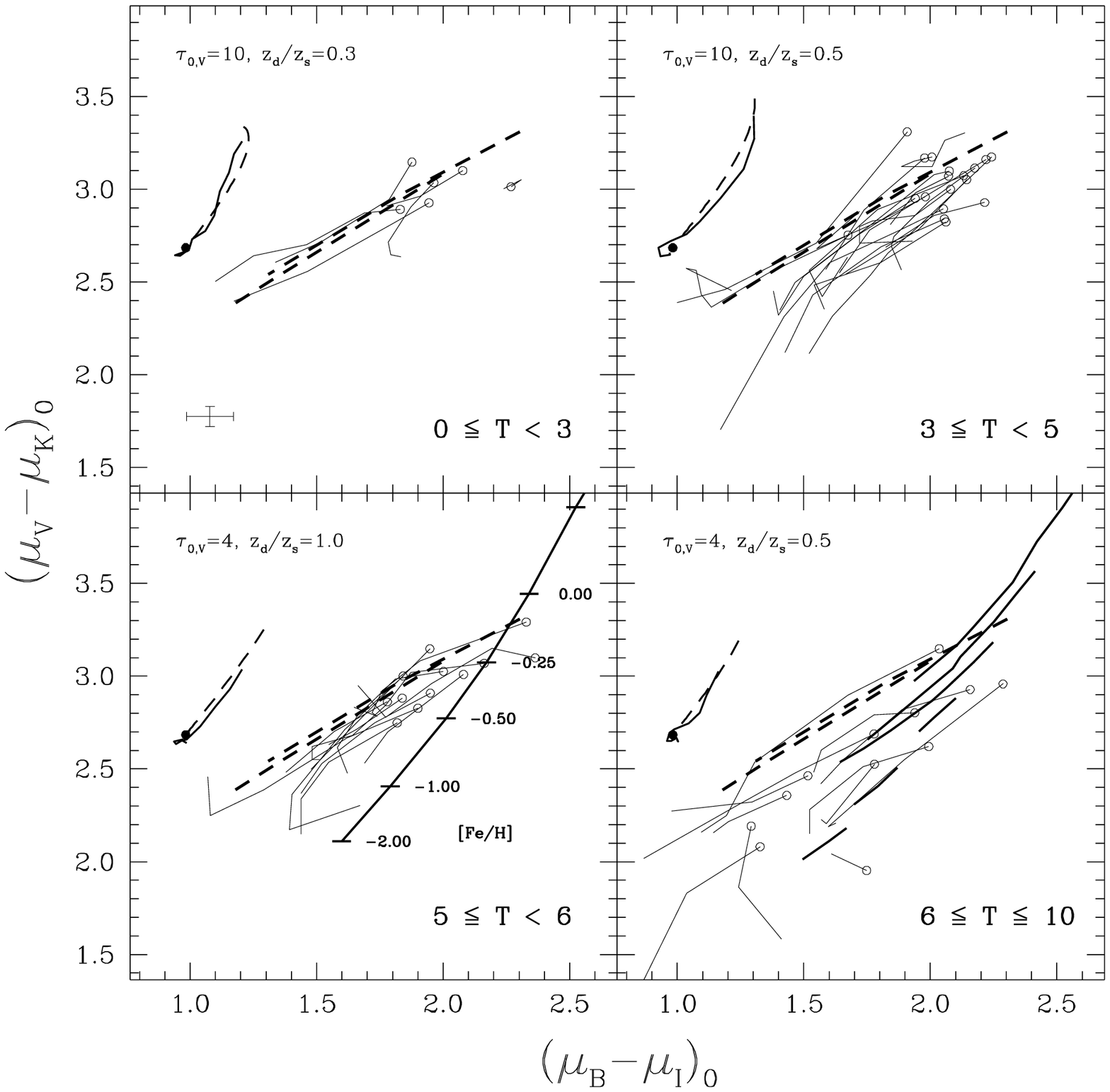}}
\caption[]{
 Same as Fig.~\ref{datcolcol1}, but for $B$--$I$ versus $V$--$K$.
\vspace{-1.5ex}
 }
 \label{datcolcol2}
\end{figure*}

The single burst BC96 and W94 models are clearly offset from each other. 
The solar metallicity BC96 model lies somewhere in between the [Fe/H]
$\! = \!-0.25$ and [Fe/H] $\! = \!0.0$ W94 models.  These are the kinds
of differences between models that were studied in detail by CWB,
resulting mainly from uncertainties in stellar evolutionairy tracks. 
The relative trends agree reasonably well, confirming that it is allowed
to look the relative trends in color space.  The BC96 models will be
mainly used to investigate trends in SFH, because contrary to the W94
models, they incorporate the very early stages of stellar evolution.  To
compare with the data, I will connect the 8\,Gyr points of the different
SFH BC96 models and do the same for the 17\,Gyr points.  These connected
data points will indicate the color trends for equally old populations
with different SFHs.  The W94 models have to be used to look at the
effects caused by metallicity, because different metallicities are not
available in the BC96 models.

\subsection{Color gradients; measurements versus models}
\label{datvermod}

In this section I first use color--color diagrams to display the
observations.  I then try to explain the observed color gradients by
comparing these measurements with the dust models, the different SFH
synthesis models, and finally the full population synthesis models, that
incorporate both age and metallicity effects. 

\subsubsection{The measurements in color--color diagrams}

The color--color profiles of the galaxies are presented in
Figs~\ref{datcolcol1} and~\ref{datcolcol2}.  The data were smoothed to
reduce the noise in the profiles.  The first data point (at the open
circle) is the average over the inner half scalelength (in the $K$
passband) of the luminosity profiles, the other points are averages
moving outward in steps of one $K$ scalelength.  The uncertainty of the
inner point is dominated by the zero-point uncertainty of the
calibration and is indicated at the bottom of the top-left panels of
Figs~\ref{datcolcol1} and~\ref{datcolcol2}.  In the blue direction of
the color--color profiles, lower surface brightnesses are traced (see
Fig.~\ref{mucol}) and errors are dominated by sky background
uncertainties.  Strange kinks at the blue ends of the profiles should
thus not be trusted. 

The profiles of galaxies with T$<$6 are confined to a small region in
the color--color plots. The colors between different passband
combinations are strongly correlated. The scatter is slightly larger than the
average zero-point error. The central colors of the galaxies become on
average a little bit bluer going from T=0 to T=6, but they follow the
main trend.

The galaxies with morphological classification T$\ge$6 clearly deviate
from the main trend.  Their central colors range from the normal red to
extreme blue, even bluer than the bluest outer parts of the earlier type
galaxies.  The spread in the color--color diagram is also significantly
larger for the late-type galaxies.  Some of the late-type galaxies have
very blue $V$--$K$ and $R$--$K$ colors for their $B$--$V$ and $B$--$I$
colors, especially when compared to the earlier types. 

When comparing models with the data one should realize that as soon as
a model for a galaxy has been chosen, it should be applied to all color
combinations. In particular, the same model should be used in
both Figs~\ref{datcolcol1} {\em and} \ref{datcolcol2}. It is tempting to
propose one single model for {\em all} galaxies, because the profiles
are confined to a small region in the diagrams. We only need to explain
the offsets from the main trends with additional parameters.

\subsubsection{Measurements versus dust models}

The reddening profiles produced by the dust models are indicated on the
left in the panels of Figs~\ref{datcolcol1} and \ref{datcolcol2}.  The
color of the underlying stellar population is arbitrary and thus the
dust profile can be placed anywhere in the diagram.  The dust models
have a distinct direction in the color--color diagrams independent of the
dust configuration, as explained in Sect.~\ref{dusmod}.  This direction
is clearly different from the general trend of the data and therefore
the whole gradient cannot be produced by the dust reddening alone.  A
small fraction of the color gradients could be due to dust reddening,
but an additional component is needed to explain the full gradient. 

This does not mean that there could not be large amounts of dust, but
rather that the color gradients are not mainly caused by dust reddening. 
If the dust is not diffuse, but strongly clumped into clouds the amount
of reddening is strongly reduced.  It could be that a large fraction of
the most luminous stars is embedded in dust clouds.  This will not
induce a color gradient or an inclination dependent extinction effect,
but will give the total color profile an offset in the general direction
of the calculated dust models.  The ``dusty nucleus'' models of Witt et
al.~(\cite{Witt92}) give an even better indication of the expected
offset vector if the luminous stars are embedded in dust clouds. 

\subsubsection{Measurements versus metallicity effects}

To what extent can metallicity differences in the stellar populations
account for the color gradients? The 12-Gyr-old W94 single burst models
for different metallicities are connected in the 5$\le$T$<$6 panels of
Figs~\ref{datcolcol1} and~\ref{datcolcol2}.  Models at other ages follow
the same direction in color space.  The single-age,
different-metallicity model does not match the data for most of the
galaxies.  Again another component is needed to explain the color
gradients; radial metallicity differences alone are not sufficient. 
It is important to note that population models cannot be arbitrarily
shifted, they predict fixed colors for a given SFH and metallicity. 

\subsubsection{Measurements versus star formation history}

The BC96 models are indicated by dashed lines in Figs~\ref{datcolcol1}
and~\ref{datcolcol2}.  The 8\,Gyr points of the models have been
connected as well as the 17\,Gyr points.  The reddest ends of these
lines indicate single burst models, the bluest ends represent constant
SFR models.  In between is a model with an exponentially decreasing SFR
with a time scale of 5\,Gyr.  The color vector of the T$<$6 systems seem
to be reasonably well matched by the BC96 models, but most galaxies are
offset to the blue in the $V$--$K$ and $R$--$K$ colors.  The offset is
the smallest for the 8\,Gyr model, but then most of the galaxy centers
are even redder than predicted by the single burst model.  Because we
know that galaxies still have star formation in their centers, this is
an unlikely situation.  As mentioned before, model uncertainties allow
for some shifts in these diagrams, but the most likely conclusion has to
be that although SFH variations as function of radius are a good
driving force for radial color gradients, alone they cannot explain the
full color gradients.  This is especially true for the galaxies with T$\ge$6,
which certainly have $V$--$K$ and $R$--$K$ colors too blue to be
explained by the solar metallicity BC96 models. 

\subsubsection{Measurements versus age and metallicity effects}

The W94 models in the 6$\le$T$\le$10 panel indicate that the very blue
galaxies in this panel can be described very well by low-metallicity
population synthesis models.  As with the earlier type galaxies, the radial
color trends are reasonably well described by age differences, but at
all radii lower metallicities are needed for most of the galaxies.  A
number of them are so blue in all color combinations that their stellar
components must be young and of low metallicity. 

Galaxies are known to have radial metallicity gradients in their current
gas content (Villa-Costas \& Edmunds~\cite{VilEdm92}, hereafter VE;
Zaritsky et al.~\cite{Zar94}, hereafter ZKH), and have radial SFRs that
are not linearly correlated with their radial stellar surface
brightness, which means they do not have one SFH as function of radius
(Ryder \& Dopita~\cite{RydDop94}).  As long as there are no consistent
stellar population synthesis models that incorporate very young stellar
evolutionary stages at all metallicities, it is difficult to make
quantitative statements about the observed colors and color gradients of
the galaxies. 

Limits can be set on the models using the metallicity measurements
collected by VE and ZKH.  Recall that metallicity measurements yield the
current gas metallicities in \hii\ regions, so that the underlying
stellar component could have completely different metallicity values. 
For galaxies T$<$6 the 12+log(O/H) values run from $\sim$9.3 in the
center to $\sim$8.6 at $R_{25}$.  The O/H indices of later type galaxies
are a few tenths lower, from about 8.9 to 8.2, but with more or less the
same gradient across the disk.  Using log(O/H)$_\odot \!\simeq \!-3.08$
(Grevesse \& Anders~\cite{GreAnd89}) and [O/Fe]\,$\simeq \!0.1$$\sim$0.5
(e.g.\ Wyse \& Gilmore~\cite{WysGil88}) the O/H values can be
transformed to [Fe/H] values and used with the W94 models.  The [Fe/H]
values run from about a central 0.2 to --0.6 in the outer regions in
galaxies with T$<$6, and from approximately --0.1 to --1.1 in the later
type galaxies. 

The metallicities just calculated should not be taken too literally in
the comparison of the W94 models with the data in Figs~\ref{datcolcol1}
and \ref{datcolcol2}.  As noted before, models should be used to
indicate trends in color--color space but cannot be expected to give
absolute colors in an individual galaxy.  Still, the $B$--$V$ and
$B$--$I$ colors are far too red at each radius using the 12\,Gyr W94
models in the metallicity range determined from the \hii\ regions
([Fe/H]\,=\,0.2 to -0.6).  This must mean that either the underlying
stellar populations have a much lower average metallicity than the
surrounding gas or that at each radius a much younger stellar population
is present, making the average age lower than 12\,Gyr.  In the center,
the contribution of the young stars cannot be very large, because the
color vector of SFH (indicated for solar metallicity by the BC96 models)
is in the wrong direction from the (t=12\,Gyr, [Fe/H]\,=\,0.25) point in
the 5$\le$T$<$6 panel.  Therefore the center probably contains a mix of
populations of different metallicities, with an average metallicity far
lower than the current metallicity of the surrounding gas
([Fe/H]\,$\simeq$\,0.2), but on average higher than the gas metallicity
of the outer regions.  Towards the outer parts of galaxies, the SFH
color vector gives an excellent description of the observed color trend. 
In the outer regions it is unlikely that the metallicity of the stars is
much higher than that of the gas for all galaxies, and the metallicity
of the stars must be of the order [Fe/H]\,=\,-0.6 (-1.1 for the late
types) or less. Thus the color gradients are probably driven by a
combination of SFH and metallicity differences as function of radius. 
The outer parts of galaxies are clearly younger on average than the
central regions.  If part of the color gradients is also caused by
reddening, the age effect must even be larger, because this is the only
effect that has a color vector that can compensate the dust color
vector. 

In conclusion, the centers of spiral galaxies with T$<$6 contain
probably a relatively old population of stars, with a range in
metallicities.  The populations in the outer regions are on average much
younger and their metallicity is probably lower, as the gas metallicity
is much lower than the metallicity needed to explain the central colors. 
Overall, later type galaxies have lower metallicity, and a number of
them are dominated by very young, low-metallicity populations.  The
observed color gradients cannot be caused by reddening alone, if the
dust properties used in the models are more or less correct.  Likewise,
metallicity gradients cannot cause the color gradients on their own, the
color vectors are in the same direction as the dust model vectors, but
as metallicity gradients are observed in the gas, it is likely they are
present in the stars as well.  Age gradients across the disk have to be
even larger if reddening or metallicity gradients are important.  It
should be noted that the population models can predict {\em both} the
right colors and the right color gradients within one system of models,
while the dust models can at best explain only the color gradients.

\section{Colors and the structural galaxy parameters}
 \label{struccol}

Most of the information that can be extracted from the galaxy colors of
this data set is contained in the previous sections, but for most data
sets such detailed radial color information is not available.  In order
to allow comparisons with these other data sets, a number of
relationships between colors and fundamental galaxy parameters are shown
in this section. 

In the previous section I argued that color gradients derived from
different passband combinations are correlated in such a way that
stellar population differences seem to be the most reasonable
explanation of the phenomenon.  Figure~\ref{mucol} suggests that the
slope of the color gradient has a universal value for all galaxies, but
this is an oversimplification. 

\begin{figure}
 \mbox{\epsfxsize=8.6cm\boundboxo{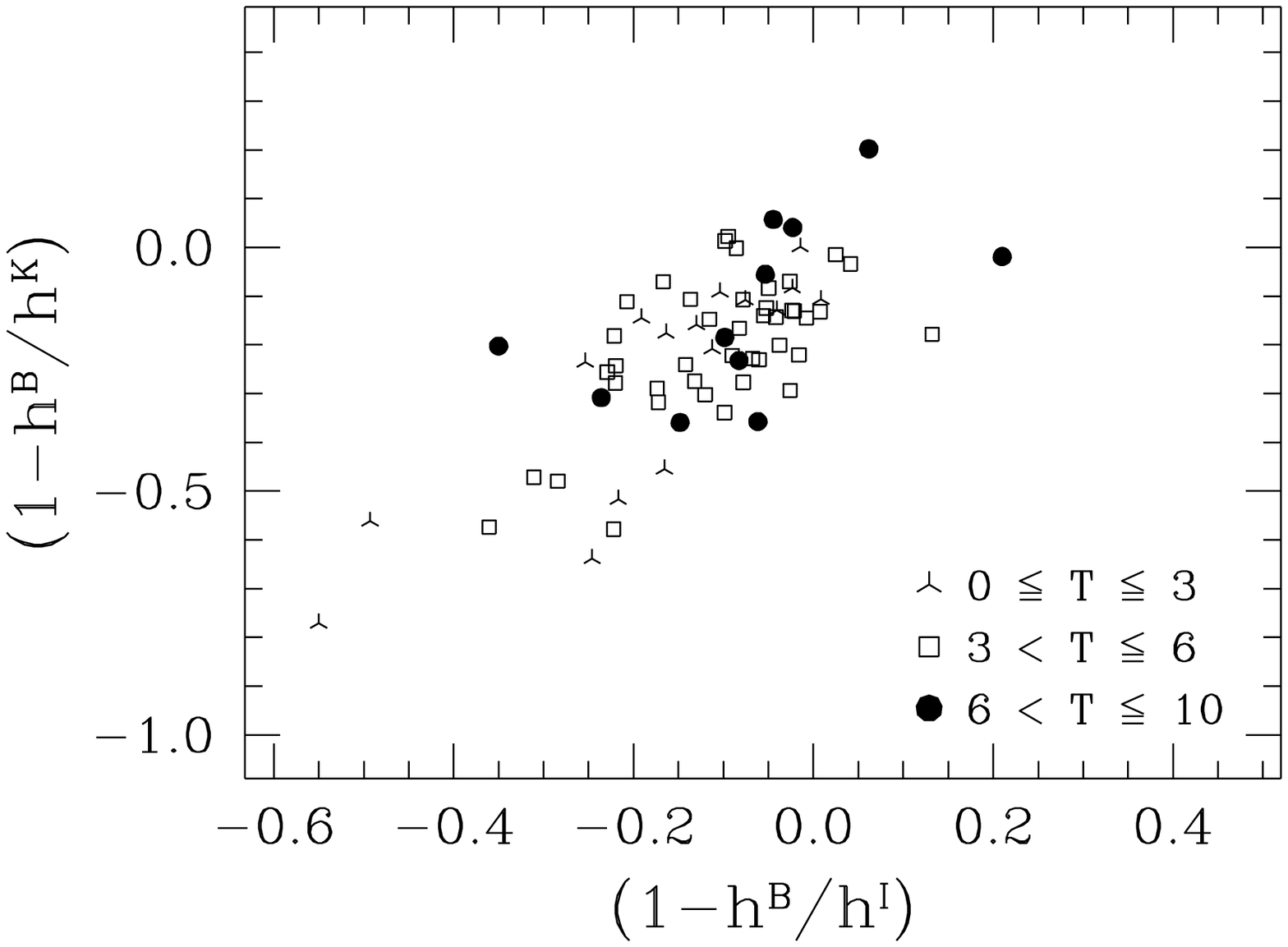}}
 \caption[]{
 The difference in scalelength between the different passbands.  Only
points with errors smaller than 0.15 are plotted.  Different symbols
are used to denote the indicated morphological types. 
 }
 \label{hBIhVK}
 \end{figure}

The change in scalelength ($h^\lambda$) as function of passband can be
used to parameterise color gradients in the disk.  The bulge/disk
decomposition technique and the method used to determine the central
surface brightnesses and scalelengths for the current data set were
described in Paper~II.  A trivial recalculation shows that each axis of
Fig.~\ref{hBIhVK} indicates approximately the color change per
scalelength in the relevant passband combinations.  This figure
illustrates again that color gradients are correlated, but also that
there is no universal value for the gradient for all galaxies.  The
values range from about 0.1 to --0.8 mag per scalelength in $B$--$K$.

The change in scalelength was used to find correlations between the
steepness of the color gradient and other structural galaxy parameters.
A large number of structural parameters were investigated, but
none of them showed a correlation with color gradient.  The
parameters investigated include: inclination, morphological type,
central surface brightness, scalelength, bulge-to-disk ratio,
integrated magnitude, integrated colors, bulge color, bar versus
non-bar, \hi\ and CO fluxes, far-infrared IRAS fluxes and colors, group
membership and rotation velocity.  Fluxes at the different wavelengths
normalized by area or integrated $K$ passband flux were also
used in the correlations, but with a negative result.  A weak
correlation was found only between the steepness of the color
gradient and the central color of the disk.

\begin{figure}
\mbox{\epsfxsize=8.8cm\boundboxo{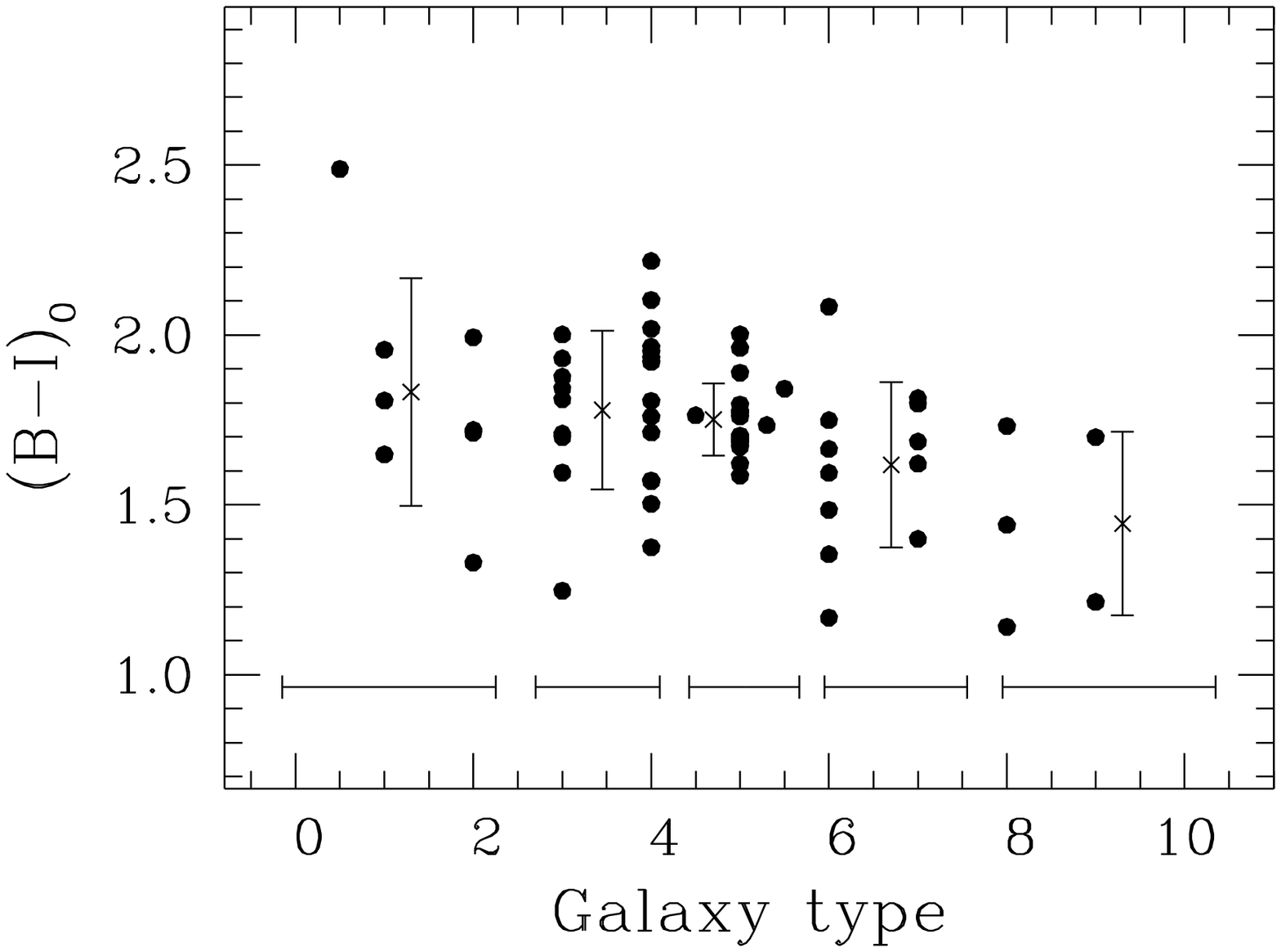}}
 \caption[]{
The Galactic reddening corrected integrated $B$--$I$ colors of the galaxies as
function of morphological type index. The crosses show the values
averaged over the bins indicated by the horizontal bars. The vertical
bars are the standard deviations on the mean values. Only the galaxies
with an error of less than 0.5 mag in their color were used. 
\vspace{1.ex}
}
 \label{type_colfig}
 \end{figure}

In the literature integrated colors of galaxies are usually used to
determine galaxy properties.  The integrated colors as function of type
are presented in Table~\ref{type_coltab} for this galaxy sample and the
$B$--$I$ colors are plotted in Fig.~\ref{type_colfig}.  There is a clear
correlation between type and color, but the scatter is large.  The
integrated color of a galaxy is dominated by the color of the central
region and even though color correlates with (central) surface
brightnesses (Fig.~\ref{mucol}), one should note that each morphological
type comes in a range of central surface brightnesses (Paper~III) which
then explains the large scatter in the integrated colors.  It is better
to determine and compare the colors of galaxies at a fixed isophote when
looking for correlations. 

\begin{figure}
 \mbox{\epsfxsize=8.6cm\boundboxo{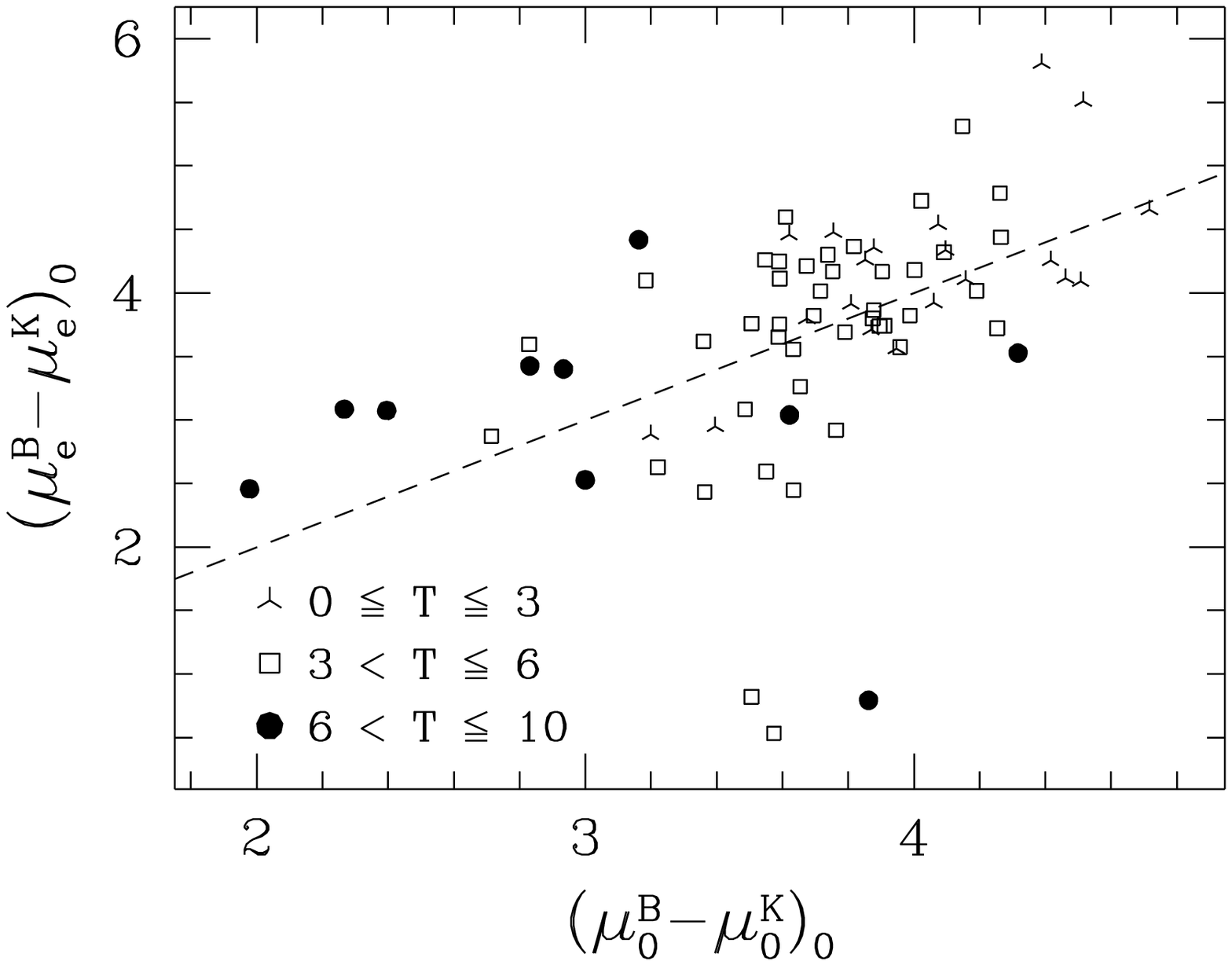}}
 \caption[]{
The central surface brightness $B$--$K$ color of the disk versus the effective
surface brightness $B$--$K$ color of the bulge. Different symbols are
used to denote the indicated morphological type ranges. The dashed line 
indicates the line of equality.
}
 \label{colmuomue}
 \end{figure}

In Fig.~\ref{colmuomue}, the central bulge and disk colors of the
galaxies are compared.  The colors of bulge and disk are clearly
corre\-lated.  This could be expected, as most color profiles show no
clear changes in color gradient in the bulge region.  Excluding the
three deviant points with bulge $B$--$K$ colors bluer than 2\,mag, the
bulge is on average 0.14$\pm$0.55\,mag redder in \mbox{$B$--$K$} than
the central disk color, which means that the stellar populations
probably do not differ very much.

\section{Discussion}
 \label{discus4}

An important consequence of the color differences in and among
galaxies is the implied change in the $M/L_\lambda$ values. The comparison
of the data with the dust and stellar population models indicates that
the color gradients in the galaxies of this sample result mainly from
population changes and that dust reddening only plays a minor role. The
$M/L_\lambda$ values corresponding to the synthesis models presented in
Figs~\ref{datcolcol1} and~\ref{datcolcol2} are listed in
Table~\ref{m_l}. 

{\tabcolsep=1.75mm
\begin{table*}

\begin{tabular}{l@{\hspace{0.7cm}}cc@{\hspace{0.7cm}}cc@{\hspace{0.7cm}}cc@{\hspace{0.7cm}}cc@{\hspace{0.7cm}}cc}
\hline
\hline
\ \ \ \ RC3 type & \# &\multicolumn{1}{l}{$B$--$V$}
         & \# &\multicolumn{1}{l}{$B$--$R$}
         & \# &\multicolumn{1}{l}{$B$--$I$}
         & \# &\multicolumn{1}{l}{$B$--$H$}
         & \# &\multicolumn{1}{l}{$B$--$K$}\\
\hline
\ \ \ \ 0$\le$T$\le$2 &  7 & 0.81 $\pm$ 0.19 & 10 & 1.35 $\pm$ 0.22 &  8 & 1.83 $\pm$ 0.31 & 6 & 3.46 $\pm$ 0.21 & 10 & 3.75 $\pm$ 0.28\ \ \\
\ \ \ \ 2$<$T$\le$4   & 14 & 0.74 $\pm$ 0.13 & 26 & 1.20 $\pm$ 0.19 & 24 & 1.78 $\pm$ 0.23 &  6 & 3.14 $\pm$ 0.34 & 26 & 3.53 $\pm$ 0.28\\
\ \ \ \ 4$<$T$<$6     & 19 & 0.67 $\pm$ 0.15 & 19 & 1.20 $\pm$ 0.13 & 20 & 1.75 $\pm$ 0.10 & 10 & 3.28 $\pm$ 0.14 & 21 & 3.51 $\pm$ 0.26\\
\ \ \ \ 6$\le$T$<$8   & 11 & 0.59 $\pm$ 0.21 & 11 & 1.06 $\pm$ 0.26 & 12 & 1.61 $\pm$ 0.23 &  3 & 2.96 $\pm$ 0.62 &  8 & 3.18 $\pm$ 0.45\\
\ \ \ \ 8$\le$T$\le$10&  4 & 0.69 $\pm$ 0.14 &  5 & 1.10 $\pm$ 0.10 &  5 & 1.44 $\pm$ 0.24 &  0 & --- &  4 & 2.99 $\pm$ 0.09\\
\hline
\hline
 \end{tabular}
 \caption[]{
 The integrated colors for different morphological type ranges using
only the photometric observations of Paper~I with errors in the colors
less than 0.5 mag. The \# lists the number of galaxies in each subsample.}
\label{type_coltab}

 \end{table*}
}
%======================================================================

In every passband, young populations have much lower $M/L_\lambda$
values than old populations.  Young populations still contain very
massive stars, which are very luminous for their mass, but expend their
energy quickly.  Continually adding in young populations, as done in the
exponentially declining and constant SFR BC96 models, decreases the
$M/L_\lambda$ values drastically.  Young massive stars are blue so that
this effect is most pronounced in the $B$ passband.  The change in
$M/L_B$ is about a factor of 6-11 between the 2 and 17\,Gyr for both the
BC96 and the W94 models and $M/L_B$ changes a factor of 3-5 between the
single burst and the constant SFR BC96 models. 

Metallicity has an entirely different effect on the $M/L_\lambda$ ratios.
The $M/L_B$ values increase with metallicity, while the $M/L_K$ values
decrease with metallicity. The turnover point is somewhere between the $I$
and the $J$ passband. This was the motivation for W94 to recommend the
$I$ passband for standard candle work and for studies of $M/L$ in
galaxies. 

The recommendation of the $I$ passband is not entirely obvious,
because the choice of optimum passband depends on whether one expects
extinction, age or metallicity to have the largest
influence on the photometry. If extinction is expected to play a
role, the $K$ passband should be used, as the extinction in the $K$
passband is $\sim$4 times less than the extinction in the $I$ passband.
The $K$ passband should also be preferred if differences in SFR are
expected to be important; see for instance the change in $M/L_\lambda$
ratios of the different BC96 models. One should turn to the $I$ and $J$
passbands only when the metallicity differences among the different
objects are large.

{\tabcolsep=0.59mm
\begin{table}
\begin{tabular}{rrrr@{\ \ \ }rrrrrr}
\hline
\hline
\multicolumn{1}{c}{age} & \multicolumn{3}{c}{BC models} & \multicolumn{6}{c}{W94 models, {[Fe/H]=}} \\
Gyr & s.b.&exp.&cnst.& --2 & --1 & --0.5 & --0.25 & \multicolumn{1}{c}{0.0} & 0.25\\
\hline
\multicolumn{9}{c}{$B$ passband}\\
 2 & 0.90 & 0.82 & 0.30 & --- \ & ---  & ---  & 1.52 & 1.73 & 2.42\\
 5 & 3.13 & 1.02 & 0.67 & --- \ & ---  & ---  & 3.32 & 3.95 & 5.26\\
 8 & 4.33 & 1.42 & 1.02 & 2.60  & 3.39 & 4.16 & 5.19 & 5.96 & 8.05\\
12 & 5.70 & 2.25 & 1.43 & 3.71  & 4.81 & 6.26 & 7.41 & 8.70 &11.95\\
17 & 9.91 & 4.02 & 1.88 & 5.01  & 6.45 & 8.53 &12.28 &10.54 &17.15\\
\hline
\multicolumn{9}{c}{$I$ passband}\\
 2 & 1.27 & 1.25 & 0.79 & --- \ & ---  & ---  & 1.09 & 1.03 & 1.24\\
 5 & 2.26 & 1.29 & 0.91 & --- \ & ---  & ---  & 1.86 & 2.00 & 2.21\\
 8 & 2.82 & 1.57 & 1.28 & 2.34  & 2.53 & 2.45 & 2.67 & 2.77 & 3.08\\
12 & 3.23 & 2.07 & 1.68 & 3.04  & 3.30 & 3.50 & 3.58 & 3.60 & 4.17\\
17 & 4.71 & 2.95 & 2.13 & 3.84  & 4.10 & 4.39 & 4.73 & 4.74 & 5.54\\
\hline
\multicolumn{9}{c}{$K$ passband}\\
 2 & 0.43 & 0.65 & 0.24 & --- \ & ---  & ---  & 0.52 & 0.42 & 0.42\\
 5 & 0.80 & 0.62 & 0.44 & --- \ & ---  & ---  & 0.78 & 0.72 & 0.60\\
 8 & 1.13 & 0.74 & 0.62 & 1.70  & 1.50 & 1.16 & 1.12 & 0.86 & 0.75\\
12 & 1.15 & 0.87 & 0.78 & 2.10  & 1.88 & 1.81 & 1.35 & 1.06 & 0.87\\
17 &\ 1.54 &\ 1.13 &\ 0.93 &\ 2.59  &\ 2.21 &\ 1.90 & 1.69 & 1.29 & 1.04\\
\hline
\hline
 \end{tabular}
 \caption[]{
 The $M/L_\lambda$ values in solar units for the models presented in
Section~\ref{popmod}.  For the BC models single burst models (s.b.),
exponential declining SFH models (exp.) and constant SFR models (cnst.)
are listed.  The single burst models of W94 are listed for several
different metallicities. 
 \label{m_l}
}

\end{table}
}

What influence do the current observations have on studies depending on
$M/L_\lambda$ ratios? I address two issues here: rotation curve fitting
and the TF-relation.

The principle of rotation curve fitting is simple.  One tries to explain
the distribution of the dynamical mass of a galaxy by assigning masses
to its known ``luminous'' (at whatever wavelength) components.  The
dynamical mass distributions can be determined from optical rotation
curves along the major axis (e.g.\ Rubin et al.~\cite{Rub85}; Mathewson
et al.~\cite{Mat92}) or, more sophisticatedly, from \hi\ or other
two-dimensional velocity fields (e.g.\ Bosma~\cite{BosPhD};
Begeman~\cite{BegPhD}; Broeils~\cite{BroPhD}).  The mass assignment to
the \hi\ and other gas components is relatively straightforward, but the
mass assignment to the stellar components has proven troublesome. 
Generally, $M/L$ values have been assigned to the different stellar
components (bulge and disk) using the maximum disk hypothesis (van
Albada et al.~\cite{Alb85}; van Albada \& Sancisi~\cite{AlbSan86}). 
This exercise revealed the ``missing light'' problem: the dynamical mass
of a galaxy is much larger than the maximum ``luminous'' mass, with the
main discrepancy in the outer regions. 

In the game of rotation curve fitting, one often encounters the use of
$B$ or at best $R$ passband luminosity profiles.  Table~\ref{m_l}
clearly illustrates the danger of using these passbands, especially if
one considers the color gradients observed here.  Independent of
whether the observed color gradients are caused by age or metallicity
gradients (or by reddening for that matter), the $M/L_B$ ratios will be
much higher in the center than in the outer regions.  Consequently the
``missing light'' discrepancy between inner and outer regions is even
larger than estimated by the use of the $B$ passband profiles together
with a constant $M/L$.  So what is the optimum passband to be used for
rotation curve fitting?

Following the conclusions of Sect.~\ref{datvermod}, let us assume that
the central region of a T$<$6 galaxy consists of old stellar populations
with a range in relatively high metallicities, say on average
t\,=\,12\,Gyr and [Fe/H]\,=\,0.  It is not likely that the metallicity
of the stars in the outer regions is higher than that of the gas.  The
outer populations are younger than the inner populations and let us
assume that their average parameters are [Fe/H]\,=\,--0.5 and
t\,=\,8\,Gyr.  Table~\ref{m_l} shows that $M/L_B$ changes by a factor
2.09 for those two populations, $M/L_I$ by a factor 1.47, and $M/L_K$ by
a factor 0.91.  Repeating this exercise for lower-metallicity late-type
systems gives similar results.  As long as the outer regions of galaxies
are younger than the inner regions (and the example used was not very
extreme), the $K$ passband is the optimum choice for rotation curve
fitting.  This is especially true if extinction plays a role, which is
expected to be the case for the highly inclined galaxies normally used
for rotation curve fitting. 

The $M/L$ effects on integrated magnitudes (as used in the TF-relation)
are less trivial.  In Sect.~\ref{Intro4} it was argued that the
integrated colors are dominated by the central colors of galaxies.  In
Figs.~\ref{datcolcol1} and~\ref{datcolcol2} the open circles indicate
the central colors of the galaxies and are therefore representative of
the integrated colors.  For the T$<$6 galaxies, the central colors
follow the same trend as the color gradients within galaxies (as might
be expected from Fig.~\ref{mucol}) and therefore the same argument as
for rotation curve fitting can be applied to recommend the $K$ passband
for TF-relation work.  Note that the differences in central color in the
bins of T$<$6 are much smaller than the differences within the
galaxies themselves and a small color correction term should be
sufficient to translate all galaxies to a common $M/L$ scale.  The
central colors of T$\ge$6 galaxies show a different distribution in
color--color space.  They come in a wide range of ages and metallicities
and their $M/L_K$ values can easily differ by a factor of two and by a
factor of four in $M/L_I$.  This would introduce an uncertainty of
$\sim$0.75 $K$-mag or $\sim$1.5 $I$-mag in the TF-relation respectively,
if one would simply assume that the TF-relation is tracing the
connection between luminous mass and dynamical stellar mass.  A
two-color correction might then be needed to reduce the scatter in the
TF-relation if late-type galaxies are also included in the sample. 
Because the scatter in the TF-relation is often much smaller than the
indicated values, one may conclude that the TF-relation is not simply
tracing the connection between luminous mass and dynamical stellar mass,
and that the amount of dark matter is varying systematicly with galaxy
color.  But in short, a shift along one vector in Figs~\ref{datcolcol1}
and~\ref{datcolcol2} is sufficient to bring the centers of most T$<$6
galaxies to one point, at least two vectors are needed to bring the
centers of the T$\ge 6$ galaxies to one point. 

A large number of galaxy formation and evolution theories predict
metallicity gradients (for references see e.g.\ VE and
Matteucci~\cite{Mat89}, \cite{Mat92}) and age gradients
(Kennicutt~\cite{Ken89}; Dopita \& Ryder~\cite{DopRyd94} and references
therein) in galaxies.  Several predict both at the same time, like
viscous galaxy evolution models (Lin \& Pringle~\cite{LinPri87};
Sommer-Larsen \& Yoshii~\cite{SomYos90}), the models of Wyse \&
Silk~(\cite{WysSil89}) and various gas infall and galaxy merger models. 
The young, low metallicity late-type systems observed here would be
ideal merger candidates to replenish the outer regions of large galaxies
with nearly unprocessed gas from which new stars can be formed.  A
detailed study of all possible models, investigating time scales and
metallicity ranges involved, is beyond the scope of this work. 

A key question that is not addressed in this investigation is whether
the color gradients originate in the arm or in the inter-arm region, or
are present in both regions.  High resolution and high signal-to-noise
observations are needed to solve this question, and the present data set
is not suited for such an investigation.  A detailed study of a few
large nearby galaxies is in preparation (Beckman et al., private
communication). 

Another question not addressed in this paper is the vertical structure
of dust and stellar populations. In a similar approach as followed here,
color maps of edge-on galaxies are very well suited for such studies
(e.g.~Wainscoat et al.~\cite{Wai89}, Just et al.~\cite{Jus96}).
Unfortunately, so far most of these studies have not included scattering
in the dust models, but the models of Just et al.\ are most promising, as
they do include a link between the stellar and dynamical evolution of
the galaxies.

\section{Conclusions}
 \label{concl4}

The stellar and the dust content of a large sample of galaxies was
investigated using the color profiles of these galaxies.  Data in four
optical and two near-IR passbands were combined simultaneously to derive
structural properties of the sample as a whole, rather than for
individual galaxies.  The main conclusions are:

 \begin{itemize} 
 \item Almost all spiral galaxies become bluer with increasing radius.
 \item The colors of galaxies correlate strongly with surface
brightness, both within and among galaxies. The morphological type is
an additional parameter in this relationship, because at the same
surface brightness late-type galaxies are bluer than early-type
galaxies.
 \item Realistic 3D radiative transfer modeling indicates that
reddening due to dust extinction cannot be the major cause of the color
gradients in face-on galaxies. The predicted color vectors in
color--color space are not compatible with the data, unless the assumed
scattering properties of the dust are entirely wrong.
 \item The color gradients in the galaxies are best explained by
differences in SFH as function of radius, with the outer parts of galaxies
being on average much younger than the central regions. This implies that
the stellar scalelength of galaxies is still growing. The central
stellar populations in a galaxy must have a range in metallicities
to explain the red central colors of the galaxies.
 \item A consequence of the population changes implied by the color
differences in and among galaxies is that there are large changes in
$M/L$ values in and among galaxies.  These changes in $M/L$ make the
missing light problem in spiral galaxies as derived from rotation curve
fitting even more severe. 
 \item The $H$ and $K$ passbands are recommended for standard candle
work and for studies depending on $M/L$ ratios in galaxies.
 \end{itemize}

 \begin{acknowledgements}
 Many thanks to Stephane Charlot and Guy Worthey for providing machine
readable versions of their stellar popu\-la\-tion synthesis results.  I
thank Edwin Huizinga for discussing and checking the dust models.  I
would like to thank Erwin de Blok, Thijs van der Hulst, Piet van der
Kruit, Ren\'e Oudmaijer, Penny Sackett and Edwin Valentijn for their
stimulating discussions and the many useful suggestions on earlier
versions of the manuscript.  This research was supported under grant
no.~782-373-044 from the Netherlands Foundation for Research in
Astronomy (ASTRON), which receives its funds from the Netherlands
Foundation for Scientific Research (NWO). 
 \end{acknowledgements}

%\newpage
%\onecolumn
\appendix
\section{Monte Carlo radiative transfer simulations of
light and dust in exponential disks}
\label{apmodel}

The modeling of dust extinction in extragalactic systems has a long
history. The effects of scattering are often ignored or are assumed to be
no more than a scaling factor (Disney et al.~\cite{DDP89};
Huizinga~\cite{HuiPhD}), which is correct as long as one is not looking
at wavelength dependent effects. Therefore, scattering is included in
most studies investigating the wavelength dependent effects of
extinction (e.g.\ Kylafis \& Bahcall~\cite{KylBah87}; Bruzual et
al.~\cite{Bru88}, Witt et al.~\cite{Witt92}, Byun et al.~\cite{Byun94})

The Monte Carlo approach followed here is a somewhat ``brute force'' method,
but it has the advantage that it can be used for luminous and dust
geometries containing little symmetry. Even though smooth light and dust
configurations are applied here, this method can easily be extended to
include dust clouds and spiral structure.

 \subsection{The mathematical method}
 A Monte Carlo radiative transfer code was used to calculate the light
and color distribution of disk galaxies as seen by a distant observer.
The Monte Carlo principle as applied to extinction in gaseous nebulae
is described in detail by Witt (\cite{Witt77}).  In the computer model
the paths of many ($\sim10^6$) photons were followed as they
traveled through the absorbing and scattering dusty medium.  At
great distance the photons were collected to produce a galaxy image
which was used for further study. 

The trajectory of a photon through a dusty medium is determined by
random processes.  One can characterize these processes by a probability
function $p(x)$ on an interval $(a,b)$ such that
 \begin{equation}
\int_{a}^{b} p(\xi ) d\xi = 1. 
 \end{equation}
Using a random number generator, which produces a random number
$R$ distributed uniformly in the interval $0 \!\le \!R \!\le \!1$, one can
simulate an event with frequency $p(x)dx$ in the interval $(x,x + dx)$
by requiring 
 \begin{equation} 
\int_{a}^{x} p(\xi ) d\xi = R.
\label{Rint}
 \end{equation}
 This equation was used to simulate the birthplace and direction of
photons, as well as the scattering properties of the dust. In the
following sections, each $R$ will denote a new random number in the interval
$0\!\le \!R \!\le \!1$.

\subsection{The creation of photons}

This section describes the creation of photons and the random processes
involved in this creation. Rather than creating photons in space with
a certain density distribution such as in a real galaxy, the photons
were instead created uniformly in space and given an initial intensity
weight to produce the exponential light profile. These initial weights
were reduced by absorption as the photon moved through the dusty
medium.

\begin{figure*}
 \mbox{\epsfxsize=17.5cm\epsfbox[10 340 500 630]{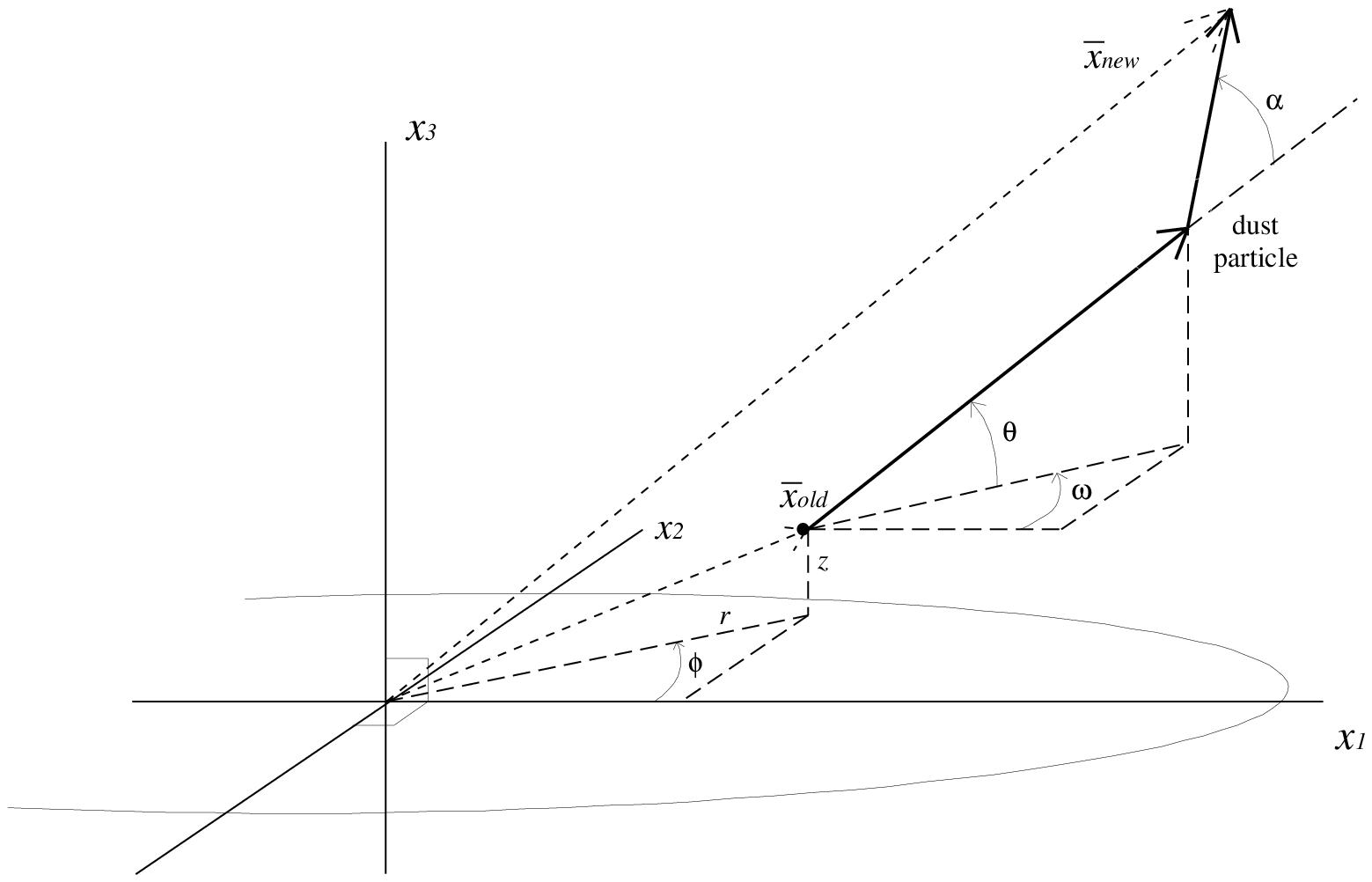}}
 \caption[]{
 The definition of the coordinate system.  A trajectory with a
scattering is indicated for one step size.  A ``photon beam''
starts at $\overline{x}_{\rm old}$ in the direction defined by
($\omega$,$\theta$), scatters off a dust particle over an angle $\alpha$,
and ends at $\overline{x}_{\rm new}$. 
}
 \label{coorfig}
 \end{figure*}

The models consisted of three-dimensional distributions of stellar light
and dust that were chosen independently.  For the stellar light, an
axisymmetric disk-like distribution was used, with an exponential
intensity behavior in both radial ($r$) and vertical ($z$) directions:
 \begin{equation}
I(r,z,\phi) = {\rm e}^{-(r/h_{\rm s}+|z|/z_{\rm s})},
\label{inint}
 \end{equation}
 where $h_{\rm s}$ and $z_{\rm s}$ are the scalelength and the
scaleheight of the stellar distribution respectively, with the
coordinate system is as defined in Fig.~\ref{coorfig}.  The luminosity was
truncated at seven scalelengths and heights.  The exponential behavior
of the radial light distribution of disks in spiral galaxies is well
established, but the vertical light distribution is more controversial,
because extinction effects make measurements difficult.  I have used
the exponential law vertically instead of (for instance) the sech or
sech$^2$ laws (van der Kruit~\cite{Kru88}), because the nearly
unobscured light distribution as obtained with near-IR observations
often can be well described by such an exponential law (Wainscoat et
al.~\cite{Wai89}; Aoki et al.~\cite{Aoki91}).  It is not expected that
the results obtained here will change significantly if other
plausible vertical light distributions are used.  Each ``photon beam''
that was created received an initial weight according to the
$I(r,z,\phi)$ of Eq.~(\ref{inint}). 

Photons were created at a certain position in the model galaxy using the
distribution functions
 \begin{equation}
r = 7 R h_{\rm s} \ \ \ \ 
z = 7 (R-0.5)2 z_{\rm s} \ \ \ \ 
\phi = 2 R \pi,
\label{inpos}
 \end{equation}
and thus their initial position in cartesian coordinates was
 \begin{equation}
\overline{x} = \left( \begin{array}{c} r \cos(\phi) \\
                                       r \sin(\phi) \\
                                       z
                      \end{array} \right).
 \end{equation}
 The distribution functions of Eq.~(\ref{inpos}) create photons
uniformly in cylindrical coordinates, but not in cartesian space.  The
density of created photons is much higher near the center than in the
outer regions and a correction factor is needed.  To still get an
exponential behavior in flux, an additional weight factor of $2 \pi r$
was given to the intensity of each created ``photon beam''.  These
distribution functions were chosen, because the final model images were
azimuthally averaged just as the real observations and in the absence of
dust give these distribution functions an equal number of photons at
each radius in the face-on case.

At creation, the initial flight direction of the photons was specified
by the functions
 \begin{equation}
\omega = (2 R - 1) \pi \ \ \ \  
\theta = (R - 0.5) \pi .
\label{indir}
 \end{equation}
The corresponding directional cosines are
 \begin{equation}
\Delta \overline{x} = \left( \begin{array}{c} \cos(\omega) \cos(\theta) \\
                                              \sin(\omega) \cos(\theta) \\
                                              \sin(\theta) 
                             \end{array} \right).
\label{dircos}
 \end{equation}
 Again, since this would not give a uniform distribution of flux
density in each direction, an extra weight of the form $\cos(\theta)$
was introduced. The initial intensity of a ``photon beam'' at creation
was thus
 \begin{equation}
 I_0(r,z,\phi,\omega,\theta) = 2 \pi r \cos(\theta) {\rm e}^{-(r/h_{\rm s}+|z|/z_{\rm s})}.
 \end{equation}

\onecolumn

\subsection{The dust properties}

Let us first define the notation for the basic equations of radiative
transfer.  A beam of light with intensity $I$ is followed while
traveling along a straight line through a dusty medium.  Since this medium
removes a fraction $\kappa\, d s$ of the incident intensity for each
distance $d s$ traveled, the full equation is
 \begin{equation}
\frac{d I(s)}{d s}  =  -\kappa(s) I(s) + J(s),
 \end{equation}
 where $J(s)$ are the new sources of intensity at point $s$ in the
travel direction. $\kappa$ is the extinction coefficient and
consists of an absorption and a scattering part, and their relative
importance is normally expressed by the albedo ($a$)
 \begin{eqnarray}
\kappa & = & \kappa_{a}+\kappa_{s} \\
a & = & \kappa_{s}/\kappa.
 \end{eqnarray}
  While following the light beam through the dusty medium, no new photons
were added to the initial beam of photons that was created with
intensity $I(0)$, because they are accounted for by the repeating Monte
Carlo principle, and $J(s) \!= \!0$.  For this single ``photon beam''
the radiative transfer equation is now
 \begin{equation}
\frac{d I(s)}{d s}  =  -\kappa(s) I(s)
 \end{equation}
with formal solution
 \begin{equation}
I(x)  =  I(0) {\rm e}^{-\int_{0}^{x} \kappa(s) ds}.
\label{inteq}
 \end{equation}
The integral in Eq.~(\ref{inteq}) defines optical thickness
 \begin{equation}
\tau \equiv \int_{0}^{x} \kappa(s) \,ds.
 \end{equation}
 Since the beam was followed while it scattered through the medium (so
no longer on a straight line), there were no losses by scattering along
the path of the beam.  One still can use Eq.~(\ref{inteq}), but with
$\kappa_{a}(s) \!= \! (1 \!- \!a)\kappa(s)$ instead of $\kappa(s)$ as
long as the line integral is taken along the path traveled.  The
extinction coefficient and the albedo are wavelength dependent, which
creates reddening of stars and could cause the observed color gradients
in the galaxies. 

To calculate the absorption along the path traveled, the dust
distribution must be defined.  The dust extinction coefficient in the
models was taken to have a similar distribution as the stellar light,
but with a scalelength and scaleheight that were chosen independently:
 \begin{equation}
\kappa_{\lambda}(r,z,\phi) = \kappa_{0,\lambda} {\rm e}^{-(r/h_{\rm d}+|z|/z_{\rm d})}
 \end{equation}
 where $\kappa_{\lambda}(r,z,\phi)$ is the local extinction coefficient at
wavelength $\lambda$ and $\kappa_{0,\lambda}$ is the extinction
coefficient at $(r\!=\!0, z\!=\!0)$.  In 
this article {\em the} optical depth of a system is denoted by the
integration of $\kappa_V$ ($\kappa$ in the $V$-passband) along the
symmetry axis from $(r\!=\!0,
z\!=\!-\infty)$ to $(r\!=\!0, z\!=\!\infty)$
 \begin{equation}
\tau_{0,V} = \int_{-\infty}^{\ \infty} \kappa_{0,V} {\rm e}^{-|z|/z_{\rm d}} \,dz 
      = 2 \kappa_{0,V} z_{\rm d}.
 \label{optdepth}
 \end{equation}
 Using Eq.~(\ref{inteq}) one can calculate that a point source located
behind the center of the galaxy will have suffered an extinction in the
$V$-passband for an observer located at its other pole of
 \begin{equation}
I(z=\infty)  =  I(0) {\rm e}^{-\tau_{0,V}},
\label{exflux}
 \end{equation}
assuming that all absorbed {\em and} scattered photons are lost for a
point source.

%\onecolumn
Three parameters are needed to describe dust properties,
the relative extinction $(\tau_\lambda/\tau_V)$, the albedo
($a_\lambda$) and the scattering phase function $\Phi_\lambda(\alpha)$,
and all are dependent on wavelength $(\lambda)$.  As there are almost no direct
measurements of the dust properties in other galaxies, I decided to use
the (also poorly determined) Galactic values.  Studies by Knapen et
al.\ (\cite{Kna91}) and Jansen et al.~(\cite{Jan94}) seem to indicate
that at least the Galactic extinction curve is applicable to some other
galaxies, the other two dust properties have never been measured in
extragalactic systems.

For relative extinction, the Galactic extinction properties of Rieke
\& Lebofsky (\cite{RieLeb85}) were used. The values for the albedo
were drawn from Bruzual et al.~(\cite{Bru88}), as were the values for
the scattering asymmetry parameter $g_\lambda$. This asymmetry
parameter enters in the scattering phase function suggested by Henyey
\& Greenstein (\cite{Henyey})
 \begin{equation}
\Phi_\lambda(\cos(\alpha),g_\lambda) = [(1- g_\lambda^2)/4\pi] (1+g_\lambda^2-2g_\lambda\cos(\alpha))^{-3/2}
\label{HenGre}
\end{equation}
 where $\alpha$ is the scattering angle between the incident and the
deflected photon.  The function is such that $g_\lambda \!= \!\langle
\cos(\alpha) \rangle$, and thus $-1 \!\leq \!g_\lambda \!\leq \!1$. 
Forward scattering dominates for $g_\lambda \!> \!0$ and $g_\lambda \!=
\!0$ results in isotropic scattering.
 To determine a random deflection angle
for a photon with the probability function $\Phi_\lambda$ 
Eq.~(\ref{Rint}) can be used and one finds (Witt \cite{Witt77})
 \begin{equation}
\alpha  =  \arccos(\{(1+g_\lambda^2)- [(1-g_\lambda^2)/(1-g_\lambda+2g_\lambda R)]^2\}/2g_\lambda).
\label{HenGreWit}
 \end{equation} 
 The values used for $\tau_\lambda/\tau_V$, $a_\lambda$, and $g_\lambda$
for the different photometric passbands are listed in
Table~\ref{dusprop}.

%\onecolumn

\subsection{The numerical method}

Each photon beam that was created using Eqs.~(\ref{inpos}) and
(\ref{indir}) was stepped through the dusty medium. The step size was
determined by the local dust density and was chosen in such
way that it would give an optical thickness (absorption plus
scattering) of 0.03 if there was a constant dust density along the
step. This means that the chance of multiple scattering is of order
$(0.03 a_\lambda)^2$ along a step. This is negligible, as it should be,
because only single scattering was incorporated along each step.  It is
trivial to show that the step size has to be of order
 \begin{equation}
\Delta s = 0.03 / \kappa(\overline{x})
         = 0.03 / (\kappa_0 {\rm e}^{-(r/h_{\rm s}+|z|/z_{\rm s})})
\label{step}
 \end{equation}
and the new position is (if there is no scattering)
 \begin{equation}
\overline{x}_{\rm new} = \overline{x}_{\rm old} + \Delta \overline{x} \Delta s.
 \end{equation}

For the actual absorption along a step, one has to use Eq.~(\ref{inteq})
with $\kappa_a \!= \!\kappa(1-a)$ instead of $\kappa$.  The integral was
at each step approximated by four-point Gauss-Legendre quadrature:
 \begin{equation}
\tau_{\Delta s}  =      \int_{\overline{x}}^{\overline{x}+\Delta \overline{x} \Delta s} \kappa_a(\overline{s})\,d\overline{s} 
                \ \ \approx \ \ 0.5 \Delta s \sum_{j=1}^4 w_j
\,\kappa(\overline{x}+0.5(1+x_j)\Delta\overline{x})(1-a),
\label{GauLeg}
 \end{equation}
where the values for $w_j$ and $x_j$ can be found in any handbook on
numerical analysis.
 When there was no scattering during the step, the
$\kappa(\overline{x}_4)$ calculated in Eq.~(\ref{GauLeg}) was used to
determine the step size with Eq.~(\ref{step}) in the next step.

If scattering did occur, the step proceeded slightly differently. The
chance of scattering during a step can be simulated by requiring
 \begin{equation}
{\rm e}^{-\tau_{\Delta s} a_\lambda} < R .
\label{scatchan}
 \end{equation}
In that case the directional cosines of Eq.~(\ref{dircos}) were changed
by an angle given by Eq.~(\ref{HenGreWit}). This left another angle of
freedom for the azimuthal change of the direction which was generated by
 \begin{equation}
\beta = 2\pi R.
 \end{equation}
Rotating the initial directional angles $\omega,\theta$ over the
scattering angles $\alpha,\beta$ gives the new directional cosines
 \begin{equation}
\Delta \overline{x}_{\rm new} =\left( \begin{array}{rr}
\cos(\omega) \cos(\theta) \cos(\beta)&+( \sin(\omega) \cos(\alpha)+\cos(\omega) \sin(\theta) \sin(\alpha) ) \sin(\beta) \\
\sin(\omega) \cos(\theta) \cos(\beta)&+(-\cos(\omega) \cos(\alpha)+\sin(\omega) \sin(\theta) \sin(\alpha) ) \sin(\beta) \\
\sin(\theta)              \cos(\beta)&                                          -    \cos(\theta) \sin(\alpha) \sin(\beta) \\
                     \end{array} \right).
 \end{equation} 
 When Eq.~(\ref{scatchan}) was satisfied and scattering occurred in a step,
the photon proceeded a fraction $f \!= \!R$ of step size $\Delta s$ in the
old direction, before following the new direction
 \begin{equation}
\overline{x}_{\rm new} = \overline{x}_{\rm old} + (f \Delta
\overline{x}_{\rm old}  + (1-f) \Delta \overline{x}_{\rm new}) \Delta s.
 \end{equation}
%\twocolumn

\subsection{Projection on the sky}

The steps in the previous paragraph were repeated until
$|\overline{x}_{\rm new}| \!> \!10 h_{\rm s}$. All $\tau_{\Delta s}$ of
the different steps were added, and the final intensity of the photon
beam was $I_{\rm end} \!= \!I_0 {\rm e}^{-\tau_{\rm tot}}$. The final
intensities were projected on the sky as if the galaxy was being observed
from infinity. Due to the axisymmetric nature of the models, one can
ignore the $\omega$ dependence of the exit direction and rotate all
photons over angle $\omega$ as if they leave the galaxy in the same
direction. The projection on the $(y_1,y_2)$-plane becomes
 \begin{equation}
\overline{y} = \left( \begin{array}{rrr}
-\sin(\omega)            & \cos(\omega)            & 0\\
-\sin(\theta)\cos(\omega)&-\sin(\theta)\sin(\omega)&\cos(\theta)
\end{array} \right) \overline{x}
 \end{equation}
 This procedure of creating photons, stepping through the medium and
projecting the exiting photons on the sky was repeated for two million
photons.  The photons were binned into pixel images in $(y_1,y_2)$
direction and the binning in viewing angle $\theta$ direction resulted
in ten model images from edge-on to face-on.  The binning of photons in
the $\theta$ direction was in equal steps of $b/a$, where $b/a$ is the
axial ratio of an inclined circle.  This has the advantage that each
image has approximately equal flux, at least in the case of no
extinction.  For the analyses discussed in this paper only the face-on
images were used, which contained the photons that exited with an angle
between 90\degr\ and 71.8\degr\ (i.e.\ $1 \!\ge \!b/a \!> \!0.95$). 

%\twocolumn
\subsection{Testing and the results}

The computer program was tested extensively to ensure that the results
were reliable.  First, a point source emitting all its photons to a pole
was placed at the center of the galaxy and the exiting flux was
correctly described by Eq.~(\ref{exflux}).  Tests showed that in the
case of no absorption all images at different viewing angle had an equal
flux.  Tests were made to ensure that the scattering angle phase
function was produced correctly
 %(Fig.~\ref{phidis}) 
 and that the angle between the incident and the deflected photon was
indeed the required angle.  One of the most important tests was the
independence of the number of scatterings on step size.  If the step
size had been too large or if the calculation $\tau_{\Delta s}$ had been
incorrect this could not have been the case.  The model profiles were
tested against the analytic profiles of DDP for the cases with and
without scattering.  In the case of no scattering (all scattered photons
were removed to mimic absorption) the agreement was perfect. 
Figure~\ref{duscolcol} shows an excellent agreement between the two
models even with scattering, as long as the extinction is high.  All
test results were correct to within the statistical noise. 

For the extraction of the (color)profiles, the same programs as in
Paper~I were used and the model results can be compared directly with
the data. Once a $\tau_{0,V}$ and a set of scalelengths and heights have
been chosen for a particular model, $\kappa_{0,\lambda}$ is determined
by Eq.(~\ref{optdepth}) and the corresponding optical depths for the other
passbands are determined by Table~\ref{dusprop}.  This model is also
useful to study inclination-dependent effects of extinction.

\end{document}